\documentclass[12pt]{article} 
\usepackage{a41}
\usepackage{cite}
\usepackage{epsfig}
\usepackage{graphicx}
\usepackage{amsmath}
\usepackage{amssymb}         
\usepackage{color}  

\usepackage[rflt]{floatflt}              
\usepackage{float}
\usepackage{slashed}
\usepackage{epstopdf}

\def\asr{\left( \frac{\alpha_s}{4 \pi} \right)}
\def\b0{\beta_0}

\usepackage{array}

\newcommand{\HA}{{\rm H}}

\usepackage[english]{babel}
\usepackage[latin1]{inputenc}
\usepackage[T1]{fontenc}
\usepackage{ae}

\usepackage{url}

\usepackage{amsmath, amsthm, amssymb}
\newtheorem{thm}{Theorem}[section]

\newtheorem{definition}[thm]{Definition}

\newcommand{\Q}{\mathbb Q}

\newcommand{\Li}{{\rm Li}}

\newcommand{\ep}{\varepsilon}

\newcommand{\cM}{\cal M}
\newcommand{\cI}{\cal I}

\usepackage{rotating}

\usepackage{graphicx}

\newcounter{mmacnt}
\def\restartmma{\setcounter{mmacnt}{0}}
\restartmma \catcode`|=\active
\def|#1|{\mathrm{#1}}
\catcode`|=12
\newenvironment{mma}{
 \par\smallskip
 \catcode`|=\active
 \parskip=0pt\parindent=0pt 
 \small
 \def\In##1\\{%
\def\linebreak{\hfill\break\null\qquad}%
\refstepcounter{mmacnt}
\hangindent=2.5em\hangafter=0
\leavevmode
\llap{\tiny\sffamily n[\arabic{mmacnt}]:=\kern.5em}%
\mathversion{bold}\footnotesize$\displaystyle##1$\normalsize
\mathversion{normal}\par
 }%
 \def\Print##1\\{%
\def\linebreak{\hfill\break}%
\hangindent=2.5em\hangafter=0
\leavevmode ##1\par}%
 \def\Out##1\\{%
\def\linebreak{$\hfill\break\null\hfill$}%
\kern\abovedisplayskip\par
\hangindent=2.5em\hangafter=0
\leavevmode
\llap{\tiny\sffamily Out[\arabic{mmacnt}]=\kern.5em}
\footnotesize$\displaystyle##1$\normalsize\hfill\null\par
\kern\belowdisplayskip
 }%
 \def\Warning##1##2\\{%
\def\linebreak{\hfill\break}%
\hangindent=2.5em\hangafter=0
\leavevmode
{\scriptsize##1 : ##2}\par}%
}{%
 \par\smallskip
}

\usepackage{color}

\newenvironment{fshaded}{%
\MakeFramed {\FrameRestore}
}%
{\endMakeFramed}

\allowdisplaybreaks[4]

\begin{document}
\setlength{\baselineskip}{0.515cm}
\sloppy
\thispagestyle{empty}
\begin{flushleft}
DESY 18--053 
\\
DO--TH 18/09
\\
TIF-UNIMI-2018-8
\end{flushleft}

\mbox{}
\vspace*{\fill}
\begin{center}

{\LARGE\bf Automated Solution of First Order Factorizable}

\vspace*{4mm}
{\LARGE\bf  Systems of Differential Equations}

\vspace*{3mm}
{\LARGE\bf in One Variable} 

\vspace*{3mm} 

\vspace{3cm}
\large
{\large 
J.~Ablinger$^a$,
J.~Bl\"umlein$^b$, 
P.~Marquard$^b$,
N.~Rana$^{b,c}$
and 
C.~Schneider$^a$
}

\vspace{1.cm}
\normalsize
{\it $^a$~Research Institute for Symbolic Computation (RISC),\\
  Johannes Kepler University, Altenbergerstra{\ss}e 69,
  A--4040, Linz, Austria}

\vspace*{3mm}
{\it  $^b$ Deutsches Elektronen--Synchrotron, DESY,}\\
{\it  Platanenallee 6, D--15738 Zeuthen, Germany}

\vspace*{3mm}
{\it  $^c$ INFN, Sezione di Milano, Via Celoria 16, I--20133 Milano, Italy}


\end{center}
\normalsize
\vspace{\fill}
\begin{abstract}
\noindent
 We present an algorithm which allows to solve analytically linear systems of differential equations 
 which factorize to first order. The solution is given in terms of iterated integrals over an alphabet
 where its structure is implied by the coefficient matrix of the differential equations. These systems
 appear in a large variety of higher order calculations in perturbative Quantum Field Theories. We apply 
 this method to calculate the master integrals of the three--loop massive form factors for different currents,
 as an illustration, and present the results for the vector form factors in detail. Here the solution space 
 emerging is given by the cyclotomic harmonic polylogarithms and their associated special constants. No 
 special basis representation of the master integrals is needed. The algorithm can be applied as well to 
 more general cases factorizing at first order, which are based on more general alphabets, iterated
 integrals and associated constants.
\end{abstract}

\vspace*{\fill}
\noindent
\newpage 

\section{Introduction}
\label{sec:1}

\vspace*{1mm}
\noindent
The fundamental objects in any gauge theory are the scattering amplitudes or correlation functions,
as they allow to compute the scattering cross sections for collider experiments at large facilities like the 
Large Hadron 
Collider (LHC) at CERN. Computations of such objects are mostly using the diagrammatic approach. Especially,
in the case of perturbative Quantum Chromodynamics (QCD), one calculates these objects by obtaining all  
Feynman diagrams at each order in the expansion coefficient, the strong coupling constant $\alpha_s$.
Decades of dedicated work have made it possible to partially automate this procedure, from generating Feynman 
diagrams to the momentum-integral structure. The main remaining step consists in the  
computation of the integrals over loop momenta and to perform the associated Feynman parameter integrals. 

Through the reduction 
of the whole problem by integration-by-parts (IBP) techniques \cite{Lagrange:IBP,Gauss:IBP, Green:IBP, 
Ostrogradski:IBP,Chetyrkin:1981qh,Laporta:2001dd,REDUZE,CRUSHER} one obtains master integrals (MIs). 
One method to solve these integrals is the method of differential equations
\cite{Kotikov:1990kg,Remiddi:1997ny,Henn:2013pwa,Ablinger:2015tua}. 
Differentiating with respect to a parameter in the system one obtains coupled systems of ordinary differential 
equations of master integrals in the uni-variate case, 
with which we deal with in the 
following.\footnote{For a recent survey on the calculation methods for multi-loop integrals, see 
Ref.~\cite{Blumlein:2018cms}.} 
In the case where these systems factorize at 
first order, the complete solution can be constructed algorithmically. This has been done before in 
Ref.~\cite{Ablinger:2015tua} mapping to systems of difference equations, which also factorize to first order.
The solution has then been performed using difference ring and field technologies 
\cite{Karr:1981,Bron:00,Schneider:01,Schneider:04a,Schneider:05a,Schneider:05b,Schneider:07d,Schneider:10b,
Schneider:10c,Schneider:15a,Schneider:08c,Schneider:08d,Schneider:08e}, implemented in the package {\tt Sigma} 
\cite{Schneider:2007a,Schneider:2013a}.

In the present paper, we present an algorithm operating on uni-variate systems of differential equations,
which are factorizing at first order, directly. In the case where the factorization of the system leads to higher 
order sub-systems, elliptic and even more involved structures will appear, cf.~e.g.~\cite{Laporta:2004rb,
Bloch:2013tra,Adams:2015gva,Adams:2014vja,Adams:2016xah,Ablinger:2017bjx,Broedel:2017kkb,Broedel:2018rwm,BOOK}. Here 
still iterative solutions can 
be found. However, the corresponding integrals contain also letters, which are given by non-iterative integrals and 
therefore these solutions are given by {\it iterative non-iterative} integrals \cite{Blumlein:2016}.

The solution in the first-order factorizing case is given by iterative integrals over a certain alphabet 
${\mathfrak A} = \{f_1(x), \dots f_m(x)\}$ together with special constants. We will present the algorithm
for solving these systems, which does not require a special choice of a basis for the MIs, like the case in 
\cite{Henn:2013pwa}.

As an illustration, we employ this method of integration for computing the set of MIs which contribute 
to both the color--planar and complete light quark non--singlet three-loop contributions to the heavy-quark 
form factors for different currents, namely the vector, axial-vector, scalar and pseudo-scalar currents.
The massive form factors for vector and axial-vector currents play an important role in the forward-backward 
asymmetry of 
bottom or top quark pair production at electron-positron and hadron colliders.
The scalar and pseudo-scalar ones contribute to 
the decay of a Higgs boson to a pair of heavy quarks. They are also of importance to scrutinize the 
properties of the top quark \cite{Abe:1995hr, D0:1995jca} during the high luminosity phase of the LHC \cite{HLHC} 
and experimental precision studies at future high energy $e^+ e^-$ colliders \cite{Accomando:1997wt}.
The perturbative QCD contributions to these massive form factors at two loops were first 
computed in \cite{Bernreuther:2004ih,Bernreuther:2004th,Bernreuther:2005rw,Bernreuther:2005gw}.
Later an independent computation was performed in \cite{Gluza:2009yy} for the vector form factors,
additionally including ${\mathcal O}(\ep)$ terms in the dimensional parameter $\ep = (4-D)/2$.
Recently, the two-loop contributions up to 
${\mathcal O}(\ep^2)$ for all the massive form factors were obtained in \cite{Ablinger:2017hst}.
At three-loop level, the color--planar contributions to the vector form factors have been computed 
in \cite{Henn:2016tyf,Henn:2016kjz} and the complete light quark contributions in \cite{Lee:2018nxa}. 
Using the method described in this paper, we have obtained both the color--planar and complete 
light quark contributions to the three-loop form factors for the other three currents, namely
axial-vector, scalar and pseudo-scalar currents in \cite{Ablinger:2018yae}. In a parallel 
and independent computation in \cite{Lee:2018rgs} the same results have been obtained.
The asymptotic behaviour of the heavy quark form factors has been studied in \cite{Blumlein:2018tmz,Ahmed:2017gyt} 
recently, see also Refs.~\cite{Gluza:2009yy}.
The large $\beta_0$ limit for massive form factors has been considered in \cite{Grozin:2017aty}
and in \cite{Archambault:2004zs}, where the three-loop scalar and pseudo-scalar form factors were computed 
in the static limit.

The paper is organized as follows. In Section~2 we describe the algorithm to solve first-order factorizing 
single-variate differential equation systems and present an illustrative example. In Section~3 we consider 
the massive three-loop vector form factors in an arbitrary basis and present the corresponding analytic results in 
Section~4. In Section~5 a numerical representation for the cyclotomic harmonic polylogarithms (HPLs) up to
weight {\sf w = 6} is given, to allow the numerical evaluation of the massive three-loop form factors. Section~6 
contains the conclusions. The complete expressions for the vector form factors, which are very large, are given 
in ancillary files together with the code {\tt CPOLY.f} for the cyclotomic harmonic polylogarithms and 
other material, attached to this paper.
\section{Description of the method}
\label{sec:2}

\newcommand{\KK}{\mathbb K}

\vspace*{1mm}
\noindent
We consider $n$ master integrals (MIs) ${\cal I} = (I_1,\ldots,I_n)$ which belong to the same topology and 
are functions of the dimensional parameter $d = (4 - 2 \ep)$  and the variable $x$
\begin{equation}
\label{EQ:1}
s = \frac{q^2}{m^2} = -\frac{(1-x)^2}{x}.
\end{equation}
Here $q^2$ denotes the virtuality of the current and $m$ is the heavy quark mass.
One obtains an $n \times n$ system of coupled linear differential equations by taking the derivative for $x$ of 
each of the MIs followed by the IBP reduction,
\begin{equation}\label{Equ:InputSystem}
 \frac{d}{dx} {\cal I} = {\cal M ~ I + R}.
\end{equation}
Here the $n\times n$ matrix ${\cal M}$ consists of entries from the rational function field $\KK(d,x)$ (or equivalently from $\KK(\ep,x)$) where $\KK$ is a field of characteristic $0$; in the examples below the entries are even from $\Q(d,x)$ (or equivalently from $\Q(\ep,x)$). Furthermore,
the inhomogeneous part ${\cal R}=({\cal R}_1,\dots,{\cal R}_n)$ is composed by simpler master integrals whose evaluations are immediate or can be carried out by other methods, like symbolic summation and integration; see~\cite{Ablinger:2015tua} for details and references therein. In simpler situations ${\cal R}$ turns out to be just the $0$-vector. For more involved applications we will assume that each entry ${\cal R}_i$ is expanded into a Laurent series\footnote{In the following $f^{(k)}$ does not denote the $k$th derivative of $f$.} in $\ep$
$${\cal R}_i = \sum_{j=-k}^{\infty} \ep^j {\cal R}_i^{(j)}$$
up to a certain order in terms of special functions. More precisely, we assume that the first coefficients ${\cal R}_i^{(j)}$ are given as polynomial expressions with coefficients from $\KK$ in terms hyperexponential functions
and iterative integrals over such functions; for a detailed definition see below. Furthermore, we assume that 
the unknown integrals ${\cal I}_i$ can be expanded in an $\ep$-expansion
\begin{equation}\label{Equ:ExpansionIR}
{\cal I}_i = \sum_{j=-k}^{\infty} \ep^j {\cal I}_i^{(j)}.
\end{equation}
This applies to the case that ${\cal M}$ has no poles in $\ep$. If this is not the case, according index-shifts
have to be performed. Here it may happen that part of the equations which have to be solved, are no differential 
equations but are algebraic.
Given such a coupled system, we seek for the first coefficients ${\cal I}_i^{(j)}$ in the form of polynomial expressions in terms of hyperexponential functions and iterative integrals over such functions.
\medskip

\noindent\textbf{Definition.}
A function $f(x)$ is called \textit{hyperexponential} if $\frac{\frac{d}{dx}f(x)}{f(x)}=r(x)$ is a rational function in $\KK(x)$. Such a function may be given in the form
$$f(x)=e^{\int_l^x\!\!dy\,r(y)}$$
for some properly chosen $l\in\KK$. An \textit{iterative integral over hyperexponential functions} is an integral of the form
\begin{equation}\label{Equ:FirstParticularRep}
\int_{l_0}^x\!\!\! dx_1\, f_1(x_1)\int_{l_1}^{x_1}\!\!\!dx_2\,f_2(x_2)\dots\int_{l_{\lambda-1}}^{x_{\lambda-1}}\!\!\!dx_{\lambda}\,f_{\lambda}(x_{\lambda})
\end{equation}
where $f_1(x),\dots,f_{\lambda}(x)$ are hyperexponential functions and the lower bounds $l_0,\dots,l_{\lambda-1}\in\KK$ are appropriately chosen.

The class of hyperexponential functions covers functions of the form $q(x)^{\mu}$ where $q(x)\in\KK(x)$ and 
$\mu\in\KK$. Note that in all our calculations that arose so far, we only dealt with the special case $\mu\in\Q$. In the following we will use the property that $f(x)\,g(x)$, $\frac1{f(x)}$ with $f\neq0$ and $\frac{d}{dx}f(x)$ are hyperexponential functions provided that $f(x)$ and $g(x)$ are hyperexponential functions. In addition, $\frac{d}{dx}$ acting on the iterative integral~\eqref{Equ:FirstParticularRep} simply removes the outermost integral. As a consequence, applying the derivative to a polynomial expression in terms of hyperexponential functions and iterative integrals over such functions will lead again to a polynomial expression in terms of such functions.  
Furthermore, the multiplication of two iterative integrals over hyperexponential functions can be written as a 
linear combination of iterative integrals over hyperexponential functions due to its shuffle algebra~\cite{Blumlein:2003gb}. 
Consequently also a polynomial expression in terms of hyperexponential functions and iterative integrals over hyperexponential functions can be always written as a linear combination of the form
$h_1(x)I_1(x)+\dots h_{\lambda}(x)I_{\lambda}(x)$
where the $I_i(x)$ are iterative integrals over hyperexponential functions and the $h_i(x)$ are hyperexponential functions.

A general assumption of our method will be that the degree of uncoupling will be of first order. More precisely, we will apply internally Z\"urcher's algorithm 
\cite{Zuercher:94,BCP13,NewUncouplingMethod} implemented in the package {\tt OreSys} \cite{ORESYS}
in order to decompose the system into one scalar linear differential equation (sometimes also several such equations) determining all unknown functions.
In the case that the scalar equations (evaluated at $\ep=0$) are first-order factorizable, we proceed. 

In general, the dimension $n$ of the system~\eqref{Equ:InputSystem} is rather high (e.g., $n=100$). In this matter we note that the MIs can be distinguished sector-wise. 
A sector is defined by a set of maximum number of non-vanishing 
propagators in a single Feynman graph. Correspondingly, the absence of some propagators defines 
sub-sectors. The differential equation of a MI hence only contains integrals from the same sector or its 
sub-sectors. Thus, organizing the integrals in a way such that integrals with a minimum number of 
propagators are kept at the end of the list, provides an upper-block-triangular form of ${\cal M}$,
i.e. the diagonal elements of ${\cal M}$ are square matrices of not only rank one but higher.
Each such square matrix represents a completely coupled set of integrals and we call them sub-systems of 
${\cal M}$. The advantage of arranging the system in this way is that now we can solve the system in a
bottom-up approach, \textit{i.e.} we first solve for the last set of coupled integrals in the list which 
depend on themselves only (plus inhomogeneous parts from ${\cal R}$ that are already expanded in terms of 
special functions), and then solve the second last set of coupled integrals, which depends on themselves 
and the last integrals and thus going up in the list. 

\vspace*{4mm}
\noindent
We will now elaborate the different steps of our proposed algorithm. Similar ideas have been utilized already 
in~Refs.~\cite{BKSF:12,Ablinger:2015tua,Bluemlein:2014qka,CoupledSys:15} in order to find solutions in terms of 
iterative sums over hypergeometric products.

\vspace*{1mm} \noindent
{\bf 1.}~Let us consider $m$ integrals ${\cal \tilde I}=(\tilde I_1,\ldots, \tilde I_m)$ 
which constitute a coupled sub-system, 
\begin{equation} \label{eq:detilde}
 \frac{d}{dx} {\cal \tilde I} = {\cal \tilde M ~ \tilde I + \tilde R}, 
\end{equation}
where the non-diagonal elements of ${\cal \tilde M}$ are mostly non-zero and are rational functions from $\KK(d,x)$, or 
equivalently from $\KK(\ep,x)$. In particular, we may assume that ${\cal \tilde M}$ is an invertible matrix; if not, one can derive an alternative system by simple row operations with this property (the new system consists of less unknown integrals and the redundant integrals, that are removed from the system, can be expressed trivially by the integrals that arise in the new system). The inhomogeneity ${\cal \tilde R}$ is formed by contributions from integrals belonging to sub-sectors and the components of ${\cal R}$. By construction we succeeded already in calculating the first coefficients of the $\ep$-expansions of these integrals in terms of iterative integrals over hyperexponential functions. Consequently, plugging these results into ${\cal R}$ yields the $\ep$-expansions 
$${\cal \tilde R}_i = \sum_{j=-k}^{\infty} \ep^j {\cal \tilde R}_i^{(j)}$$
for $1\leq i\leq m$
where the first coefficients ${\cal \tilde R}_i^{(j)}$ are given explicitly in terms of iterative integrals over hyperexponential functions.\\
Now we exploit the fact that for a certain topology and kinematics, the order $k$ of highest pole of an integral in 
${\cal\tilde I}$ is well-defined,
as \textit{e.g.} the integrals arising in three-loop massive form factors can have at most a pole of $1/\ep^3$. 
Hence, one has the following Laurent expansions
\begin{equation}
 {\cal \tilde I}_i = \sum_{j=-k}^{\infty} \ep^j {\cal \tilde I}_i^{(j)}.
\end{equation}
In order to determine the first coefficients ${\cal \tilde I}_i$ in terms of iterative integrals, we proceed 
as follows.
We plug in~\eqref{Equ:ExpansionIR} with undetermined coefficients ${\cal \tilde I}_i$ into~\eqref{eq:detilde}, perform the series expansion in $\ep$ and consider the 
coefficient of $\ep^{k}$ :
\begin{equation} \label{eq:detildek}
 \frac{d}{dx} {\cal \tilde I}^{(k)} = {\tilde \cM^{(0)} ~ \tilde \cI^{(k)} 
 + \Big( \tilde \cM^{(1)} ~ \tilde \cI^{(k-1)} + \tilde \cM^{(2)} ~ \tilde \cI^{(k-2)} + \cdots 
+ \tilde \cM^{(k+l)} ~ \tilde \cI^{(-l)} \Big) + \tilde {\cal R}^{(k)}}, 
\end{equation}
for $k=-l,-l+1, etc.$ 

\vspace*{2mm}
\noindent
{\bf 2.}~At each order in the $\ep$-expansion we have now functions of a single variable $x$ only. Solving order 
by order, one obtains ${\cal \tilde I}$ as a Laurent series expansion in $\ep$. To accomplish that we start with 
the coefficient of the leading pole 
$\ep^{-l}$. The corresponding sub-system is
\begin{equation} \label{eq:detildeml}
 \frac{d}{dx} {\cal \tilde I}^{(-l)} = {\tilde \cM^{(0)} ~ \tilde \cI^{(-l)} + \tilde {\cal R}^{(-l)}} \,. 
\end{equation}
To solve Eq.~(\ref{eq:detildeml}), a natural first step is to reduce 
this $m \times m$ system to a higher order differential equation for a single integral. We will refer 
to this procedure as `uncoupling' from now on. By using the package {\tt OreSys} one obtains 
\begin{equation} 
\label{eq:MAST}
\sum_{k=0}^{m} p_k(x)\frac{d^k}{dx^k} {\cal \tilde I}^{(-l)}_1(x) = r(x).
\end{equation}
Here $p_l(x)$ are rational functions in $\KK(x)$ and 
\begin{equation} \label{eq:X1}
r(x)=\sum_{i=0}^{\lambda}\sum_{j=1}^m r_{i,j}(x)\,\frac{d^i}{dx^i}{\cal \tilde R}^{(-l)}_j(x)
\end{equation}
for some integer $\lambda$.
Since the differentiation of iterative integrals over hyperexponential functions yields again iterative integrals over hyperexponential functions, the inhomogeneous part $r(x)$ can be given explicitly in terms of iterative integrals over hyperexponential functions. Besides this scalar differential equation, the package {\tt 
OreSys} provides in addition the solutions $\left. {\cal \tilde I}^{(-l)}_k(x)\right|_{k=2}^m$ in terms of linear combinations of ${\cal \tilde I}^{(-l)}_1(x)$ and its derivatives:
\begin{equation}\label{Equ:LinComb}
{\cal \tilde I}^{(-l)}_k(x)=\sum_{i=0}^{m-1}a_{k,i}(x)\,\frac{d^i}{dx^i}{\cal \tilde I}^{(-l)}_k(x)+\rho_k(x)
\end{equation}
with $a_{k,i}\in\KK(x)$. Like the $r(x)$ in~\eqref{eq:X1} the $\rho_k(x)$ can be given in such a form. 
Consequently, also the $\rho_k(x)$ can be expressed explicitly in terms of iterative integrals over hyperexponential 
functions.
In other words, if one succeeds in solving the linear differential equations~\eqref{eq:MAST} and obtains the solution for
${\cal \tilde I}^{(-l)}_1(x)$ in terms of iterative integrals over hyperexponential functions, one can plug  this closed form into~\eqref{Equ:LinComb} and can extract such an integral representation of the remaining functions $\left. {\cal \tilde I}^{(-l)}_k(x)\right|_{k=2}^{m}$. 
Concerning the uncoupling we remark that it is not advised using the classical 
cyclic vector algorithm to achieve the uncoupling, since generally this method provides uncoupled 
equations with large coefficients. Moreover, it is beneficial that Z\"urcher's algorithm may find several linear differential equations for several of the unknown functions: they have usually smaller orders than the cyclic vector algorithm (which always finds only one differential equation). The solving tools are now applied to each of the found equations. For simplicity we assume in the following that only one scalar differential equation for ${\cal \tilde I}^{(-l)}_1(x)$ is produced.

\vspace*{1mm}
\noindent
{\bf 3.} From Eq.~(\ref{eq:detildek}) it is evident that the homogeneous solutions are always the same for
any order in $\ep$. The inhomogeneous solutions are different, however, by action of the inhomogeneities.
We first consider the homogeneous 
solutions of Eq.~(\ref{eq:MAST}). First we check if the differential equation can be factorized into first-order factors of the form
\begin{equation}
\label{eq:FACT}
\left(\frac{d}{dx} - \hat{p}_1(x)\right)
\left(\frac{d}{dx} - \hat{p}_2(x)\right) 
{\dots} 
\left(\frac{d}{dx} - \hat{p}_m(x)\right) y_1(x) = 0,
\end{equation}
with $\hat{p}_k$ being rational functions in $\KK(x)$
by using algorithms from~\cite{Singer:91,Bronstein:92,Hoeij:97}; for more details see~\cite[Chapter~4]{PutSinger:03}. 
If this is possible, we proceed as follows. Define for $1\leq k\leq m$ the hyperexponential functions
\begin{equation}
h_k(x)=e^{\int_{l'_k}^x\!\!dy\,\hat{p}_k(y)}
\end{equation}
for some appropriate lower bounds $l'_k\in\KK$, that are solutions of the $k$th first-order factor, i.e., 
\begin{equation}
\left(\frac{d}{dx} - \hat{p}_k(x)\right) h_k(x)=0\quad\Longleftrightarrow\quad\frac{\frac{d}{dx}h_k(x)}
{h_k(x)}=\hat{p}_k(x).
\end{equation}
Finally, one can read off from the factorization~\eqref{eq:FACT} the $m$ solutions
\begin{equation}\label{Equ:HomSol}
\begin{split}
y_1(x)=&h_1(x),\\
y_2(x)=&h_1(x)\int_{l_0}^x\!\!\! dx_1\, \frac{h_2(x_1)}{h_1(x_1)},\\
\vdots&\\
y_m(x)=&h_1(x)\int_{l_0}^x\!\!\! dx_1\, \frac{h_2(x_1)}
{h_1(x_1)}\int_{l_1}^{x_1}\!\!\!dx_2\,\frac{h_3(x_2)}{h_2(x_2)}
\dots\int_{l_{m-2}}^{x_{m-2}}\!\!\!dx_{m-1}\,\frac{h_m(x_{m-1})}{h_{m-1}(x_{m-1})},
\end{split}
\end{equation}
where the lower bounds $l_0,\dots,l_{m-2}$ are chosen accordingly. These solutions, also called 
d'Alembertian solutions~\cite{Alembertian:94}, form iterative integrals over hyperexponential functions $\frac{h_k(x)}{h_{k-1}(x)}$. Since they are linearly independent over $\KK$, see~\cite[Thm.~5]{Alembertian:94}, 
$$\{C_1\,y_1(x)+\dots+C_m\,y_m(x)\,\mid\,C_1,\dots,C_m\in\KK\}$$ 
yield the full solution space of the homogeneous recurrence~\eqref{eq:FACT}.
In the calculations presented below the hyperexponential functions $h_k(x)$ can be simplified all to rational functions in $\KK$. Even more is true after further simplifications: The integrands can be decomposed into the form 
\begin{equation}\label{Equ:PFD}
\frac{h_k(x)}{h_{k-1}(x)} =  q_{k} + \sum_l q_{k,l} \phi_l(x),~~~~q_k, q_{k,l} \in\KK
\end{equation}
by partial fractioning with $\phi_l(x)=\frac{\alpha_l(x)}{\beta_l(x)^{e_l}}$ where the $\beta_l(x)$ are irreducible polynomials in $\KK[x]$ and $\alpha_l(x)$ are polynomials in $\KK[x]$ with 
$\deg(\alpha_l(x))<\deg(\beta_l(x))$ and $e_l \in \mathbb{N} \backslash \{0\}$.

Let us consider a typical example. Usually the functions $\hat{p}_k(x)$ in Eq.~(\ref{eq:FACT}) are rational 
functions, which factor into the letters $f_l(x)$ of an alphabet ${\mathfrak A}'$.
If these letters are the ones of the Kummer-Poincar\'e type \cite{KUMPO} the ratios $h_m/h_{m-1}$ in 
Eqs.~(\ref{Equ:HomSol}) are again Kummer-Poincar\'e letters after partial fractioning. This representation holds 
as
well for cyclotomic harmonic polylogarithms, since these have complex representations by Kummer-Poincar\'e
letters.

By linearity one can now apply the integration sign to each of the summands in~\eqref{Equ:PFD} and eliminate algebraic relations among the arising integrals utilizing their shuffle relations; these ideas have been 
elaborated in detail for the sum case~\cite{Blumlein:2003gb,AS:18}. In particular, the multiplicity $e_l$ 
can be reduced 
upon noting
\begin{equation} 
\int dx \frac{1}{x^k} = \frac{-1}{(k-1)} \frac{1}{x^{k-1}},~~
\int dx \frac{1}{(1 \pm x)^k} = \frac{\mp 1}{(k-1)} \frac{1}{(1 \pm x)^{k-1}}.
\end{equation}
Likewise, the cyclotomic letters integrate to structures like
\begin{eqnarray} 
\label{eq:CHP1}
\int dx \frac{1}{(1-x+x^2)^3} &=& \frac{2x-1}{6 (1-x+x^2)^2} +\frac{2x-1}{3 (1-x+x^2)} + \frac{2}{3}
\int dx\,\frac{1}{1-x+x^2}
\\
\label{eq:CHP2}
\int dx \frac{x}{(1+x+x^2)^2} &=& -\frac{x+2}{3 (1+x+x^2)} - \frac{1}{3}
\int dx \,\frac{1}{1+x+x^2},~~{\rm etc.}
\end{eqnarray}
In the course of these simplifications also products of these functions and corresponding harmonic polylogarithms 
arise that can be joined using shuffle relations \cite{Blumlein:2003gb}. In addition to the cyclotomic harmonic polylogarithms also their corresponding values at 
$x=1$ contribute through partial integration. In the case of the harmonic polylogarithms these are the multiple 
zeta values (MZVs) \cite{Blumlein:2009cf}. In the cyclotomic case they are given by the cyclotomic 
constants \cite{Broadhurst:1998rz,Kalmykov:2010xv,Ablinger:2011te,Ablinger:2017tqs}. In \cite{Henn:2015sem} 
relations beyond those known from \cite{Broadhurst:1998rz,Kalmykov:2010xv,Ablinger:2011te,Ablinger:2017tqs}
already have been conjectured using {\tt PSLQ} \cite{PSLQ}.
More generally, this tactic can be applied in combination with the Almkvist--Zeilberger algorithm~\cite{AZ:90} 
if in addition hyperexponential functions arise that cannot be handled by the simplifications described above. In summary, 
a homogeneous linear differential equation stored in the 
variable \texttt{de} in terms of the unknown function $f(x)$ can be solved in terms of d'Alembertian solutions by executing the \texttt{HarmonicSums} command 
$$\texttt{SolveDE[de,f[x],x]}.$$
In particular, if a factorization of the form~\eqref{eq:FACT} exists for the given differential operator, it will be computed and the full set of solutions~\eqref{Equ:HomSol} will be produced where all the simplifications described above are applied.\\
In all our applications so far, 
the homogeneous solutions $y_i (x), i =1, \ldots, m$ could be expressed in terms of iterative integrals
\begin{equation}\label{Equ:HIntegral}
{\cal H}_{b,\vec{a}}(x) = \int_0^x dy f_b(y) {\cal H}_{b,\vec{a}}(y),~~~{\cal H}_\emptyset = 1,
\end{equation}
where $f_b(y)\in\mathfrak{A}$ are rational functions (or roots of rational functions) taken from a finite 
alphabet $\mathfrak{A}$. For instance, for the massive three-loop form factor discussed below, the alphabet can 
be 
chosen by 
\begin{equation}\label{Equ:Alphabet}
\mathfrak{A} = \left\{\tfrac{1}{x},~\tfrac{1}{1-x},
~\tfrac{1}{1+x}, 
~\tfrac{1}{1+x^2}, 
~\tfrac{x}{1+x^2}, 
~\tfrac{1}{1+x+x^2},~\tfrac{x}{1+x+x^2},
~\tfrac{1}{1-x+x^2},~\tfrac{x}{1-x+x^2} \right\},
\end{equation}
where $f_b(x)$ corresponds to the $b$th entry.
Summarizing, in our concrete application below the arising d'Alembertian solutions~\eqref{Equ:HomSol} will be simplified to expressions in terms of the class of
harmonic polylogarithms \cite{Remiddi:1999ew} and the cyclotomic harmonic 
polylogarithms \cite{Ablinger:2011te}.

\vspace*{1mm}
\noindent
{\bf 4.} The solution of the inhomogeneous differential
equation (\ref{eq:MAST}) can be given explicitly by the following iterative integral~\cite{Alembertian:94} 
\begin{equation}
g(x)=h_1(x)\int_{l_0}^x\!\!\! dx_1\, \frac{h_2(x_1)}{h_1(x_1)}\int_{l_1}^{x_1}\!\!\!dx_2\,\frac{h_3(x_2)}{h_2(x_2)}\dots\int_{l_{m-2}}^{x_{m-2}}\!\!\!dx_{m-1}\,\frac{h_m(x_{m-1})}{h_{m-1}(x_{m-1})}\int_{l_{m-1}}^{x_{m-1}}\!\!\!dx_{m}\,\frac{r(x_{m})}{h_{m}(x_{m})}.
\end{equation}
Consequently,
\begin{equation}\label{Equ:CompleteSolution}
{\cal \tilde I}^{(-l)}_1(x)=g(x)+C_1\,y_1(x)+\dots+C_m\,y_m(x)
\end{equation}
where the constants $C_i$ are implied by (physical) boundary conditions and 
they are usually determined by separate calculations. Since the inhomogeneous part $r(x)$ can be given in terms of iterative integrals over hyperexponential functions, also $g(x)$ and thus ${\cal \tilde I}^{(-l)}_1(x)$  can be expressed in terms of iterative integral over hyperexponential functions. 
Furthermore, using our simplification tools from above, these integrals can be simplified further. E.g., within all our calculations we end up at alphabets of the form~\eqref{Equ:Alphabet} or variants involving also rooted letters.\\
We want to emphasize an alternative approach to find a particular solution $g(x)$ of~\eqref{eq:MAST}. If
\begin{equation}
W(x) = \left|
\begin{array}{ccc}    
        y_1 & \hdots & y_m\\
        \frac{d}{dx}y_1^{(1)} & \hdots & \frac{d}{dx}y_m\\
        \vdots &  & \vdots\\
        \frac{d^{m-1}}{dx^{m-1}}y_1 &  & \frac{d^{m-1}}{dx^{m-1}}y_m\\
\end{array} \right|
\end{equation}
is the Wronskian of the linear differential equation~\eqref{eq:MAST} and 
\begin{equation}
W_i(x) = (-1)^{i+m} \left|
\begin{array}{cccccc}    
        y_1         & \hdots & y_{i-1}         & y_{i+1}         & \hdots & y_m\\
        \frac{d}{dx}y_1   & \hdots & \frac{d}{dx}y_{i-1}   & \frac{d}{dx}y_{i+1}   & \hdots & \frac{d}{dx}y_m\\
        \vdots      &        & \vdots          & \vdots          &        & \vdots    \\
        \frac{d^{m-2}}{dx^{m-2}}y_1 & \hdots & \frac{d^{m-2}}{dx^{m-2}}y_{i-1} & \frac{d^{m-2}}{dx^{m-2}}y_{i+1} & \hdots & \frac{d^{m-2}}{dx^{m-2}}y_m\\
\end{array} \right|, \end{equation} 
then 
\begin{equation}\label{Equ:SecondParticularRep}
 g(x) = \sum_{i=1}^m y_i (x)\int_l^{x} d\tilde{x} \frac{r(\tilde{x}) W_i(\tilde{x})}{W(\tilde{x})}
\end{equation}
for some appropriately chosen $l\in\KK$
yields another particular solution. Note that by (a mild generalization) of Abel's theorem we have that $W(x)$ itself can be written as a hyperexponential function
$$W(x)=c\,e^{-\int_{l}^x\!\!dy \frac{p_{m-1}(y)}{p_m(y)}}$$
for some constant $c\in\KK$ and an appropriately chosen lower bound $l\in\KK$; the polynomials $p_m(x),p_{m-1}(x)\in\KK[x]$ come from the linear differential equation~\eqref{eq:MAST}.
Furthermore, the $W_i(x)$ are given by polynomial expressions in terms of the homogeneous solutions $y_i(x)$. As a consequence
$\frac{r(\tilde{x}) W_i(\tilde{x})}{W(\tilde{x})}$ in~\eqref{Equ:SecondParticularRep} forms a polynomial expression in terms of hyperexponential functions and iterative integrals over such functions. In particular, $g(x)$ yields such a representation.\\ 
The following extra bonus often makes the formula~\eqref{Equ:SecondParticularRep} superior to~\eqref{Equ:FirstParticularRep}: By reusing the simplified homogeneous solutions 
$y_1(x),\dots,y_m(x)$ for~\eqref{Equ:SecondParticularRep}, it is much easier to obtain a simplification of~\eqref{Equ:SecondParticularRep} than of~\eqref{Equ:FirstParticularRep} in terms of iterative integrals of the form~\eqref{Equ:HIntegral} with alphabets like~\eqref{Equ:Alphabet}.

\vspace*{1mm}
\noindent
{\bf 5.} 
Now we plug this representation of ${\cal \tilde I}^{(-l)}_1(x)$ in terms of iterative integrals 
into~\eqref{Equ:LinComb} for $k=2,\dots,m$. Since the derivation of iterative integrals over hyperexponential functions yields again iterative integrals over hyperexponential functions, all entries in the vector ${\cal \tilde I}^{(-l)}_1(x)$ can be given within the class of iterative integrals over hyperexponential functions.

\vspace*{1mm}
\noindent
{\bf 6.} Finally, we plug this representation of ${\cal \tilde I}^{(-l)}(x)$ in terms of iterative integrals 
into~\eqref{eq:detildek} for $k=-l+1$ and obtain a new system of the form~\eqref{eq:detildeml} for the 
$\ep^{-l+1}$-coefficient ${\cal \tilde I}^{(-l+1)}=({\cal \tilde I}_1^{(-l+1)}(x),\dots,{\cal \tilde I}_m^{(-l+1)}(x))$. Thus we repeat the game for $\ep^{-l+1}$ and the remaining coefficients in~\eqref{Equ:ExpansionIR} by induction/recursion. We note once more that the formula~\eqref{Equ:CompleteSolution} remains the same, except that in~\eqref{Equ:SecondParticularRep} the function $r(x)$ changes. As a consequence one can again reuse the already simplified homogeneous solutions $y_1(x),\dots,y_m(x)$ and just needs to simplify $g(x)$ in~\eqref{Equ:SecondParticularRep} with the updated function $r(x)$.

\vspace*{1mm}
\noindent
Let us illustrate the above algorithm by an example, which concerns the solution of a sub-system in the 
calculation of the three loop massive form factors in the color planar limit.

\vspace*{5mm}
\noindent
\textbf{Example.} \\
We consider the following system of differential equations for the integrals 
$\{J_1,J_2,J_3\} \in 
{\cal I}$:
\begin{align}
\label{EQ:DI}
  \begin{split}
    \frac{d}{d x} \left(
    \begin{array}{l}
      J_{1}\\
      J_{2}\\
      J_{3}
    \end{array}
    \right)&=
    \left[
      \begin{array}{ccc}
        c_{11} & c_{12} & c_{13}\\
        c_{21} & c_{22} & c_{23}\\
        c_{31} & c_{32} & c_{33}\\
      \end{array}
      \right]
    \left(
    \begin{array}{l}
      J_{1}\\
      J_{2}\\
      J_{3}
    \end{array}
    \right)
    +\left(
    \begin{array}{l}
      R_{1}(\epsilon,x)\\
      R_{2}(\epsilon,x)\\
      R_{3}(\epsilon,x)
    \end{array}
    \right),
  \end{split}
\end{align}
where  $c_{ij}$'s are rational functions in $d$, resp. $\ep$, and $x$ as given by
\begin{align}
 c_{11} &= \frac{\big(7+6 x+7 x^2-2 d \big(1+x+x^2\big)\big)}{x (1-x^2)} \,,
 c_{12} = \frac{(-4+d) (-10+3 d)}{2 (-3+d)^2 (1-x^2)}\,,
 \nonumber \\ 
 c_{13} &=  \frac{\big(
        d^2 \big(15+8 x+15 x^2\big)
        +8 \big(20+9 x+20 x^2\big)
        -2 d \big(49+24 x+49 x^2\big)\big)}{4 (-3+d)^2 x (1-x^2)} \,,
 \nonumber \\
c_{21} &= \frac{(-3+d)^2 \big(  d (-3+x) (-1+3 x) -2 \big( 5-18 x+5 x^2\big)\big)}{(-10+3 d) x (1-x^2)} \,,
c_{22} = \frac{(-7+2 d) \big(  1+x^2\big)}{x (1-x^2)}\,,
\nonumber \\
c_{23} &= \frac{\big( -30  +188 x -30 x^2 +d^2 \big( -3+16 x-3 x^2\big) +d \big( 19-110 x+19 x^2\big)\big)}{(-10+3 d) x (1-x^2)} \,,
\nonumber\\
c_{31} &= -\frac{(-3+d)^2 (1+x)}{x(1-x)} \,,
c_{32} = 0 \,,
c_{33} = \frac{2 (-3+d) \big( 1+x+x^2\big)}{x (1-x^2)} \,.
\end{align}
The functions $R_{i}(\ep,x)$ contain the inhomogeneous contributions from sub-sectors. They can be 
expanded into a Laurent series expansion in $\ep$ up to the required order and read 
\begin{align}
R_1(\ep,x) &= \frac{1}{3(1-x^2)} \frac{1}{\ep^3} - \frac{1 - x}{6 x (1 + x)} \frac{1}{\ep^2}
- \left[\frac{2 + 11 x + 2 x^2}{3 x (1-x^2)} - \frac{9 \zeta_2}{2(1-x^2)}\right] \frac{1}{\ep}
\nonumber\\ &
-\frac{1 - 4 x + 188 x^2 - 4 x^3 + x^4}{6 x^2(1-x^2)}
 - \frac{(1 - 34 x + x^2) \zeta_2}{4 x(1-x^2)} + \frac{31 \zeta_3}{3(1-x^2)}
+ \frac{2}{x} \HA_0 + \frac{1 - x}{2x(1+x)} \HA_0^2
\nonumber\\ &
- \frac{2}{3(1-x^2)} \HA_0^3
- \frac{8 \left(\tfrac{1}{2} \HA_0^2 \HA_1 - \HA_0 \HA_{0,1} + \HA_{0,0,1}\right)}
  {1-x^2} + O(\ep)
\\
R_2(\ep,x) &= \frac{1+x}{6x(1-x)} \frac{1}{\ep^3}
+ \frac{3 + 2 x + 3 x^2}{6 x(1-x^2)}  \frac{1}{\ep^2}
+\Biggl[-\frac{1 - 15 x + 16 x^2 - 15 x^3 + x^4}{6 x^2 (1 - x^2)} 
+ \frac{(1 + x) \zeta_2}{4x(1-x)} + \frac{\HA_0}{x} 
\nonumber\\ &
- \frac{(1 + 4 x + x^2)}{2x(1-x^2)} \HA_0^2\Biggr] \frac{1}{\ep}
-\frac{(1-x) (9-59 x+9 x^2)}{6 x^2 (1+x)}
-\frac{(1-2 x-7 x^2) \zeta_2}{4 x(1-x^2)}
\nonumber\\ &
+\frac{(25+110 x+25 x^2) \zeta_3}{6 x(1-x^2)}
+\left(-\frac{1-14 x+x^2}{2 x^2}+\frac{(1+4 x+x^2) \zeta_2}{x(1-x^2)}
\right) \HA_0
\nonumber\\ &
+\frac{1-3x-6x^2}{x(1-x^2)} \HA_0^2
-\frac{5+22 x+5 x^2}{6x (1-x^2)} \HA_0^3
-\frac{10}{x} \HA_{-1,0}
+\frac{4}{x} \left(\HA_0 \HA_1 - \HA_{0,1}\right)
+\frac{10+40 x+10 x^2}{x(1-x^2)} 
\nonumber\\ & \times
\left(
\HA_0 \HA_{-1,0} - 2 \HA_{-1,0,0} \right)
-\frac{4 (1+4 x+x^2)}{x(1-x^2)} \left(\HA_0 \HA_{0,1} -2 \HA_{0,0,1} \right)
+\frac{4 (1+3 x+x^2}{x(1-x^2)}
\Bigl(\frac{1}{2} \HA_0^2 \HA_1 
\nonumber\\ &
- \HA_0 \HA_{0,1} + \HA_{0,0,1}\Bigr)
+ O(\ep)
\\ 
R_3(\ep,x) &= 
\frac{1}{3(1-x^2)}\frac{1}{\ep^3} + \frac{1}{3(1-x^2)}\frac{1}{\ep^2}
+ \left[\frac{1 + 2 x - 8 x^2 + 2 x^3 + x^4}{6 x^2 (1-x^2)} + \frac{9 \zeta_2}{2(1-x^2)} \right] \frac{1}{\ep}
\nonumber\\ &
+ \frac{9+18 x-76 x^2+18 x^3+9 x^4}{6 x^2(1-x^2)}
+\frac{9 \zeta_2}{2(1-x^2)}+\frac{31 \zeta_3}{3(1-x^2)}
+\frac{1+4 x+x^2}{2 x^2} \HA_0
- \frac{1}{1-x^2} \HA_0^2
\nonumber\\ &
-\frac{2}{3 (1-x^2)} \HA_0^3
-\frac{8}{1-x^2} \left[\frac{1}{2} \HA_0^2 \HA_1 - \HA_0 \HA_{0,1} + \HA_{0,0,1}\right] + O(\ep).
\end{align}
Here we use the convention $\HA_{\vec{a}}(x) \equiv \HA_{\vec{a}}$ and $\zeta_l = \sum_{k=1}^\infty 1/k^l,~~l 
\in \mathbb{N}, l \geq 2$ denote the values of Riemann's
$\zeta$-function. The harmonic polylogarithms \cite{Remiddi:1999ew} are defined by
\begin{equation}
\HA_{b,\vec{a}}(x) = \int_0^x dy f_b(y) \HA_{\vec{a}}(y),~~\HA_\emptyset = 1,~~b, a_i \in \{-1,0,1\},
\end{equation}
and the letters $f_c$ are
\begin{equation}\label{eq:HPL1}
f_0(x) = \frac{1}{x},~~~~
f_1(x) = \frac{1}{1-x},~~~~
f_{-1}(x) = \frac{1}{1+x}.
\end{equation}
The HPLs are dual, by the Mellin transform, to the harmonic sums \cite{Vermaseren:1998uu,Blumlein:1998if}.

The solutions $J_i$ are calculated in terms of the following expansion in $\ep$
\begin{equation}
J_i(x,\ep) = 
  \frac{1}{\ep^3} J_i^{(-3)}
+ \frac{1}{\ep^2} J_i^{(-2)}
+ \frac{1}{\ep} J_i^{(-1)}
+ J_i^{(0)} + O(\ep).
\end{equation}
First, one obtains the determining equation for the leading pole $O(1/\ep^3)$:
\begin{align}
  \begin{split}
    \frac{d}{d x} \left(
    \begin{array}{l}
      J_{1}^{-3}\\
      J_{2}^{-3}\\
      J_{3}^{-3}
    \end{array}
    \right)&=
   - \left[
      \begin{array}{ccc}
        \frac{1}{x} + \frac{2}{1-x} & 0 & \frac{1}{1+x} - \frac{2}{x} - \frac{3}{1-x} \\
        -\frac{1}{x} + \frac{2}{1+x} & \frac{1}{1+x} - \frac{1}{x} - \frac{1}{1-x} & \frac{1}{x} - \frac{2}{1+x}\\
        \frac{1}{x} + \frac{2}{1-x} & 0 & \frac{1}{1+x} - \frac{2}{x} - \frac{3}{1-x}\\
      \end{array}
      \right]
    \left(
    \begin{array}{l}
      J_{1}^{-3}\\
      J_{2}^{-3}\\
      J_{3}^{-3}
    \end{array}
    \right)
    +\left(
    \begin{array}{l}
      R_{1}^{-3}(x)\\
      R_{2}^{-3}(x)\\
      R_{3}^{-3}(x)
    \end{array}
    \right).
  \end{split}   \label{eq:exepm3}
\end{align}
Using the incomplete Z\"urcher algorithm, one of four algorithms
implemented in {\tt OreSys}, we obtain for Eq.~(\ref{eq:exepm3}) two
uncoupled differential equations, one of order two for $J_3(x)$ and another of
first order for $J_2(x)$.  $J_1(x)$ can directly be obtained from the solution for $J_3(x)$. 

The second order differential equation for $J_3(x)$ 
\begin{align}
\label{eq:DEQ2}
 \bigg[ \frac{d^2}{d x^2} - \frac{2}{1-x} \frac{d}{d x} + \Big( \frac{2}{x} - \frac{2}{1+x} - \frac{2}{(1+x)^2} \Big) \bigg] J_{3}^{-3}(x)
 &= r_3^{-3} (x) 
\end{align}
with the inhomogeneous part
\begin{align}
 r_3^{-3} (x) = \frac{1}{3 x} - \frac{1}{3 (1+x)} - \frac{1}{3 (1+x)^2} 
\end{align}
needs to be solved first.  The differential operator in Eq.~(\ref{eq:DEQ2}) can be written in factorized form
\begin{eqnarray}  
D &=& \left(\frac{d}{dx} - p_1(x)\right) \circ \left(\frac{d}{dx} - p_2(x)\right), \\
p_1(x) &=& \frac{1}{1 - x} - \frac{1}{x} + \frac{1}{1 + x}, \\
p_2(x) &=& \frac{1}{1 - x} + \frac{1}{x} - \frac{1}{1 + x}. 
\end{eqnarray}
In this case the rational functions $\hat{p}_i(x)$ are linear combinations of the letters spanning the HPLs.
One now uses the method of the variation of the constants to obtain the solutions for the differential equation.
The homogeneous solutions $y_1(x), y_2 (x)$  are given by
\begin{equation}
 y_1 (x) = \frac{x}{1-x^2} \,, \quad y_2 (x) = 1 - \frac{2x}{1-x^2} \HA_0  \,,
\end{equation}
and we obtain the solution 
\begin{align}
\label{solJ3}
 & J_{3}^{-3} (x) = y_1(x) \bigg[ C_1 - 
\int dx \frac{r_3^{-3} (x) y_2(x)}{W(y_1,y_2)} \bigg]
         + y_2(x) \bigg[ C_2 + \int dx \frac{r_3^{-3} (x) y_1(x)}{W(y_1,y_2)} \bigg] \,.
\end{align}
where the constants $C_i$ can be determined from the physical boundary conditions and are known 
from a separate calculation.  With the Wronskian $W$  given by
\begin{equation}
W(y_1,y_2) = -\frac{1}{(1-x)^2} \,,
\end{equation}
the integrals in (\ref{solJ3}) are easily evaluated
\begin{eqnarray}
\int dx \frac{r_3^{-3} (x) y_2(x)}{W(y_1,y_2)} &=&
-\frac{2}{3(1+x)} - \frac{1+x^2}{3(1+x)^2} \HA_0,
\\
\int dx \frac{r_3^{-3} (x) y_1(x)}{W(y_1,y_2)} &=& - \frac{x}{3(1+x)^2} \,.
\end{eqnarray}
For the remaining constants we find 
\begin{equation}
  C_1 = - \frac{1}{3}\,,~~~C_2 = \frac{1}{6} \,,
\end{equation}
and thus 
\begin{align}
J_3^{-3} (x) &= \hspace*{3mm} \frac{1}{6} \,.
\end{align}
The solution for integral $J_1^{-3}(x)$ can directly be obtained from this result 
\begin{align}
 & J_1^{-3} (x) = \frac{x}{3 (1+x)^2} + \frac{2 \left(1+x+x^2\right)}{(1+x)^2} J_{3}^{-3} (x) - \frac{(1-x) x}{x+1} \frac{d}{d x} J_{3}^{-3} (x)  = \frac{1}{3} \,.
\end{align}
With these results at hand we can obtain the  first order differential equation for $J_{2}^{-3}(x)$
\begin{align}
 \bigg[ \frac{d}{d x} - \Big( \frac{1}{1-x} + \frac{1}{x} - \frac{1}{1+x} \Big) \bigg] J_{2}^{-3}(x) 
&= r_2^{-3} (x),
\end{align}
where the inhomogeneous part is given by
\begin{align}
 r_2^{-3} (x) = \frac{1}{3 x} - \frac{1}{3 (1+x)} + \frac{1}{3 (1-x)}.
\end{align}
The homogenous solution is given by 
\begin{equation*}
 y_3(x) = \frac{x}{1-x^2}
\end{equation*}
and thus we can obtain the full solution by evaluating 
\begin{align}
\label{solJ2}
 & J_2^{-3} (x) = y_3(x) \bigg( C_3 + \int dx \frac{r_2^{-3} (x)}{y_3(x)} \bigg) \,.
\end{align}
The integral is easily evaluated 
\begin{eqnarray}
\int dx \frac{r_2^{-3}(x)}{y_3(x)} &=& 
-\frac{1-x^2}{3 x} \,,
\end{eqnarray}
and after fixing the constant of integration 
\begin{equation}
  C_3 = 0 
\end{equation}
we obtain the final result 
\begin{align}
J_2^{-3} (x) &= -\frac{1}{3} \,.
\end{align}

To summarize, the results obtained so far, the solutions $J_1^{(-3)}, J_2^{(-3)}, J_3^{(-3)}$  are given by the  
numbers:
\begin{align}
J_1^{-3} (x) &= \hspace*{3mm} \frac{1}{3},
\\
J_2^{-3} (x) &= -\frac{1}{3},
\\
J_3^{-3} (x) &= \hspace*{3mm} \frac{1}{6}~.
\end{align}

Now, once the sub-system is solved for the highest pole, we consider the next order in $\ep$. By construction, the 
homogeneous structure of the sub-system remains the same for any order in $\ep$ and hence the uncoupling procedure. 
The 
only change that takes place for different orders in $\ep$ is in the inhomogeneous parts which also constitute the 
contributions from the already-known previous orders in $\ep$-expansion. Thus, for higher orders in $\ep$, the only 
step is to iterate Eqs.~(\ref{solJ3}) and (\ref{solJ2}). In this way we obtain the following solutions for the functions 
$J_1, J_2, 
J_3$ up to $O(\ep^0)$. Here we use the basis for the harmonic polylogarithms defined in \cite{Ablinger:2018sat}.
\begin{align}
J_1^{-2} (x) &= \frac{5}{3} 
\\ 
J_2^{-2} (x) &= -2 
\\ 
J_3^{-2} (x) &= \frac{1}{2} 
\\
J_1^{-1} (x) &= \frac{1+22 x+x^2}{6 x} +\frac{9}{2} \zeta_2  
\\ 
J_2^{-1} (x) &= -\frac{28}{3} +\frac{1}{2} \HA_{0}^2 -\frac{2 x}{3(1-x^2)} \HA_{0}^3 
-\frac{5}{2} \zeta_2 - \frac{4 x \zeta_2}{1-x^2} \HA_0 
\\ 
J_3^{-1} (x) &= \frac{1}{6} + \frac{9}{4} \zeta_2 
\\ 
J_{1}^{(0)} (x) &= \frac{13-46 x+13 x^2}{6 x}
+\frac{1-x^2}{2x} \HA_0
+\frac{1}{2} \HA_0^2
+\frac{1+3 x^2}{3 (1-x^2)} \HA_0^3
-\frac{x}{4 (1-x^2)} \HA_0^4
\nonumber\\ &
+\left(
        -2 \HA_0^2
        -\frac{4 x \HA_0^3}{3 (1-x^2)}
\right) \HA_1
+\left(
        4 \HA_0
        +\frac{2 x}{1-x^2} \HA_0^2
\right) \HA_{0,1}
-4 \HA_{0,0,1}
-\frac{4 x}{1-x^2} \HA_{0,0,0,1}
\nonumber\\ &
+\left(
        \frac{49}{2}
        +\frac{4 \big(
                1+x^2\big)}{1-x^2} \HA_0
        -\frac{2 x}{1-x^2} \HA_0^2
        -\frac{8 x}{1-x^2} \HA_0 \HA_1
        +\frac{8 x}{1-x^2} \HA_{0,1}
\right) \zeta_2
-\frac{32 x \zeta_2^2}{5 (1-x^2)}
+\frac{31}{3} \zeta_3
\\ 
J_{2}^{(0)} (x) &= -40
-\frac{16 x}{1-x^2} \big(\HA_ 0 \HA_{0,0,1} - 3 \HA_{0,0,0,1}\big)
+\frac{16 x}{1-x^2} \Biggl[\frac{1}{6} \big(
                \HA_ 0 \HA_ 1 - \HA_{0,1}\big) \HA_0^2
        -\frac{1}{3} \HA_0^2 \HA_{0,1} + \HA_0 \HA_{0,0,1}
\nonumber\\ &
        -\HA_{0,0,0,1}\Biggr]
+\frac{16 x}{1-x^2} \Biggl[\frac{1}{2} \HA_0^2 \HA_{0,1}
        -2 \HA_0 \HA_{0,0,1}
        +3 \HA_{0,0,0,1}
\Biggr]
+\frac{40 x}{1-x^2} \Biggl[
        \frac{1}{2} \HA_0^2 \HA_{-1,0}
        -2 \HA_0 \HA_{-1,0,0}
\nonumber\\ &
        +3 \HA_{-1,0,0,0}
\Biggr]
+4 \big(
        \HA_0 \HA_{0,1}
        -2 \HA_{0,0,1}
\big)
-10 \big(
        \HA_0 \HA_{-1,0}
        -2 \HA_{-1,0,0}
\big)
-2 \Biggl[
        \frac{1}{2} \HA_0^2 \HA_1
        -\HA_0 \HA_{0,1}
\nonumber\\ &
        +\HA_{0,0,1}
\Biggr]
-\Biggl[
        \frac{(3+x) \zeta_2}{1-x}
        -\frac{8 x \zeta_3}{1-x^2}
\Biggr] \HA_0
+\Biggl[
        3
        +\frac{8 x \zeta_2}{1-x^2}
\Biggr] \HA_0^2
+\frac{2 (1-2 x)}{3 (1-x)} \HA_0^3
-\frac{x}{3 (1-x^2)} \HA_0^4
\nonumber\\ &
-\frac{8 x}{1-x^2} \HA_{-1,0,0,0}
-15 \zeta_2
+\frac{16 x}{1-x^2} \big(
        \HA_0 \HA_1
        -\HA_{0,1}
\big) \zeta_2
-\frac{8 x \zeta_2}{1-x^2} \HA_{-1,0}
+\frac{96 x \zeta_2^2}{5 (1-x^2)}
-\frac{28}{3} \zeta_3
\\
J_{3}^{(0)} (x) &=
- \frac{15}{2}
-\frac{8 x}{1-x^2} \Biggl[
        \frac{1}{6} \big(
                \HA_0 \HA_1
                -\HA_{0,1}
        \big) \HA_0^2
        -\frac{1}{3} \HA_0^2 \HA_{0,1}
        +\HA_0 \HA_{0,0,1}
        -\HA_{0,0,0,1}
\Biggr]
-\frac{4 x}{1-x^2} \Biggl[
        \frac{1}{2} \HA_0^2 \HA_{0,1}
\nonumber\\ &
        -2 \HA_0 \HA_{0,0,1}
        +3 \HA_{0,0,0,1}
\Biggr]
-2 \Biggl[
        \frac{1}{2} \HA_0^2 \HA_1
        -\HA_0 \HA_{0,1}
        +\HA_{0,0,1}
\Biggr]
+\frac{1}{2} \Biggl[
        1
        -\frac{4 x \zeta_2}{1-x^2}
\Biggr] \HA_0^2
\nonumber\\ &
+\frac{1+3 x^2}{6 (1-x^2)} \HA_0^3
-\frac{x}{4 (1-x^2)} \HA_0^4
+\frac{35}{4} \zeta_2
-\frac{8 x}{1-x^2} \big(
        \HA_0 \HA_1
        -\HA_{0,1}
\big) \zeta_2
+\frac{2 (1+x^2)\zeta_2}{1-x^2} \HA_0
\nonumber\\ &
-\frac{32 x \zeta_2^2}{5 (1-x^2)}
+\frac{31}{6} \zeta_3~.
\end{align}
In total the solution of Eq.~(\ref{EQ:DI}) has to be expanded to $O(\ep^4)$. Here harmonic polylogarithms 
to weight {\sf w = 8} are contributing. The result is attached in the file {\tt exampleIntegral.m}
to this paper. The {\tt Mathematica} notebook {\tt CheckExample.nb} allows to verify the solution {\tt 
exampleIntegral.m}. It needs {\tt HarmonicSums.m}, which can be obtained from \cite{HSM}.
After the reduction to the algebraic basis, 796 to 833 different harmonic polylogarithms 
contribute. In the present example no cyclotomic harmonic polylogarithms occur, which are, however, present
in the solution of other sub-systems. The pole terms do not contain the latter functions. In assembling 
the form factors, harmonic and cyclotomic harmonic polylogarithms of up to weight {\sf w=6} contribute.
\section{Application to the heavy quark form factors}
\label{sec:3}

\vspace*{1mm}
\noindent
We consider the decay of a virtual massive vector boson 
of momentum $q$ into a pair of heavy quarks of mass $m$, momenta $q_1$ and $q_2$ and color
$c$ and $d$, through the vertex $\Gamma^{\mu}_{V,cd}$. We follow the notation used in Ref.~\cite{Ablinger:2017hst}. 
Here $q^2 = (q_1+q_2)^2$ is the center of mass energy squared.
The general form of the amplitude is given by
\begin{align}
\bar{u}_c (q_1) \Gamma_{V,cd}^{\mu} v_d (q_2) \equiv
  -i \bar{u}_c (q_1) \Big[ \delta_{cd}
 v_Q \Big( \gamma^{\mu} ~F_{V,1} 
         + \frac{i}{2 m} \sigma^{\mu \nu} q_{\nu}  ~ F_{V,2}  \Big) \Big] v_d (q_2) , 
\end{align}
where $\bar{u}_c (q_1)$ and $v_d (q_2)$ are the bi-spinors of the quark and the anti-quark, 
respectively,
$\sigma^{\mu\nu} = \frac{i}{2} [\gamma^{\mu},\gamma^{\nu}]$ and $v_Q$ is the 
Standard Model (SM) vector coupling constant.
$F_{V,1}$ and $F_{V,2}$ are the corresponding ultraviolet (UV) renormalized form factors, 
also called the electric and magnetic form factors.
They are expanded in the strong coupling constant $\alpha_s = g_s^2/(4\pi)$ as follows
\begin{equation}
 F_{V,i} = \sum_{n=0}^{\infty} \asr^n F_{V,i}^{(n)} \,.
\end{equation}
The form factors are obtained from the amplitudes by multiplying 
appropriate projectors as provided in \cite{Ablinger:2017hst} and performing the trace over the color and spinor indices.
$n_l$ and $n_h$ are the numbers of light and heavy quarks.
For convenience, we use the Landau variable \cite{Barbieri:1972as}
\begin{equation} \label{eq:varxp}
 x=\frac{\sqrt{q^2-4m^2}-\sqrt{q^2}}{\sqrt{q^2-4m^2}+\sqrt{q^2}},
\end{equation}
see also Eq.~(\ref{EQ:1}).
Particularly, we focus on the Euclidean region, $q^2<0$, corresponding to $x \in 
[0,1[$. The expressions for the other kinematic regions are obtained by the analytic continuation of the 
final result, given by HPLs \cite{Remiddi:1999ew} and cyclotomic HPLs \cite{Ablinger:2011te}.

\subsection{Details of the calculation}
\label{sec:31}

\begin{figure}[H]
\begin{center}
    \begin{minipage}[c]{0.13\linewidth}
    \includegraphics[width=1\textwidth]{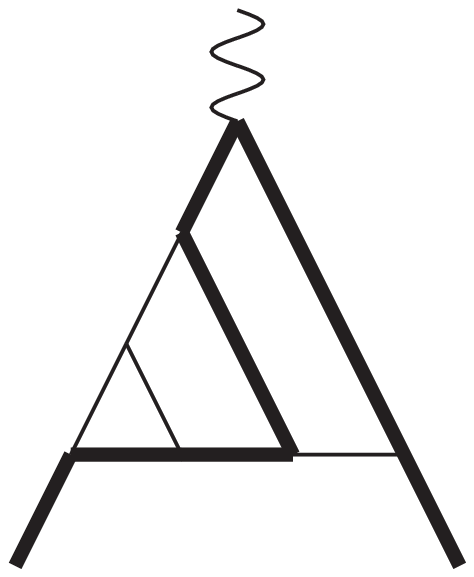}
    \vspace*{-11mm}
    \end{minipage}
        \hspace*{2mm}
    \begin{minipage}[c]{0.13\linewidth}
    \includegraphics[width=1\textwidth]{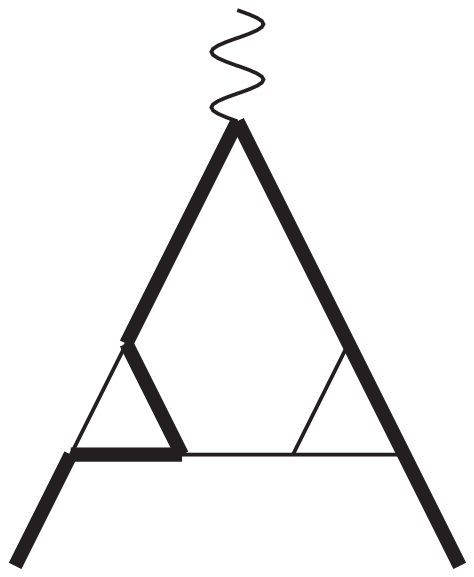}
    \vspace*{-11mm}
    \end{minipage}
        \hspace*{2mm}
    \begin{minipage}[c]{0.13\linewidth}
    \includegraphics[width=1\textwidth]{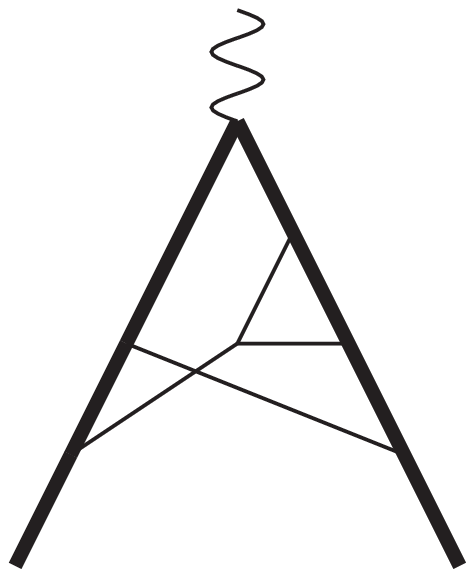}
    \vspace*{-11mm}
    \end{minipage}
\end{center}
\caption{\sf \small The $n_l$ topologies}
\label{fig:nltopol}
\end{figure}

\vspace*{1mm}
\noindent
The calculation of the three-loop massive vector form factors proceeds in a similar way as has been outlined 
in Refs.~\cite{Ablinger:2017hst,Ablinger:2018yae}. As usual the packages {\tt QGRAF} \cite{Nogueira:1991ex}, {\tt 
Color} \cite{vanRitbergen:1998pn},
{\tt Q2e/Exp} \cite{Harlander:1997zb,Seidensticker:1999bb} and {\tt FORM} \cite{Vermaseren:2000nd, Tentyukov:2007mu}
have been used to generate the Feynman diagrams, calculate their color and Dirac-structure, and 
to determine moments
for comparisons. The reduction to master integrals has been performed using {\tt Crusher} \cite{CRUSHER}.
Finally, we have obtained 109 MIs, out of which 96 appear in the color--planar case, as indicated 
in \cite{Ablinger:2018yae}.
\begin{figure}[H]
\begin{center}
\begin{minipage}[c]{0.09\linewidth}
     \includegraphics[width=1\textwidth]{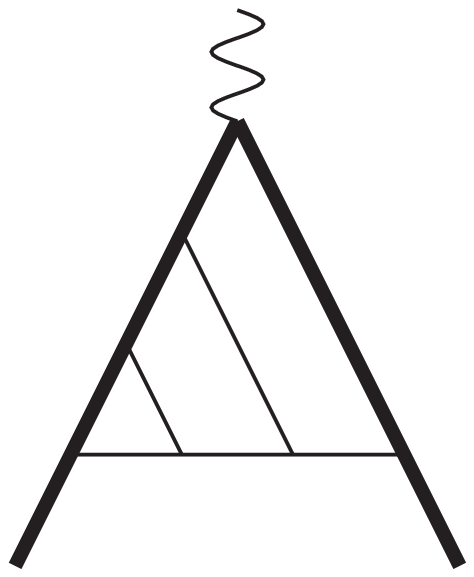}
\vspace*{-11mm}
\begin{center}
\end{center}
\end{minipage}
\hspace*{2mm}
\begin{minipage}[c]{0.09\linewidth}
     \includegraphics[width=1\textwidth]{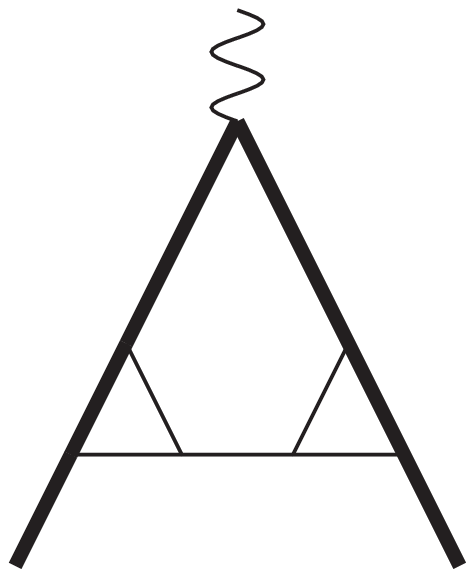}
\vspace*{-11mm}
\begin{center}
\end{center}
\end{minipage}
\hspace*{2mm}
\begin{minipage}[c]{0.09\linewidth}
     \includegraphics[width=1\textwidth]{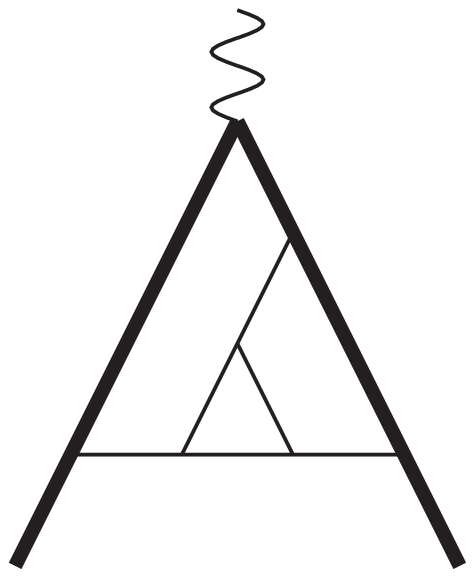}
\vspace*{-11mm}
\begin{center}
\end{center}
\end{minipage}
\hspace*{2mm}
\begin{minipage}[c]{0.09\linewidth}
     \includegraphics[width=1\textwidth]{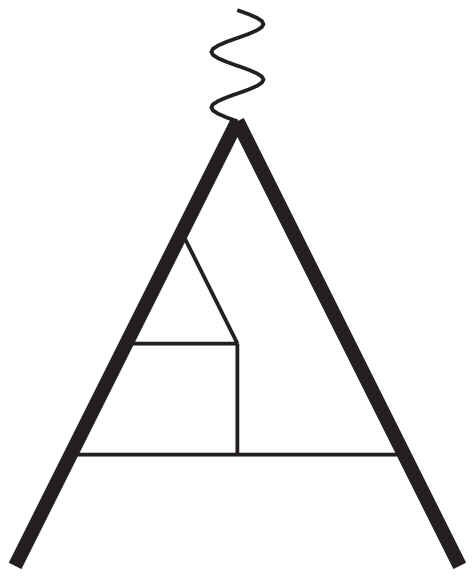}
\vspace*{-11mm}
\begin{center}
\end{center}
\end{minipage}
\hspace*{2mm}
\begin{minipage}[c]{0.09\linewidth}
     \includegraphics[width=1\textwidth]{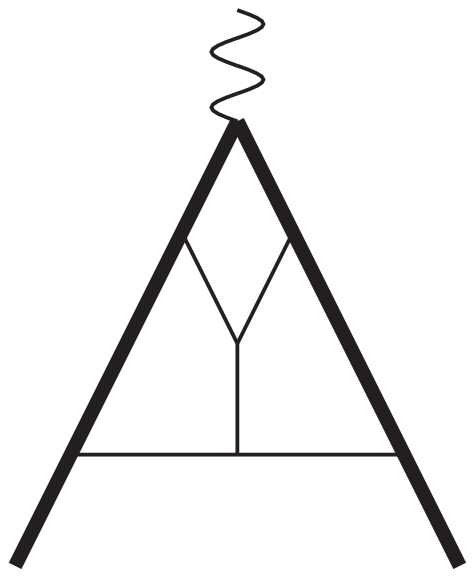}
\vspace*{-11mm}
\begin{center}
\end{center}
\end{minipage}
\hspace*{2mm}
\begin{minipage}[c]{0.09\linewidth}
     \includegraphics[width=1\textwidth]{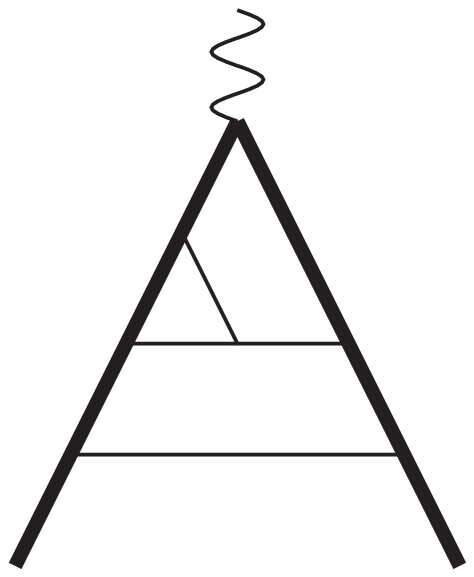}
\vspace*{-11mm}
\begin{center}
\end{center}
\end{minipage}
\hspace*{2mm}
\begin{minipage}[c]{0.09\linewidth}
     \includegraphics[width=1\textwidth]{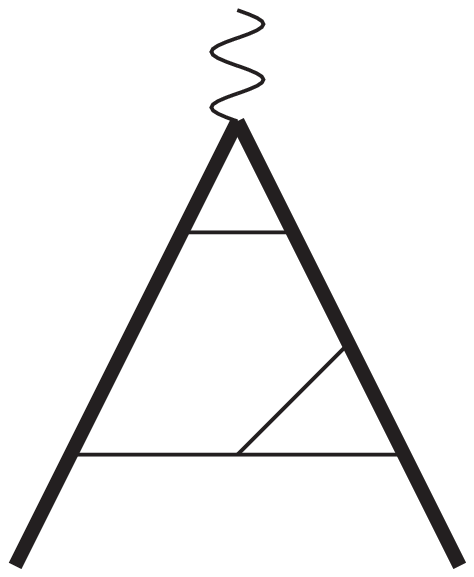}
\vspace*{-11mm}
\begin{center}
\end{center}
\end{minipage}
\hspace*{2mm}
\begin{minipage}[c]{0.09\linewidth}
     \includegraphics[width=1\textwidth]{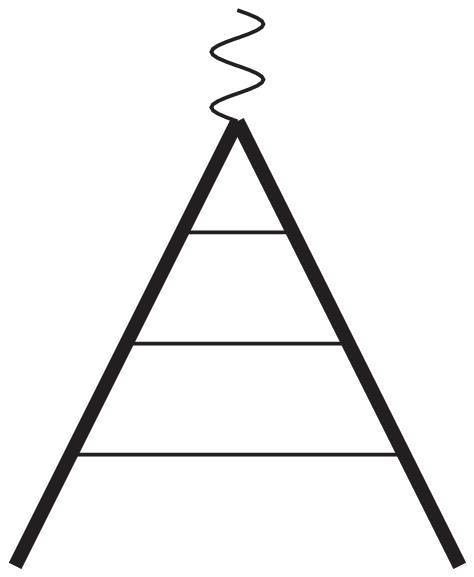}
\vspace*{-11mm}
\begin{center}
\end{center}
\end{minipage}
\end{center}
\caption{\sf \small The color--planar topologies}
\label{fig:cptopologies}
\end{figure}
\noindent
In the color--planar limit, the families of integrals can be represented by eight topologies, shown in
Figure~\ref{fig:cptopologies}, whereas for the complete light quark contributions, three more topologies are
required, cf.~Figure~\ref{fig:nltopol}. Note that, only sub-topologies with a maximum of eight propagators 
contribute in the latter scenario.
%
%
\noindent
Finally, to compute the master integrals, we implemented the algorithm described in the previous section, applying
it to all occurring systems of differential equations.
This calculation is performed by intense use of {\tt HarmonicSums}
\cite{Ablinger:2014rba, Ablinger:2010kw, Ablinger:2013hcp, Ablinger:2011te, Ablinger:2013cf, Ablinger:2014bra}, 
which uses the package {\tt Sigma} 
\cite{Schneider:2007a,Schneider:2013a}. Finally we have checked all MIs 
numerically using {\tt FIESTA} \cite{Smirnov:2008py, Smirnov:2009pb, Smirnov:2015mct}.
\subsection{Ultraviolet renormalization and infrared structure}
\label{sec:32}

\vspace*{1mm}
\noindent
The UV renormalization of the form factors has been performed in a mixed scheme.
The heavy quark mass and wave function have been renormalized in the on-shell (OS)
renormalization scheme, while the strong coupling constant is renormalized 
in the $\overline{\rm MS}$ scheme, where we set the universal factor $S_\varepsilon = 
\exp(-\varepsilon (\gamma_E - \ln(4\pi))$ for each loop order to one at the end of the calculation. The 
required renormalization constants are available and are denoted by 
$Z_{m, {\rm OS}}$ \cite{Broadhurst:1991fy, Melnikov:2000zc,Marquard:2007uj,
Marquard:2015qpa,Marquard:2016dcn}, 
$Z_{2,{\rm OS}}$ \cite{Broadhurst:1991fy, Melnikov:2000zc,Marquard:2007uj,Marquard:2018rwx} and 
$Z_{a_s}$ \cite{Tarasov:1980au,Larin:1993tp,vanRitbergen:1997va,Czakon:2004bu,Baikov:2016tgj,Herzog:2017ohr,
Luthe:2017ttg} for the heavy quark mass, wave function and strong coupling constant, respectively. 
The renormalization of the heavy-quark wave function and the
strong coupling constant are multiplicative, while the renormalization of massive fermion 
lines has been taken care of by properly considering the counter terms.

Considering the high energy limit, the universal behavior of infrared (IR) singularities of the massive form factors was first
investigated in \cite{Mitov:2006xs}. Later in \cite{Becher:2009kw},
a general argument was provided to factorize the IR singularities as a multiplicative renormalization
constant. Its structure is constrained by the renormalization group equation (RGE), as follows,
\begin{equation}
 F_{I} = Z (\mu) F_{I}^{\mathrm{fin}} (\mu)\, ,
\end{equation}
where $F_{I}^{\mathrm{fin}}$ is finite as $\ep \rightarrow 0$. The RGE for $Z(\mu)$ reads
\begin{equation} \label{eq:rgeZ}
 \frac{d}{d \ln \mu} \ln Z(\ep, x, m, \mu)  = - \Gamma (x,m,\mu).
\end{equation}
Here $\Gamma$ is the corresponding cusp anomalous dimension, which is by now available up to three--loop 
order \cite{Grozin:2014hna,Grozin:2015kna}. Note that $Z$ is independent of the current.
Both $Z$ and $\Gamma$ can be expanded in a perturbative series in $\alpha_s$ as follows
\begin{equation}
 Z = \sum_{n=0}^{\infty} \asr^n Z^{(n)} \,, \qquad
 \Gamma = \sum_{n=0}^{\infty} \asr^{n+1} \Gamma_{n},
\end{equation}
and the solution for Eq.~(\ref{eq:rgeZ}) up to ${\cal O}(\alpha_s^3)$ is 
\begin{align} \label{eq:solnZ}
 Z &= 1 + \asr \Bigg[ \frac{\Gamma_0}{2 \ep} \Bigg] 
   + \asr^2 \Bigg[ \frac{1}{\ep^2} \Big( \frac{\Gamma_0^2}{8} - \frac{\beta_0 \Gamma_0}{4} \Big) + \frac{\Gamma_1}{4 \ep} \Bigg] 
   \nonumber\\
  &+ \asr^3 \bigg[ \frac{1}{\ep^3} \left( \frac{\Gamma_0^3}{48} - \frac{\beta_0 \Gamma_0^2}{8} + \frac{\beta_0^2 \Gamma_0}{6} \right)
                 + \frac{1}{\ep^2} \left( \frac{\Gamma_0 \Gamma_1}{8} - \frac{\beta_1 \Gamma_0}{6} \right) 
                 + \frac{1}{\ep} \left( \frac{\Gamma_2}{6} \right) \bigg]
   + {\cal O} (\alpha_s^4) \,.
\end{align}
\section{The Three-Loop Vector Form Factors}
\label{sec:4}

\vspace*{1mm}
\noindent
We apply our algorithm to the single scale and first order factorizable system of differential 
equations which are relevant for the 
color--planar and the complete light quark contributions to the heavy quark form factors. 
One obtains the solutions for all contributing
integrals in Laurent series expansion up to the required order in $\ep$ .

Finally, using the results for the integrals, we obtain the color--planar and complete light quark ($n_l$) 
non--singlet contributions of
the three-loop massive form factors for the vector current. 
Due to the substantial length of the expressions, we provide them as supplemental material along with this 
publication. In the following we only present expansions of the form factors in different kinematic limits
and give numerical results for the whole kinematic region. Here
the following abbreviation is used 
\begin{equation}
 c_1 = 12 \zeta_2 \ln^2 (2) + \ln^4 (2) + 24 {\rm Li}_4 \left(\frac{1}{2}\right),
\end{equation}
as mentioned in \cite{Ablinger:2017hst} and
$\Li_k(x)$ denotes the polylogarithm \cite{DILOG4,DUDE}. 
In Figures~\ref{fig:VF12ep0} we illustrate the behavior of the $O(\varepsilon^0)$ 
parts 
of the vector form factors
as a function of $x \in [0,1]$. We also show their small- and large-$x$ expansions. The latter 
representations are obtained using {\tt HarmonicSums}. 
For the numerical evaluation of the HPLs and the cyclotomic  HPLs we use the {\tt GiNaC} package
\cite{Vollinga:2004sn,Bauer:2000cp} and the {\tt FORTRAN}-codes {\tt HPOLY.f} \cite{Ablinger:2018sat} and 
{\tt CPOLY.f}.

We present now the expansion of the form factors in different kinematic regions.
\subsection{The low energy region \boldmath $x \rightarrow 1$}
\label{sec:41}

\vspace*{1mm}
\noindent
In the static limit, i.e. $q^2 \rightarrow 0$, we define $y=1-x$ and expand the form factors around $y=0$.
In this region, the electric component of the vector form factor vanishes. 
The ${\cal O}(y^2)$ contribution of the electric component of the three loop vector form factor 
($F_{V,1}^{(3)}$) is given by
\begin{align}
{F}_{V,1}^{(3)} \simeq
y^2 &\Bigg[
N_C^3 \bigg\{
\frac{1}{\ep^3} 
\frac{121}{81}
+ \frac{1}{\ep^2} \bigg(
-\frac{1340}{243}
+\frac{44}{27} \zeta_2
\bigg)
+ \frac{1}{\ep} \bigg(
\frac{473}{54}
-\frac{680}{81} \zeta_2
+\frac{8}{3} \zeta_2^2
+\frac{10}{27} \zeta_3
\bigg)
\nonumber\\
&
+ \bigg(
\frac{4961563}{69984}
+\frac{8977}{486} \zeta_2
-\frac{280}{9} \zeta_2^2
-\frac{7127}{324} \zeta_3
+\frac{92}{3} \zeta_2 \zeta_3
+5 \zeta_5
\bigg)
\bigg\}
+ C_F^2 n_l T_F \bigg\{
\frac{1}{\ep^2} \
\frac{8}{9} 
\nonumber\\
&
- \frac{1}{\ep} \bigg(
\frac{110}{27}
-\frac{32}{9} \zeta_3 
\bigg)
+ \bigg(
\frac{3107}{162}
-\frac{64 c_1}{9}
-\frac{19676}{81} \zeta_2
+\frac{3536}{9} \ln(2) \zeta_2
+\frac{1792}{15} \zeta_2^2
-\frac{1100}{9} \zeta_3
\bigg)
\bigg\}
\nonumber\\
&
+ C_A C_F n_l T_F \bigg\{
-\frac{1}{\ep^3}
\frac{176}{81} 
+ \frac{1}{\ep^2} \bigg(
\frac{1552}{243}
-\frac{32}{27} \zeta_2
\bigg)
+ \frac{1}{\ep} \bigg(
-\frac{1556}{243}
+\frac{320}{81} \zeta_2
-\frac{112}{27} \zeta_3 
\bigg)
\nonumber\\
&
+ \bigg(
-\frac{260644}{2187}
+\frac{32 c_1}{9}
+\frac{10474}{243} \zeta_2
-\frac{1768}{9} \ln (2) \zeta_2
-\frac{1408}{45} \zeta_2^2
+\frac{1622}{27} \zeta_3 
\bigg)
\bigg\}
\nonumber\\
&
+ C_F n_l^2 T_F^2 \bigg\{
\frac{1}{\ep^3}
\frac{32}{81}
- \frac{1}{\ep^2} 
-\frac{160}{243}
- \frac{1}{\ep} 
-\frac{32}{243}
+ \bigg(
\frac{29524}{2187}
+\frac{928}{81} \zeta_2
+\frac{448}{81} \zeta_3
\bigg)
\bigg\}
\nonumber\\
&
+ C_F n_h n_l T_F^2 \bigg\{
-\frac{1}{\ep} 
\frac{16}{27} \zeta_2 
+ \bigg(
-\frac{10088}{243}
+\frac{1784}{81} \zeta_2
+16 \ln(2) \zeta_2
-\frac{724}{81} \zeta_3 
\bigg)
\bigg\}
%
%
%
%
%
\Bigg] \,.
\end{align}
Here $C_A = N_C, C_F = (N_C^2 - 1)/(2 N_C), T_F =1/2$ and $N_C$ denotes the number of colors for $SU(N_C)$ with
$N_C = 3$ in case of QCD.
The magnetic component of the vector form factor $(F_{V,2})$ is the anomalous magnetic moment of a heavy quark 
in this limit allowing for a cross-check of our computation with Ref.~\cite{Grozin:2007fh}. 
The form factor $F_{V,2}^{(3)}$ up to order $y^2$ reads
\begin{align}
{F}_{V,2}^{(3)} \simeq
&\Bigg[
N_C^3 \bigg\{
\frac{104147}{648}
+\frac{962}{9} \zeta_2
-24 \zeta_2^2
+\frac{80}{3} \zeta_3
+48 \zeta_2 \zeta_3
-20 \zeta_5
\bigg\}
\nonumber\\
&
+ C_F^2 n_l T_F \bigg\{
250
-\frac{64 c_1}{9}
-\frac{5056}{9} \zeta_2
+640 \ln (2) \zeta_2
+\frac{352}{3} \zeta_2^2
-192 \zeta_3
\bigg\}
\nonumber\\
&
+ C_A C_F n_l T_F \bigg\{
-\frac{38576}{81}
+\frac{32 c_1}{9}
+\frac{1232}{9} \zeta_2
-320 \ln (2) \zeta_2
-\frac{176}{3} \zeta_2^2
+\frac{304}{3} \zeta_3
\bigg\}
\nonumber\\
&
+ C_F n_l^2 T_F^2 \bigg\{
\frac{5072}{81}
+\frac{128}{9} \zeta_2
\bigg\}
+ C_F n_h n_l T_F^2 \bigg\{
-\frac{1952}{81}
+\frac{128}{9} \zeta_2
\bigg\}
\Bigg]
%
\nonumber\\
+
y^2 &\Bigg[
\frac{1}{\ep^2} \Bigg( -\frac{11}{18} N_C^3 + \frac{8}{9} C_F^2 n_l T_F  
\Bigg)
+ \frac{1}{\ep} \Bigg( N_C^3 \bigg\{ \frac{31}{12} 
+ \frac{2}{3} \zeta_2   \bigg\}  - \frac{16}{3}
C_F^2 n_l T_F  \Bigg)
\nonumber\\ &
+N_C^3 \bigg\{
\frac{3236461}{155520}
-\frac{22849}{720} \zeta_2
+\frac{46}{5} \zeta_2^2
-\frac{407}{36} \zeta_3
-\frac{72}{5} \zeta_2 \zeta_3
+6 \zeta_5
\bigg\}
\nonumber\\
&
+ C_F^2 n_l T_F \bigg\{
-\frac{12653}{90}
+\frac{736 c_1}{135}
+\frac{12184}{45} \zeta_2
-\frac{16096}{45} \ln (2) \zeta_2
-\frac{4048}{45} \zeta_2^2
+\frac{664}{5} \zeta_3
\bigg\}
\nonumber\\
&
+ C_A C_F n_l T_F \bigg\{
\frac{26626}{243}
-\frac{368 c_1}{135}
-\frac{916}{9} \zeta_2
+\frac{8048}{45} \ln (2) \zeta_2
+\frac{2024}{45} \zeta_2^2
-\frac{348}{5} \zeta_3
\bigg\}
\nonumber\\
&
- C_F n_l^2 T_F^2 \bigg\{
\frac{3736}{243}
+\frac{64}{27} \zeta_2
\bigg\}
+ C_F n_h n_l T_F^2 \bigg\{
\frac{11824}{243}
-\frac{80}{3} \zeta_2
-\frac{32}{3} \ln (2) \zeta_2
+\frac{56}{9} \zeta_3
\bigg\}
%
\Bigg] \,.
\end{align}

\subsection{High energy region  {\boldmath $x \rightarrow 0$}} 
Here we present the expansion of the form factors $F_{V,1}^{(3)}$ and $F_{V,2}^{(3)}$ in the asymptotic
limit i.e. for $x\rightarrow0^+$ up to ${\cal O}(x^2)$. The abbreviation $L$ has been used to indicate $\ln (x)$. 
\begin{align}
{F}_{V,1}^{(3)} & \simeq
\frac{1}{\ep^3} \bigg[
N_C^3 \bigg\{
-\frac{175}{27}
-\frac{467 L}{54}
-\frac{7 L^2}{3}
-\frac{L^3}{6}
\bigg\}
+ C_F^2 n_l T_F \bigg\{
\frac{8}{3}
+\frac{16 L}{3}
+\frac{8 L^2}{3}
\bigg\}
\nonumber\\&
+ C_A C_F n_l T_F \bigg\{
\frac{176}{27}
+\frac{176 L}{27}
\bigg\}
+ C_F n_l^2 T_F^2 \bigg\{
-\frac{32}{27}
-\frac{32 L}{27}
\bigg\}
+ x^2 \bigg(
N_C^3 \bigg\{
-\frac{467 L}{27}-\frac{28 L^2}{3}-L^3
\bigg\}
\nonumber\\&
+ C_F^2 n_l T_F 
\frac{32}{3} L (1+L)
+ C_A C_F n_l T_F
\frac{352 L}{27}
- C_F n_l^2 T_F^2 
\frac{64 L}{27}
\bigg)
\bigg]
\nonumber\\&
+ \frac{1}{\ep^2} \bigg[
N_C^3 \bigg\{
\frac{1375}{162}
-\frac{10}{9} \zeta_2
+\frac{31}{9} \zeta_3
+L \bigg(
        \frac{1645}{162}
        -\frac{29 \zeta_2}{18}
        +\zeta_3
\bigg)
+L^2 \bigg(
        -\frac{97}{36}
        -\frac{\zeta_2}{2}
\bigg)
-\frac{13 L^3}{6}
-\frac{L^4}{4}
\bigg\}
\nonumber\\&
+ C_F^2 n_l T_F \bigg\{
-\frac{16}{9}
-\frac{8}{3} \zeta_2
+L \bigg(
        -\frac{20}{9}
        -\frac{8 \zeta_2}{3}
\bigg)
+\frac{8 L^2}{9}
+\frac{4 L^3}{3}
\bigg\}
+ C_A C_F n_l T_F \bigg\{
-\frac{1192}{81}
+\frac{16}{9} \zeta_2
\nonumber\\&
-\frac{16}{9} \zeta_3
+L \bigg(
        -\frac{1336}{81}
        +\frac{16 \zeta_2}{9}
\bigg)
\bigg\}
+ C_F n_l^2 T_F^2 \bigg\{
\frac{160}{81}
+\frac{160 L}{81}
\bigg\}
+ x \bigg(
N_C^3 \bigg\{
-\frac{14}{3}-\frac{10 L}{3}+\frac{11 L^2}{6}
\nonumber\\&
+\frac{L^3}{2}
\bigg\}
+ C_F^2 n_l T_F \bigg\{
-\frac{8}{3} (-2+L) (1+L)
\bigg\}
%
%
%
%
\bigg)
+ x^2 \bigg(
N_C^3 \bigg\{
        -\frac{103}{18}
        -\frac{172 L^3}{27}
        -\frac{5 L^4}{6}
        +L \bigg(
                \frac{8983}{324}
\nonumber\\&
                +\frac{34 \zeta_2}{3}
                +6 \zeta_3
        \bigg)
        +L^2 \bigg(
                -\frac{41}{36}
                +\zeta_2
        \bigg)
        +\frac{104}{9} \zeta_2
        +\frac{124}{9} \zeta_3
\bigg\}
+ C_F^2 n_l T_F \bigg\{
        -\frac{4}{3}
        +\frac{32 L^2}{9}
        +\frac{16 L^3}{3}
\nonumber\\&
        +L \bigg(
                -\frac{28}{9}
                -\frac{32 \zeta_2}{3}
        \bigg)
        -\frac{16}{3} \zeta_2
\bigg\}
+ C_A C_F n_l T_F \bigg\{
        \frac{32}{9}
        -\frac{32}{9} \zeta_2
        -\frac{64}{9} \zeta_3
        -\frac{3104 L}{81}
        -\frac{32 L^2}{9}
        -\frac{32 L^3}{27}
\bigg\}
\nonumber\\&
+ C_F n_l^2 T_F^2 \bigg\{
\frac{320 L}{81}
\bigg\}
\bigg)
\bigg]
+ \frac{1}{\ep} \bigg[
N_C^3 \bigg\{
\frac{637}{54}
-\frac{161}{108} \zeta_2
+\frac{64}{15} \zeta_2^2
-\frac{550}{27} \zeta_3
-\frac{7}{3} \zeta_2 \zeta_3
+6 \zeta_5
+L \bigg(
        \frac{4369}{216}
\nonumber\\&
        -\frac{161}{108} \zeta_2
        +\frac{64}{15} \zeta_2^2
        -\frac{281}{18} \zeta_3
\bigg)
+L^2 \bigg(
        \frac{463}{27}
        +\frac{19 \zeta_2}{12}
        -\frac{11 \zeta_3}{2}
\bigg)
+L^3 \bigg(
        \frac{14}{9}
        -\frac{3 \zeta_2}{4}
\bigg)
-\frac{10 L^4}{9}
-\frac{5 L^5}{24}
\bigg\}
\nonumber\\&
+ C_F^2 n_l T_F \bigg\{
-\frac{470}{27}
+8 \zeta_2
-\frac{16}{3} \zeta_3
+L \bigg(
        -\frac{1198}{27}
        +\frac{4 \zeta_2}{3}
        -\frac{16 \zeta_3}{3}
\bigg)
+L^2 \bigg(
        -\frac{962}{27}
        -\frac{20 \zeta_2}{3}
\bigg)
\nonumber\\&
-\frac{82 L^3}{9}
-\frac{4 L^4}{9}
\bigg\}
+ C_A C_F n_l T_F \bigg\{
\frac{356}{81}
-\frac{160}{27} \zeta_2
+\frac{496}{27} \zeta_3
+L \Bigl(
        \frac{836}{81}
        -\frac{160 \zeta_2}{27}
        +\frac{112 \zeta_3}{9}
\Bigr)
\bigg\}
\nonumber\\&
+ C_F n_l^2 T_F^2 \bigg\{
\frac{32}{81}
+\frac{32 L}{81}
\bigg\}
+ C_F n_h n_l T_F^2 \bigg\{
\frac{16}{9} \zeta_2
+\frac{16}{9} L \zeta_2
\bigg\}
+ x \bigg(
N_C^3 \bigg\{
        \frac{109}{2}
        -\frac{469}{6} \zeta_2
        -\frac{102}{5} \zeta_2^2
\nonumber\\&
        +124 \zeta_3
        +L \bigg(
                \frac{457}{12}
                -\frac{323}{3} \zeta_2
                -\frac{102}{5} \zeta_2^2
                +169 \zeta_3
        \bigg)
        +L^2 \bigg(
                -\frac{41}{6}
                -\frac{71 \zeta_2}{2}
                +48 \zeta_3
        \bigg)
        +L^3 \bigg(
                \frac{40}{3}
                -6 \zeta_2
        \bigg)
\nonumber\\&
        +\frac{3 L^4}{4}
\bigg\}
+ C_F^2 n_l T_F \bigg\{
        -40
        -\frac{8}{3} \zeta_2
        +L \bigg(
                -\frac{32}{3}
                -\frac{8 \zeta_2}{3}
        \bigg)
        +\frac{92 L^2}{3}
        +\frac{4 L^3}{3}
\bigg\}
%
%
%
%
\bigg)
+ x^2 \bigg(
N_C^3 \bigg\{
        -\frac{84209}{216}
\nonumber\\&
        +\frac{32975}{54} \zeta_2
        +\frac{3509}{15} \zeta_2^2
        -\frac{59627}{54} \zeta_3
        -14 \zeta_2 \zeta_3
        +20 \zeta_5
        +L \bigg(
                -\frac{33271}{216}
                +\frac{2572}{3} \zeta_2
                +\frac{3661}{15} \zeta_2^2
\nonumber\\&
                -\frac{14450}{9} \zeta_3
        \bigg)
        +L^2 \bigg(
                \frac{19889}{108}
                +\frac{988 \zeta_2}{3}
                -\frac{1631 \zeta_3}{3}
        \bigg)
        +L^3 \bigg(
                -\frac{31061}{324}
                +\frac{1001 \zeta_2}{18}
        \bigg)
        -\frac{17 L^4}{6}
        -\frac{151 L^5}{180}
\bigg\}
\nonumber\\&
+ C_F^2 n_l T_F \bigg\{
        12
        +\frac{112}{3} \zeta_2
        +\frac{32}{3} \zeta_3
        +L \bigg(
                -\frac{2540}{27}
                +40 \zeta_2
        \bigg)
        +L^2 \bigg(
                -\frac{3884}{27}
                -\frac{80 \zeta_2}{3}
        \bigg)
        -\frac{332 L^3}{9}
\nonumber\\&
        -\frac{16 L^4}{9}
\bigg\}
+ C_A C_F n_l T_F \bigg\{
        -\frac{320}{27}
        +\frac{320}{27} \zeta_2
        +\frac{640}{27} \zeta_3
        +L \bigg(
                \frac{3112}{81}
                +\frac{224 \zeta_3}{9}
        \bigg)
        +\frac{320 L^2}{27}
        +\frac{320 L^3}{81}
\bigg\}
\nonumber\\&
+ C_F n_l^2 T_F^2 \bigg\{
\frac{64 L}{81} 
\bigg\}
+ C_F n_h n_l T_F^2 \bigg\{
\frac{32}{9} L \zeta_2
\bigg\}
\bigg)
\bigg]
%
+  \bigg[
N_C^3 \bigg\{
-\frac{554267}{2916}
+\frac{23773}{216} \zeta_2
-\frac{1727}{30} \zeta_2^2
\nonumber\\&
+\frac{8156}{315} \zeta_2^3
+\frac{33197}{162} \zeta_3
+\frac{113}{3} \zeta_2 \zeta_3
-\frac{16}{3} \zeta_3^2
-\frac{875}{3} \zeta_5
+L \bigg(
        -\frac{669127}{5832}
        -\frac{31609}{324} \zeta_2
        -\frac{71}{10} \zeta_2^2
\nonumber\\&
        +\frac{4297}{54} \zeta_3
        -\frac{13}
        {6} \zeta_2 \zeta_3
        +15 \zeta_5
\bigg)
+L^2 \bigg(
        \frac{1535}{36}
        +\frac{593}{72} \zeta_2
        -4 \zeta_2^2
        -\frac{629}{36} \zeta_3
\bigg)
+L^3 \bigg(
        \frac{6373}{324}
        +\frac{37 \zeta_2}{12}
\nonumber\\&
        -\frac{49 \zeta_3}{6}
\bigg)
+L^4 \bigg(
        \frac{1141}{432}
        -\frac{2 \zeta_2}{3}
\bigg)
-\frac{3 L^5}{8}
-\frac{L^6}{8}
\bigg\}
+ C_F^2 n_l T_F \bigg\{
-\frac{2011}{81}
-\frac{64 c_1}{9}
-\frac{962}{3} \zeta_2
\nonumber\\&
+\frac{448}{3} \ln (2) \zeta_2
+\frac{12232}{45} \zeta_2^2
+\frac{2752}{9} \zeta_3
-48 \zeta_2 \zeta_3
+40 \zeta_5
+L \bigg(
        -\frac{18812}{81}
        -\frac{682}{9} \zeta_2
        +\frac{392}{15} \zeta_2^2
\nonumber\\&
        +\frac{1976}{9} \zeta_3
\bigg)
+L^2 \bigg(
        -\frac{18817}{81}
        -\frac{100 \zeta_2}{3}
        +\frac{232 \zeta_3}{9}
\bigg)
+L^3 \bigg(
        -\frac{2032}{27}
        -\frac{58 \zeta_2}{9}
\bigg)
-\frac{355 L^4}{27}
-L^5
\bigg\}
\nonumber\\&
+ C_A C_F n_l T_F \bigg\{
\frac{259150}{729}
+\frac{32 c_1}{9}
+\frac{3008}{81} \zeta_2
-\frac{224}{3} \ln (2) \zeta_2
-\frac{7288}{45} \zeta_2^2
-\frac{31120}{81} \zeta_3
+\frac{8}{3} \zeta_2 \zeta_3
\nonumber\\&
+\frac{596}{3} \zeta_5
+L \bigg(
        \frac{309838}{729}
        +\frac{11728}{81} \zeta_2
        -\frac{88}{15} \zeta_2^2
        -\frac{1448}{9} \zeta_3
\bigg)
+L^2 \bigg(
        \frac{11752}{81}
        +\frac{32 \zeta_2}{3}
        -16 \zeta_3
\bigg)
\nonumber\\&
+L^3 \bigg(
        \frac{1948}{81}
        -\frac{16 \zeta_2}{9}
\bigg)
+\frac{44 L^4}{27}
\bigg\}
+ C_F n_l^2 T_F^2 \bigg\{
-\frac{29344}{729}
-\frac{976}{81} \zeta_2
+\frac{928}{45} \zeta_2^2
+\frac{256}{9} \zeta_3
\nonumber\\&
+L \bigg(
        -\frac{39352}{729}
        -\frac{608 \zeta_2}{27}
        -\frac{64 \zeta_3}{27}
\bigg)
+L^2 \bigg(
        -\frac{1624}{81}
        -\frac{32 \zeta_2}{9}
\bigg)
-\frac{304 L^3}{81}
-\frac{8 L^4}{27}
\bigg\}
\nonumber\\&
+ C_F n_h n_l T_F^2 \bigg\{
-\frac{5072}{27}
+\frac{1552}{81} \zeta_2
-\frac{64}{3} \zeta_2^2
-\frac{416}{9} \zeta_3
+L \bigg(
        -\frac{7408}{81}
        -48 \zeta_2
        -\frac{416 \zeta_3}{27}
\bigg)
\nonumber\\&
+L^2 \bigg(
        -\frac{3248}{81}
        -\frac{64 \zeta_2}{9}
\bigg)
-\frac{608 L^3}{81}
-\frac{16 L^4}{27}
\bigg\}
+ x \bigg(
N_C^3 \bigg\{
        -\frac{149093}{324}
        +\frac{122195}{108} \zeta_2
        +\frac{5291}{60} \zeta_2^2
\nonumber\\&
        -\frac{4794}{35} \zeta_2^3
        -\frac{24647}{9} \zeta_3
        -278 \zeta_2 \zeta_3
        +18 \zeta_3^2
        +2154 \zeta_5
        +L \bigg(
                -\frac{1451}{81}
                +\frac{8221}{12} \zeta_2
                +\frac{401}{2} \zeta_2^2
                -\frac{3391}{3} \zeta_3
\nonumber\\&
                +12 \zeta_2 \zeta_3
                +96 \zeta_5
        \bigg)
        +L^2 \bigg(
                -\frac{13525}{108}
                +\frac{22}{3} \zeta_2
                +\frac{306}{5} \zeta_2^2
                -130 \zeta_3
        \bigg)
        +L^3 \bigg(
                -\frac{3095}{108}
                -\frac{197 \zeta_2}{12}
                +40 \zeta_3
        \bigg)
\nonumber\\&
        +L^4 \bigg(
                \frac{1319}{72}
                -9 \zeta_2
        \bigg)
        +\frac{5 L^5}{8}
\bigg\}
+ C_F^2 n_l T_F \bigg\{
        -\frac{1676}{9}
        +\frac{128 c_1}{3}
        +\frac{6800}{9} \zeta_2
        -3328 \ln (2) \zeta_2        
        -\frac{2096}{45} \zeta_2^2
\nonumber\\&
        +\frac{4528}{9} \zeta_3
        +128 \zeta_2 \zeta_3
        -\frac{1312}{3} \zeta_5
        +L \bigg(
                \frac{790}{9}
                -\frac{5836}{9} \zeta_2
                +\frac{1376}{15} \zeta_2^2
                +\frac{224}{3} \zeta_3
        \bigg)        
        +L^2 \bigg(
                \frac{3416}{9}
                -\frac{356 \zeta_2}{9}
\nonumber\\&
                +\frac{16 \zeta_3}{3}
        \bigg)
        +L^3 \bigg(
                \frac{3436}{27}
                -16 \zeta_2
        \bigg)
        +\frac{352 L^4}{27}
        +\frac{4 L^5}{9}
\bigg\}
+ C_A C_F n_l T_F \bigg\{
        \frac{97384}{81}
        -\frac{64 c_1}{3}
        -\frac{33976}{27} \zeta_2
\nonumber\\&
        +1664 \ln (2) \zeta_2
        +\frac{3472}{45} \zeta_2^2
        +\frac{14440}{9} \zeta_3
        -64 \zeta_2 \zeta_3
        -\frac{1840}{3} \zeta_5
        +L \bigg(
                -\frac{34664}{81}
                -\frac{1576}{3} \zeta_2
                +\frac{944}{15} \zeta_2^2
\nonumber\\&
                +560 \zeta_3
        \bigg)
        +L^2 \bigg(
                \frac{104}{27}
                -\frac{1040 \zeta_2}{9}
                +\frac{232 \zeta_3}{3}
        \bigg)
        +L^3 \bigg(
                -\frac{160}{27}
                -8 \zeta_2
        \bigg)
        -\frac{116 L^4}{27}
        -\frac{2 L^5}{9}
\bigg\}
\nonumber\\&
+ C_F n_l^2 T_F^2 \bigg\{
        -\frac{11296}{81}
        -\frac{1952}{27} \zeta_2
        -\frac{128}{9} \zeta_3
        +L \bigg(
                \frac{8720}{81}
                +\frac{64 \zeta_2}{9}
        \bigg)
        +\frac{592 L^2}{27}
        +\frac{32 L^3}{27}
\bigg\}
\nonumber\\&
+ C_F n_h n_l T_F^2 \bigg\{
        -\frac{5408}{81}
        +\frac{11392}{27} \zeta_2
        +\frac{1664}{9} \zeta_3
        +L \bigg(
                \frac{15136}{81}
                +\frac{896 \zeta_2}{9}
        \bigg)
        +\frac{3104 L^2}{27}
        +\frac{448 L^3}{27}
\bigg\}
\bigg)
\nonumber\\&
+ x^2 \bigg(
N_C^3 \bigg\{
        \frac{521947}{216}
        -\frac{6272521}{648} \zeta_2
        -\frac{120907}{90} \zeta_2^2
        +\frac{516347}{315} \zeta_2^3
        +\frac{1655092}{81} \zeta_3
        +\frac{24502}{9} \zeta_2 \zeta_3
\nonumber\\&
        +93 \zeta_3^2
        -18684 \zeta_5
        +L \Bigl(
                \frac{724091}{2916}
                -\frac{1172257}{216} \zeta_2
                -\frac{12376}{5} \zeta_2^2
                +\frac{352654}{27} \zeta_3
                +\frac{223}{3} \zeta_2 \zeta_3
                -\frac{5998}{3} \zeta_5
        \bigg)
\nonumber\\&
        +L^2 \bigg(
                \frac{24851}{72}
                -\frac{35311}{72} \zeta_2
                -\frac{8579}{15} \zeta_2^2
                +\frac{38353}{18} \zeta_3
        \bigg)
        +L^3 \bigg(
                \frac{1165405}{1944}
                -\frac{43 \zeta_2}{54}
                -\frac{4373 \zeta_3}{9}
        \bigg)
\nonumber\\&
        +L^4 \bigg(
                -\frac{61187}{432}
                +\frac{1711 \zeta_2}{18}
        \bigg)
        -\frac{541 L^5}{540}
        -\frac{71 L^6}{180}
\bigg\}
+ C_F^2 n_l T_F \bigg\{
        -\frac{25951}{9}
        -\frac{256 c_1}{3}
        +\frac{52714}{9} \zeta_2
\nonumber\\&
        +7680 \ln (2) \zeta_2
        -\frac{3824}{45} \zeta_2^2
        -\frac{17152}{3} \zeta_3
        -\frac{3008}{3} \zeta_2 \zeta_3
        +\frac{16928}{3} \zeta_5
        +L \bigg(
                \frac{116420}{81}
                +\frac{43234}{9} \zeta_2
\nonumber\\&
                -\frac{10208}{15} \zeta_2^2
                -1520 \zeta_3
        \bigg)
        +L^2 \bigg(
                -\frac{127414}{81}
                +\frac{4160 \zeta_2}{9}
                -\frac{848 \zeta_3}{9}
        \bigg)
        +L^3 \bigg(
                -\frac{13786}{27}
                +\frac{440 \zeta_2}{3}
        \bigg)
\nonumber\\&
        -\frac{2398 L^4}{27}
        -\frac{124 L^5}{15}
\bigg\}
+ C_A C_F n_l T_F \bigg\{
        -\frac{299576}{81}
        +\frac{128 c_1}{3}
        +\frac{163612}{27} \zeta_2
        -3840 \ln (2) \zeta_2
\nonumber\\&
        -\frac{20584}{45} \zeta_2^2
        -\frac{1148368}{81} \zeta_3
        +\frac{4544}{9} \zeta_2 \zeta_3
        +\frac{20224}{3} \zeta_5
        +L \bigg(
                \frac{1311029}{729}
                +\frac{109804}{27} \zeta_2
                -552 \zeta_2^2
\nonumber\\&
                -\frac{18704}{3} \zeta_3
        \bigg)
        +L^2 \bigg(
                -\frac{5182}{9}
                +1080 \zeta_2
                -\frac{7792 \zeta_3}{9}
        \bigg)
        +L^3 \bigg(
                -\frac{22322}{243}
                +\frac{2456 \zeta_2}{27}
        \bigg)
        +\frac{232 L^4}{9}
\nonumber\\&
        +\frac{364 L^5}{135}
\bigg\}
+ C_F n_l^2 T_F^2 \bigg\{
        \frac{8296}{81}
        +\frac{8896}{81} \zeta_2
        +\frac{1856}{45} \zeta_2^2
        +\frac{2816}{27} \zeta_3
        +L \bigg(
                -\frac{139040}{729}
                -\frac{1408 \zeta_2}{27}
\nonumber\\&
                -\frac{128 \zeta_3}{27}
        \bigg)
        +L^2 \bigg(
                -\frac{4160}{81}
                -\frac{64 \zeta_2}{9}
        \bigg)
        -\frac{704 L^3}{81}
        -\frac{16 L^4}{27}
\bigg\}
+ C_F n_h n_l T_F^2 \bigg\{
        \frac{25040}{81}
        -\frac{146656}{81} \zeta_2
\nonumber\\&
        +\frac{256}{3} \zeta_2^2
        -\frac{25856}{27} \zeta_3
        +L \bigg(
                -\frac{4448}{3}
                -\frac{4768 \zeta_2}{9}
                +\frac{1472 \zeta_3}{27}
        \bigg)
        +L^2 \bigg(
                -\frac{53968}{81}
                +\frac{256 \zeta_2}{9}
        \bigg)
\nonumber\\&
        -\frac{7072 L^3}{81}
        +\frac{64 L^4}{27}
\bigg\}
\bigg)
\bigg]  \,.
%
\end{align}
In this limit, the electric component of the vector form factor i.e. $F_{V,1}$ satisfies 
the Sudakov evolution equation. This behavior has been studied in detail in \cite{Blumlein:2018tmz,Mitov:2006xs, 
Ahmed:2017gyt} accounting for the components known. The complete three-loop result has been given in 
\cite{Blumlein:2018tmz} 
and partial four--loop results in \cite{Blumlein:2018tmz,Ahmed:2017gyt}.
\begin{align}
{F}_{V,2}^{(3)} & \simeq
 \frac{1}{\ep^2} \bigg[
%
 x \bigg(
N_C^3 \bigg\{
-\frac{14 L}{3}-\frac{17 L^2}{3}-L^3
\bigg\}
+ C_F^2 n_l T_F
\frac{16}{3} L (1+L)
\bigg)
\bigg]
+ \frac{1}{\ep} \bigg[
%
 x \bigg(
N_C^3 \bigg\{
        -6
        +\frac{85}{3} \zeta_2
\nonumber\\&
        -44 \zeta_3
        +L \bigg(
                -\frac{15}{2}
                +\frac{112 \zeta_2}{3}
                -42 \zeta_3
        \bigg)
        +L^2 \bigg(
                -\frac{17}{3}
                +9 \zeta_2
        \bigg)
        -\frac{23 L^3}{3}
        -\frac{3 L^4}{2}
\bigg\}
+ C_F^2 n_l T_F \bigg\{
         \frac{16}{3} \zeta_2
\nonumber\\&
        +L \bigg(
                -32
                +\frac{16 \zeta_2}{3}
        \bigg)
        -\frac{104 L^2}{3}
        -\frac{8 L^3}{3}
\bigg\}
\bigg)
+ x^2 \bigg(
N_C^3 \bigg\{
        \frac{466}{3}
        -516 \zeta_2
        -\frac{816}{5} \zeta_2^2
        +944 \zeta_3
\nonumber\\&
        +L \bigg(
                66
                -752 \zeta_2
                -\frac{816}{5} \zeta_2^2
                +1328 \zeta_3
        \bigg)
        +L^2 \bigg(
                -\frac{16}{3}
                -284 \zeta_2
                +384 \zeta_3
        \bigg)
        +L^3 \bigg(
                84
                -48 \zeta_2
        \bigg)
\bigg\}
\nonumber\\&
+ C_F^2 n_l T_F \bigg\{
\frac{32}{3} (-1+L) (1+L)
\bigg\}
\bigg)
\bigg]
+  \bigg[
%
 x \bigg(
N_C^3 \bigg\{
        \frac{548}{3}
        -\frac{39005}{54} \zeta_2
        +\frac{329}{15} \zeta_2^2
        +\frac{4697}{3} \zeta_3
        +154 \zeta_2 \zeta_3
\nonumber\\&
        -732 \zeta_5
        +L \bigg(
                \frac{2255}{162}
                -\frac{4100}{9} \zeta_2
                -\frac{246}{5} \zeta_2^2
                +\frac{754}{3} \zeta_3
        \bigg)
        +L^2 \bigg(
                \frac{7201}{54}
                +\frac{133 \zeta_2}{6}
                -13 \zeta_3
        \bigg)
\nonumber\\&
        +L^3 \bigg(
                \frac{277}{27}
                +\frac{19 \zeta_2}{2}
        \bigg)
        -\frac{227 L^4}{36}
        -\frac{5 L^5}{4}
\bigg\}
+ C_F^2 n_l T_F \bigg\{
        \frac{16}{3}
        -\frac{256 c_1}{9}
        -\frac{5440}{9} \zeta_2
        +\frac{5632}{3} \ln (2) \zeta_2
\nonumber\\&
        +\frac{2624}{15} \zeta_2^2
        -\frac{3328}{9} \zeta_3
        +L \Bigl(
                -\frac{4172}{9}
                +\frac{3176 \zeta_2}{9}
                +\frac{416 \zeta_3}{3}
        \Bigr)
        +L^2 \Bigl(
                -\frac{2936}{9}
                -\frac{56 \zeta_2}{3}
        \Bigr)
        -\frac{832 L^3}{9}
\nonumber\\&
        -\frac{80 L^4}{9}
\bigg\}
+ C_A C_F n_l T_F \bigg\{
        -\frac{496}{3}
        +\frac{128 c_1}{9}
        +\frac{16160}{27} \zeta_2
        -\frac{2816}{3} \ln (2) \zeta_2
        -96 \zeta_2^2
        -\frac{2192}{3} \zeta_3
\nonumber\\&
        +L \bigg(
                \frac{44320}{81}
                +\frac{2336 \zeta_2}{9}
                -\frac{512 \zeta_3}{3}
        \bigg)
        +L^2 \bigg(
                \frac{1256}{27}
                +\frac{80 \zeta_2}{3}
        \bigg)
        +\frac{64 L^3}{27}
\bigg\}
+ C_F n_l^2 T_F^2 \bigg\{
         \frac{1600}{27} \zeta_2
        +\frac{256}{9} \zeta_3
\nonumber\\&
        +L \bigg(
                -\frac{10144}{81}
                -\frac{128 \zeta_2}{9}
        \bigg)
        -\frac{800 L^2}{27}
        -\frac{64 L^3}{27}
\bigg\}
+ C_F n_l n_h T_F^2 \bigg\{        
        -\frac{2240}{27}
        -\frac{7424}{27} \zeta_2
        -\frac{256}{9} \zeta_3
\nonumber\\&
        +L \bigg(
                -\frac{20288}{81}
                -\frac{256 \zeta_2}{9}
        \bigg)
        -\frac{1600 L^2}{27}
        -\frac{128 L^3}{27}
\bigg\}
\bigg)
+ x^2 \bigg(
N_C^3 \bigg\{
        -\frac{71695}{27}
        +\frac{73108}{9} \zeta_2
        +\frac{13537}{15} \zeta_2^2
\nonumber\\&
        -\frac{41788}{35} \zeta_2^3
        -\frac{156496}{9} \zeta_3
        -1956 \zeta_2 \zeta_3
        -168 \zeta_3^2
        +16112 \zeta_5
        +L \bigg(
                \frac{19669}{27}
                +\frac{36889}{9} \zeta_2
                +\frac{8344}{5} \zeta_2^2
\nonumber\\&
                -\frac{32768}{3} \zeta_3
                +48 \zeta_2 \zeta_3
                +1656 \zeta_5
        \bigg)
        +L^2 \bigg(
                -\frac{10907}{9}
                +733 \zeta_2
                +\frac{2142}{5} \zeta_2^2
                -1130 \zeta_3
        \bigg) 
        +L^3 \bigg(
                -\frac{593}{3}
\nonumber\\&
                -126 \zeta_2
                +332 \zeta_3
        \bigg)
        +L^4 \bigg(
                126
                -72 \zeta_2
        \bigg)
\bigg\}
+ C_F^2 n_l T_F \bigg\{
        \frac{6112}{3}
        +\frac{512 c_1}{9}
        -\frac{51680}{9} \zeta_2
        -\frac{17408}{3} \ln (2) \zeta_2
\nonumber\\&
        +\frac{2240}{3} \zeta_2^2
        +\frac{11264}{3} \zeta_3
        +512 \zeta_2 \zeta_3
        -3968 \zeta_5
        +L \bigg(
                -\frac{15040}{9}
                -\frac{31168}{9} \zeta_2
                +\frac{9856}{15} \zeta_2^2
                +\frac{18880}{9} \zeta_3
        \bigg)
\nonumber\\&
        +L^2 \bigg(
                \frac{4432}{3}
                -\frac{6496 \zeta_2}{9}
                +\frac{704 \zeta_3}{3}
        \bigg)
        +L^3 \bigg(
                \frac{11344}{27}
                -\frac{320 \zeta_2}{3}
        \bigg)
        +\frac{928 L^4}{27}
        +\frac{16 L^5}{9}
\bigg\}
\nonumber\\&
+ C_A C_F n_l T_F \bigg\{
        \frac{63008}{27}
        -\frac{256 c_1}{9}
        -\frac{48448}{9} \zeta_2
        +\frac{8704}{3} \ln (2) \zeta_2
        +\frac{1936}{5} \zeta_2^2
        +\frac{102880}{9} \zeta_3
        -256 \zeta_2 \zeta_3
\nonumber\\&
        -4672 \zeta_5
        +L \bigg(
                -\frac{15200}{27}
                -\frac{32096}{9} \zeta_2
                +\frac{8128}{15} \zeta_2^2
                +\frac{41920}{9} \zeta_3
        \bigg)
        +L^2 \bigg(
                \frac{6640}{9}
                -\frac{8272 \zeta_2}{9}
                +\frac{1568 \zeta_3}{3}
        \bigg)
\nonumber\\&
        +L^3 \bigg(
                \frac{736}{27}
                -\frac{224 \zeta_2}{3}
        \bigg)
        -\frac{464 L^4}{27}
        -\frac{8 L^5}{9}
\bigg\}
+ C_F n_l^2 T_F^2 \bigg\{
        -\frac{3968}{27}
        -\frac{256}{9} \zeta_2
        +\frac{3200 L}{27}
        +\frac{128 L^2}{9}
\bigg\}
\nonumber\\&
+ C_F n_l n_h T_F^2 \bigg\{
        \frac{512}{27}
        +\frac{11776}{9} \zeta_2
        +\frac{2048}{3} \zeta_3
        +L \bigg(
                \frac{30976}{27}
                +\frac{1024 \zeta_2}{3}
        \bigg)
        +\frac{5120 L^2}{9}
        +\frac{512 L^3}{9}
\bigg\}
\bigg)
\bigg].
%
\end{align}

\subsection{Threshold region  \boldmath $x \rightarrow -1$}
\noindent
We consider the parameter 
\begin{equation}
\beta = \sqrt{1 -\frac{4 m^2}{q^2}} 
\end{equation}
to perform the expansion of the form factors
in the threshold region $q^2 \sim 4 m^2$ or $x \rightarrow -1$.
In $\beta$, the limit translates to $\beta \rightarrow 0$ and we expand the form factors up to ${\cal 
O}(\beta^4)$. Here the physical quantities like the decay rates and production cross sections, get 
contributions 
only from the form factors, as the contributions from real radiation are suppressed. One obtains 
\begin{align}
{F}_{V,1}^{(3)} & \simeq
\frac{1}{\ep^3} \bigg[
N_C^3 \bigg\{
 \frac{11 \zeta_2}{4 \beta ^2}
+\frac{9}{2} \zeta_2
+ i \pi \bigg(
 \frac{\zeta_2}{8 \beta ^3}
+\frac{1}{\beta } \left( -\frac{121}{54}+\frac{3 \zeta_2}{8} \right)
\bigg)
\bigg\}
+ C_F^2 n_l T_F \bigg\{
-\frac{4 \zeta_2}{\beta ^2}
-8 \zeta_2
%
%
\bigg\}
\nonumber\\ &
+ C_A C_F n_l T_F \bigg\{
 i \pi \bigg(
\frac{88}{27 \beta }
\bigg)
\bigg\}
+ C_F n_l^2 T_F^2 \bigg\{
 i \pi \bigg(
-\frac{16}{27 \beta }
\bigg)
\bigg\}
%
%
\bigg]
%
\nonumber\\ & 
+ \frac{1}{\ep^2} \bigg[
N_C^3 \bigg\{
-\frac{9 \zeta_2^2}{4 \beta ^3}
+\frac{1}{\beta ^2} \left( \frac{29}{12} \zeta_2
-\frac{11}{2} \zeta_2 \ln (2)
-\frac{11}{2} \zeta_2 \ln (\beta ) \right)
-\frac{27 \zeta_2^2}{4 \beta }
+\frac{31}{12} \zeta_2
-15 \zeta_2 \ln (2)
\nonumber\\ &
-11 \zeta_2 \ln (\beta )
+ i \pi \bigg(
 \frac{1}{\beta ^3} \left( \frac{3}{8} \zeta_2-\frac{3}{4} \zeta_2 \ln (2)-\frac{3}{4} \zeta_2 \ln (\beta )  
\right)
+\frac{11 \zeta_2}{4 \beta ^2} 
+\frac{1}{\beta} \Bigl( \frac{403}{324}
+\frac{9}{8} \zeta_2
-\frac{9}{4} \zeta_2 \ln (2)
\nonumber\\ &
-\frac{9}{4} \zeta_2 \ln (\beta ) \Bigr)
+\frac{11}{2} \zeta_2
\bigg)
\bigg\}
+ C_F^2 n_l T_F \bigg\{
\frac{1}{\beta ^2} \left( \frac{8}{3} \zeta_2
+8 \zeta_2 \ln (2)
+8 \zeta_2 \ln (\beta )  \right)
+\frac{16}{3} \zeta_2
+16 \zeta_2 \ln (2)
\nonumber\\ &
+16 \zeta_2 \ln (\beta )
- 4 i \pi \bigg(
\frac{\zeta_2}{\beta ^2}
-\frac{2}{3 \beta }
+2 \zeta_2
\bigg)
\bigg\}
- i C_A C_F n_l T_F  
  \frac{524 \pi}{81 \beta}
\nonumber\\ &
+ i C_F n_l^2 T_F^2 
\frac{80 \pi}{81 \beta }
%
%
\bigg]
%
+ \frac{1}{\ep} \bigg[
N_C^3 \bigg\{
 \frac{1}{\beta^3} \left(
        -\frac{27}{4} \zeta_2^2
        +\frac{27}{2} \zeta_2^2 \ln (2)
        +\frac{27}{2} \zeta_2^2 \ln (\beta )
\right)
+ \frac{1}{\beta^2} \Bigl(
        -\frac{431}{36} \zeta_2
\nonumber\\ &
        -\frac{33}{8} \zeta_2^2
        +15 \zeta_2 \ln (2)
        -\frac{11}{2} \zeta_2 \ln^2 (2)
        +\Bigl(
                15 \zeta_2
                -11 \ln (2) \zeta_2
        \Bigr) \ln (\beta )
        -\frac{11}{2} \zeta_2 \ln ^2(\beta )
\Bigr)
+ \frac{1}{\beta} \Bigl(
        \frac{303}{20} \zeta_2^2
\nonumber\\ &
        +\frac{81}{2} \zeta_2^2 \ln (2)
        +\frac{81}{2} \zeta_2^2 \ln (\beta )
\Bigr)
-\frac{1103}{12} \zeta_2
-\frac{365}{12} \zeta_2^2
+24 \zeta_2 \ln (2)
+45 \zeta_2 \ln^2 (2)
+\Bigl(
        \frac{104 \zeta_2}{3}
\nonumber\\ &
        +50 \ln (2) \zeta_2
\Bigr) \ln (\beta )
+9 \zeta_2 \ln ^2(\beta )
+ i \pi \bigg(
\frac{1}{\beta^3} \Bigl(
        \Bigl(
                -\frac{9 \zeta_2}{4}
                +\frac{9}{2} \ln (2) \zeta_2
        \Bigr) \ln (\beta )
        +2 \zeta_2
        -\frac{21}{16} \zeta_2^2
\nonumber\\ &
        -\frac{9}{4} \zeta_2 \ln (2)
        +\frac{9}{4} \zeta_2 \ln^2 (2)
        +\frac{9}{4} \zeta_2 \ln ^2(\beta )
\Bigr)
+ \frac{1}{\beta^2} \left(
        -\frac{15}{2} \zeta_2
        +\frac{11}{2} \zeta_2 \ln (2)
        +\frac{11}{2} \zeta_2 \ln (\beta )
\right)
\nonumber\\ &
+ \frac{1}{\beta} \Bigl(
        \frac{7573}{1080}
        +\Bigl(
                \frac{101 \zeta_2}{20}
                +\frac{27}{2} \ln (2) \zeta_2
        \Bigr) \ln (\beta )
        -\frac{7961}{600} \zeta_2
        -\frac{39}{16} \zeta_2^2
        +\frac{3277}{360} \zeta_3
\nonumber\\ &
        +\frac{219}{20} \zeta_2 \ln (2)
        +\frac{27}{4} \zeta_2 \ln^2 (2)
        +\frac{27}{4} \zeta_2 \ln ^2(\beta )
\Bigr)
-\frac{52}{3} \zeta_2
-25 \zeta_2 \ln (2)
-9 \zeta_2 \ln (\beta )
\bigg)
\bigg\}
\nonumber\\ &
+ C_F^2 n_l T_F \bigg\{
\frac{1}{\beta ^2}   \Bigl( \frac{328}{9} \zeta_2
+6 \zeta_2^2-48 \zeta_2 \ln (2)
+8 \zeta_2 \ln^2 (2)
+\left(
        -48 \zeta_2
        +16 \ln (2) \zeta_2
\right) \ln (\beta )
\nonumber\\ &
+8 \zeta_2 \ln ^2(\beta )  \Bigr)
+\frac{512}{9} \zeta_2
+12 \zeta_2^2
-96 \zeta_2 \ln (2)
+16 \zeta_2 \ln^2 (2)
+\left(
        -96 \zeta_2
        +32 \ln (2) \zeta_2
\right) \ln (\beta )
\nonumber\\ &
+16 \zeta_2 \ln ^2(\beta )
+ i \pi \bigg(
\frac{1}{\beta^2} \left(  24 \zeta_2 -8 \zeta_2 \ln (2) -8 \zeta_2 \ln (\beta ) \right)
+\frac{1}{\beta}  \left( -\frac{65}{9}-\frac{16 \zeta_3}{3}  \right)
+48 \zeta_2
\nonumber\\ &
-16 \zeta_2 \ln (2)
-16 \zeta_2 \ln (\beta )
\bigg)
\bigg\}
+ C_A C_F n_l T_F \bigg\{
 i \pi \bigg(
\frac{1}{\beta } \Bigl( -\frac{62}{81}+\frac{56 \zeta_3}{9}  \Bigr)
\bigg)
\bigg\}
\nonumber\\ &
+ C_F n_l^2 T_F^2 \bigg\{
 i \pi \bigg(
\frac{16}{81 \beta }
\bigg)
\bigg\}
+ C_F n_l n_h T_F^2 \bigg\{
 i \pi \bigg(
\frac{8 \zeta_2}{9 \beta }
\bigg)
\bigg\}
\bigg]
\nonumber\\ &
%
+ \bigg[
N_C^3 \bigg\{
 \frac{1}{\beta^3} \Bigl(
        -36 \zeta_2^2
        -\frac{135}{8} \zeta_2^3
        +\frac{81}{2} \zeta_2^2 \ln (2)
        -\frac{81}{2} \zeta_2^2 \ln^2 (2)
        +\left(
                \frac{81}{2} \zeta_2^2
                -81 \zeta_2^2 \ln (2)
        \right) \ln (\beta )
\nonumber\\ &
        -\frac{81}{2} \zeta_2^2 \ln ^2(\beta )
\Bigr)
+ \frac{1}{\beta^2} \Bigl(
        \Bigl(
                -104 \zeta_2
                +121 \ln (2) \zeta_2
        \Bigr) \ln ^2(\beta )
        -\frac{6625}{108} \zeta_2
        +\frac{791}{24} \zeta_2^2
        -\frac{1001}{12} \zeta_2 \zeta_3
\nonumber\\ &
        +\frac{365}{2} \zeta_2 \ln (2)
        -\frac{121}
        {4} \zeta_2^2 \ln (2)
        -104 \zeta_2 \ln^2 (2)
        +
        \frac{121}{3} \zeta_2 \ln^3 (2)
        +\Bigl(
                \frac{365}{2} \zeta_2
                -\frac{121}{4} \zeta_2^2
                -208 \zeta_2 \ln (2)
\nonumber\\ &
                +121 \zeta_2 \ln^2 (2)
        \Bigr) \ln (\beta )
        +\frac{121}{3} \zeta_2 \ln ^3(\beta )
\Bigr)
+ \frac{1}{\beta} \Bigl(
        \Bigl(
                \frac{484 \zeta_2}{3}
                -\frac{243 \zeta_2^2}{2}
        \Bigr) \ln ^2(\beta )
        +\frac{178301}{540} \zeta_2
\nonumber\\ &
        +\frac{161437}{300} \zeta_2^2
        -\frac{621}{8} \zeta_2^3
        -\frac{799}{20} \zeta_2 \zeta_3
        -\frac{3320}{9} \zeta_2 \ln (2)
        -\frac{1971}{10} \zeta_2^2 \ln (2)
        +\frac{484}{3} \zeta_2 \ln^2 (2)
        -\frac{243}{2} \zeta_2^2 \ln^2 (2)
\nonumber\\ &
        +\Bigl(
                -\frac{3320}{9} \zeta_2
                -\frac{1617}{10} \zeta_2^2
                +\frac{968}{3} \zeta_2 \ln (2)
                -243 \zeta_2^2 \ln (2)
        \Bigr) \ln (\beta )
\Bigr)
-\frac{102923}{360}
\nonumber\\ &
-\frac{495}{4} \HA_{0,0,\{3,0\},1}(1)
-\frac{495}{2} \HA_{0,0,\{3,1\},1}(1)
+\frac{432}{5} \HA_{0,0,\{6,0\},1,-1}(1)
-\frac{864}{5} \HA_{0,0,\{6,1\},1,-1}(1)
\nonumber\\ &
-\frac{557351 \zeta_2}{1125}
+64 \pi  \frac{\HA_{0,\{6,0\}}(1) \zeta_2}{\sqrt{3}}
+55 \HA_{\{6,0\},-1}(1) \zeta_2
-110 \HA_{\{6,1\},-1}(1) \zeta_2
+\frac{144}{5} \HA_{\{6,0\},1,-1}(1) \zeta_2
\nonumber\\ &
-\frac{288}{5} \HA_{\{6,1\},1,-1}(1) \zeta_2
-\frac{104039}{400} \zeta_2^2
-\frac{1473169 \zeta_3}{2160}
-\frac{17485}{48} \zeta_2 \zeta_3
-\frac{277337}{960} \zeta_5
+\frac{2215}{108} \zeta_2 \ln (2)
\nonumber\\ &
+\frac{49509}{200} \zeta_2^2 \ln (2)
-\frac{3986}{5} \zeta_2 \ln^2 (2)
-\frac{704}{3} \zeta_2 \ln^3 (2)
+\frac{17459 c_1}{1080}
+\frac{32}{5} \ln (2) c_1
+\frac{571 c_2}{300}
\nonumber\\ &
+\Bigl(
        \frac{41777}{450} \zeta_2
        +\frac{1013}{6} \zeta_2^2
        -\frac{1738}{3} \zeta_2 \ln (2)
        -310 \zeta_2 \ln^2 (2)
\Bigr) \ln (\beta )
+\Bigl(
        -\frac{2681 \zeta_2}{15}
\nonumber\\ &
        -150 \ln (2) \zeta_2
\Bigr) \ln ^2(\beta )
-\frac{22}{3} \zeta_2 \ln ^3(\beta )
+ i \pi \bigg(
\frac{1}{\beta^3} \Bigl(
        \Bigl(
                \frac{27 \zeta_2}{4}
                -\frac{27}{2} \ln (2) \zeta_2
        \Bigr) \ln ^2(\beta )
        +\frac{1}{2} \zeta_2
        -\frac{63}{16} \zeta_2^2
\nonumber\\ &
        +\frac{97}{8} \zeta_2 \zeta_3
        -12 \zeta_2 \ln (2)
        +\frac{63}{8} \zeta_2^2 \ln (2)
        +\frac{27}{4} \zeta_2 \ln^2 (2)
        -\frac{9}{2} \zeta_2 \ln^3 (2)
        +\Bigl(
                -12 \zeta_2
                +\frac{63}{8} \zeta_2^2
\nonumber\\ &
                +\frac{27}{2} \zeta_2 \ln (2)
                -\frac{27}{2} \zeta_2 \ln^2 (2)
        \Bigr) \ln (\beta )
        -\frac{9}{2} \zeta_2 \ln ^3(\beta )
\Bigr)
+ \frac{1}{\beta^2} \Bigl(
        -\frac{365}{4} \zeta_2
        -\frac{363}{8} \zeta_2^2
        +104 \zeta_2 \ln (2)
\nonumber\\ &
        -\frac{121}{2} \zeta_2 \ln^2 (2)
        +\Bigl(
                104 \zeta_2
                -121 \ln (2) \zeta_2
        \Bigr) \ln (\beta )
        -\frac{121}{2} \zeta_2 \ln ^2(\beta )
\Bigr)
+ \frac{1}{\beta} \Bigl(
        -\frac{5285431}{291600}
\nonumber\\ &
        -\frac{983849 \zeta_2}{9000}
        +\frac{50287}{400} \zeta_2^2
        +\frac{228767 \zeta_3}{5400}
        +\frac{291}{8} \zeta_2 \zeta_3
        +\frac{178301 \ln (2)}{1620}
        +\frac{15819}{100} \zeta_2 \ln (2)
\nonumber\\ &
        +\frac{165}{8} \zeta_2^2 \ln (2)
        -\frac{799}
        {60} \zeta_3 \ln (2)
        -\frac{1660}{27} \ln^2 (2)
        -\frac{303}{20} \zeta_2 \ln^2 (2)
        +\frac{484 \ln^3 (2)}{27}
        -\frac{27}{2} \zeta_2 \ln^3 (2)
\nonumber\\ &
        +\Bigl(
                \frac{178301}{1620}
                +\frac{37679}{300} \zeta_2
                +\frac{117}{8} \zeta_2^2
                -\frac{799}{60} \zeta_3
                -\frac{3320 \ln (2)}{27}
                -\frac{657}{10} \zeta_2 \ln (2)
                +\frac{484 \ln^2 (2)}{9}
\nonumber\\ &
                -\frac{81}{2} \zeta_2 \ln^2 (2)
        \Bigr) \ln (\beta )
        +\Bigl(
                -\frac{1660}{27}
                -\frac{539}{20} \zeta_2
                +\frac{484 \ln (2)}{9}
                -\frac{81}{2} \zeta_2 \ln (2)
        \Bigr) \ln ^2(\beta )
        +\Bigl(
                \frac{484}{27}
\nonumber\\ &
                -\frac{27 \zeta_2}{2}
        \Bigr) \ln ^3(\beta )
        -\frac{59 c_1}{60}
\Bigr)
-\frac{41777}{900} \zeta_2
-\frac{881}{12} \zeta_2^2
+\frac{869}{3} \zeta_2 \ln (2)
+155 \zeta_2 \ln^2 (2)
+\Bigl(
        \frac{2681 \zeta_2}{15}
\nonumber\\ &
        +150 \ln (2) \zeta_2
\Bigr) \ln (\beta )
+11 \zeta_2 \ln ^2(\beta )
\bigg)
\bigg\}
+ C_F^2 n_l T_F \bigg\{
 \frac{1}{\beta^2} \Bigl(
        \Bigl(
                208 \zeta_2
                -176 \ln (2) \zeta_2
        \Bigr) \ln ^2(\beta )
\nonumber\\ &
        +\frac{6062}{27} \zeta_2
        -\frac{236}{3} \zeta_2^2
        +\frac{364}{3} \zeta_2 \zeta_3
        -368 \zeta_2 \ln (2)
        +44 \zeta_2^2 \ln (2)
        +208 \zeta_2 \ln^2 (2)
        -\frac{176}{3} \zeta_2 \ln^3 (2)
\nonumber\\ &
        +\Bigl(
                -368 \zeta_2
                +44 \zeta_2^2
                +416 \zeta_2 \ln (2)
                -176 \zeta_2 \ln^2 (2)
        \Bigr) \ln (\beta )
        -\frac{176}{3} \zeta_2 \ln ^3(\beta )
\Bigr)
+ \frac{1}{\beta} \Bigl(
        2 \zeta_2
        +96 \zeta_2 \zeta_3
\nonumber\\ &
        -48 \zeta_2 \ln (2)
        -48 \zeta_2 \ln (\beta )
\Bigr)
+\frac{12902}{135}
-\frac{424586 \zeta_2}{3375}
-\frac{21418}{225} \zeta_2^2
+\frac{38468}{45} \zeta_3
+\frac{728}{3} \zeta_2 \zeta_3
\nonumber\\ &
+\frac{107072}{225} \zeta_2 \ln (2)
+88 \zeta_2^2 \ln (2)
+\frac{2144}{15} \zeta_2 \ln^2 (2)
-\frac{352}{3} \zeta_2 \ln^3 (2)
+\frac{764 c_1}{135}
+\Bigl(
        \frac{48112}{225} \zeta_2
        +88 \zeta_2^2
\nonumber\\ &
        +\frac{9536}{15} \zeta_2 \ln (2)
        -352 \zeta_2 \ln^2 (2)
\Bigr) \ln (\beta )
+\Bigl(
        \frac{4768 \zeta_2}{15}
        -352 \ln (2) \zeta_2
\Bigr) \ln ^2(\beta )
-\frac{352}{3} \zeta_2 \ln ^3(\beta )
\nonumber\\ &
+ i \pi \bigg(
 \frac{1}{\beta^2} \Bigl(
        184 \zeta_2
        +66 \zeta_2^2
        -208 \zeta_2 \ln (2)
        +88 \zeta_2 \ln^2 (2)
        +\Bigl(
                -208 \zeta_2
                +176 \ln (2) \zeta_2
        \Bigr) \ln (\beta )
\nonumber\\ &
        +88 \zeta_2 \ln ^2(\beta )
\Bigr)
+ \frac{1}{\beta} \Bigl(
        -\frac{1163}{54}
        +\Bigl(
                \frac{2}{3}
                -16 \ln (2)
                +32 \zeta_3
        \Bigr) \ln (\beta )
        -34 \zeta_2
        -\frac{16}{5} \zeta_2^2
        -\frac{296}{9} \zeta_3
\nonumber\\ &
        +\frac{2 \ln (2)}{3}
        +32 \zeta_3 \ln (2)
        -8 \ln^2 (2)
        -8 \ln ^2(\beta )
\Bigr)
-\frac{24056}{225} \zeta_2
+132 \zeta_2^2
-\frac{4768}{15} \zeta_2 \ln (2)
\nonumber\\ &
+176 \zeta_2 \ln^2 (2)
+\Bigl(
        -\frac{4768 \zeta_2}{15}
        +352 \ln (2) \zeta_2
\Bigr) \ln (\beta )
+176 \zeta_2 \ln ^2(\beta )
\bigg)
\bigg\}
\nonumber\\ &
+ C_A C_F n_l T_F \bigg\{
\frac{1}{\beta} \Bigl(
        -\frac{17228}{27} \zeta_2
        -\frac{704}{3} \zeta_2^2
        -112 \zeta_2 \zeta_3
        +\frac{6016}{9} \zeta_2 \ln (2)
        -\frac{704}{3} \zeta_2 \ln^2 (2)
        +\Bigl(
                \frac{6016 \zeta_2}{9}
\nonumber\\ &
                -\frac{1408}{3} \ln (2) \zeta_2
        \Bigr) \ln (\beta )        
        -\frac{704}{3} \zeta_2 \ln ^2(\beta )
\Bigr)
+\frac{12824}{45}
-\frac{938876 \zeta_2}{3375}
+\frac{76516}{225} \zeta_2^2
+\frac{3338}{45} \zeta_3
\nonumber\\ &
+\frac{13288}{25} \zeta_2 \ln (2)
+\frac{2048}{15} \zeta_2 \ln^2 (2)
-\frac{1208 c_1}{135}
+\Bigl(
        \frac{26864 \zeta_2}{75}
        -\frac{768}{5} \ln (2) \zeta_2
\Bigr) \ln (\beta )
\nonumber\\ &
-\frac{384}{5} \zeta_2 \ln ^2(\beta )
+ i \pi \bigg(
\frac{1}{\beta} \Bigl(
        \frac{111305}{729}
        -\frac{8}{3} \zeta_2
        +8 \zeta_2^2
        +\frac{76}{3} \zeta_3
        -\frac{17228 \ln (2)}{81}
        -\frac{112}{3} \zeta_3 \ln (2)
\nonumber\\ &
        +\frac{3008 \ln^2 (2)}{27}
        -\frac{704}{27} \ln^3 (2)
        +\Bigl(
                -\frac{17228}{81}
                -\frac{112}{3} \zeta_3
                +\frac{6016 \ln (2)}{27}
                -\frac{704}{9} \ln^2 (2)
        \Bigr) \ln (\beta )
\nonumber\\ &
        +\Bigl(
                \frac{3008}{27}-\frac{704 \ln (2)}{9}\Bigr) \ln ^2(\beta )
        -\frac{704}{27} \ln ^3(\beta )
\Bigr)
-\frac{13432}{75} \zeta_2
+\frac{384}{5} \zeta_2 \ln (2)
+\frac{384}{5} \zeta_2 \ln (\beta )
\bigg)
\bigg\}
\nonumber\\ &
+ C_F n_l^2 T_F^2 \bigg\{
\frac{1}{\beta} \Bigl(
        \Bigl(
                -\frac{1024 \zeta_2}{9}
                +\frac{256}{3} \ln (2) \zeta_2
        \Bigr) \ln (\beta )
        +\frac{3200}{27} \zeta_2
        +\frac{128}{3} \zeta_2^2
        -\frac{1024}{9} \zeta_2 \ln (2)
\nonumber\\ &
        +\frac{128}{3} \zeta_2 \ln^2 (2)
        +\frac{128}{3} \zeta_2 \ln ^2(\beta )
\Bigr)
-\frac{880}{27}
-\frac{128}{3} \zeta_2
+ i \pi \bigg(
\frac{1}{\beta} \Bigl(
        -\frac{24680}{729}
        -\frac{32}{27} \zeta_3
        +\frac{3200 \ln (2)}{81}
\nonumber\\ &
        -\frac{512}{27} \ln^2 (2)
        +\frac{128 \ln^3 (2)}{27}
        +\Bigl(
                \frac{3200}{81}-\frac{1024 \ln (2)}{27}+\frac{128 \ln^2 (2)}{9}\Bigr) \ln (\beta )
        +\Bigl(
                -\frac{512}{27}
\nonumber\\ &
+\frac{128 \ln (2)}{9}\Bigr) \ln ^2(\beta )
        +\frac{128 \ln ^3(\beta )}{27}
\Bigr)
\bigg)
\bigg\}
+ C_F n_l n_h T_F^2 \bigg\{
-\frac{1664}{9}
+\frac{13952}{135} \zeta_2
\nonumber\\ &
+ i \pi \bigg(
\frac{1}{\beta } \Bigl( -\frac{40 \zeta_2}{27}
-\frac{16 \zeta_3}{27}  \Bigr)
\bigg)
\bigg\}
\bigg].
\end{align}
The corresponding expansion for $F_{V,2}^{(3)}$ reads
\begin{align}
{F}_{V,2}^{(3)} & \simeq
%
%
%
%
%
%
%
%
%
%
%
%
%
 \frac{1}{\ep^2} \bigg[
N_C^3 \bigg\{
-\frac{2 \zeta_2}{\beta ^2}
+3 \zeta_2
+ i \pi \bigg(
-\frac{3 \zeta_2}{8 \beta ^3}
+\frac{1}{\beta } \Bigl( -\frac{11}{12}
-\frac{3 \zeta_2}{8} \Bigr)
\bigg)
\bigg\}
+ C_F^2 n_l T_F \bigg\{
\frac{4 \zeta_2}{\beta ^2}
+ i \pi \bigg(
\frac{4}{3 \beta}
\bigg)
\bigg\}
%
%
%
%
%
%
%
\bigg]
%
\nonumber\\ &
+ \frac{1}{\ep} \bigg[
N_C^3 \bigg\{
+\frac{27 \zeta_2^2}{4 \beta ^3}
+\frac{1}{\beta^2}  \Bigl( \frac{27}{4} \zeta_2 -\frac{23}{2} \zeta_2 \ln (2) - \frac{23}{2} \zeta_2 \ln (\beta )  \Bigr)
+\frac{387 \zeta_2^2}{20 \beta }
-\frac{69}{4} \zeta_2
+10 \zeta_2 \ln (2)
\nonumber\\ & 
-2 \zeta_2 \ln (\beta )
+ i \pi \bigg(
\frac{1}{\beta^3} \Bigl(
        -2 \zeta_2
        +\frac{9}{4} \zeta_2 \ln (2)
        +\frac{9}{4} \zeta_2 \ln (\beta )
\Bigr)
+\frac{23}{4} \frac{1}{\beta^2} \zeta_2
+ \frac{1}{\beta} \Bigl(
        -\frac{137}{120}
        -\frac{3439}{600} \zeta_2
\nonumber\\ &
        +\frac{147}{40} \zeta_3
        +\frac{171}{20} \zeta_2 \ln (2)
        +\frac{129}{20} \zeta_2 \ln (\beta )
\Bigr)
+\zeta_2
\bigg)
\bigg\}
+ C_F^2 n_l T_F \bigg\{
\frac{1}{\beta^2} \Bigl(
        -24 \zeta_2
        +8 \zeta_2 \ln (2)
\nonumber\\ &
        +8 \zeta_2 \ln (\beta )
\Bigr)
+4 \zeta_2
+ i \pi \bigg(
-\frac{4 \zeta_2}{\beta^2}
-\frac{16}{3 \beta}
\bigg)
\bigg\}
%
%
%
%
%
%
%
%
\bigg]
%
+ \bigg[
N_C^3 \bigg\{
 \frac{1}{\beta^3} \Bigl(
        36 \zeta_2^2
        -\frac{81}{2} \zeta_2^2 \ln (2)
        -\frac{81}{2} \zeta_2^2 \ln (\beta )
\Bigr)
\nonumber\\ &
+ \frac{1}{\beta^2} \Bigl(
         \frac{2263}{36} \zeta_2
        -\frac{49}{2} \zeta_2^2
        -\frac{907}{6} \zeta_2 \ln (2)
        +\frac{163}{2} \zeta_2 \ln^2 (2)
        +\Bigl(
                -\frac{907 \zeta_2}{6}
                +163 \ln (2) \zeta_2
        \Bigr) \ln (\beta )
\nonumber\\ &
        +\frac{163}{2} \zeta_2 \ln ^2(\beta )
\Bigr)
+ \frac{1}{\beta} \Bigl(
        -\frac{55247}{180} \zeta_2
        +\frac{10321}{100} \zeta_2^2
        -\frac{441}{20} \zeta_2 \zeta_3
        +\frac{712}{3} \zeta_2 \ln (2)
        -\frac{1539}{10} \zeta_2^2 \ln (2)
\nonumber\\ &
        +\Bigl(
                \frac{712 \zeta_2}{3}
                -\frac{1413 \zeta_2^2}{10}
        \Bigr) \ln (\beta )
\Bigr)
-\frac{334903}{3240}
+\frac{81}{2} \HA_{0,0,\{3,0\},1}(1)
+81 \HA_{0,0,\{3,1\},1}(1)
\nonumber\\ &
+\frac{648}{5} \HA_{0,0,\{6,0\},1,-1}(1)
-\frac{1296}{5} \HA_{0,0,\{6,1\},1,-1}(1)
-\frac{4847663 \zeta_2}{13500}
+32 \pi  \sqrt{3} \HA_{0,\{6,0\}}(1) \zeta_2
\nonumber\\ &
-18 \HA_{\{6,0\},-1}(1) \zeta_2
+36 \HA_{\{6,1\},-1}(1) \zeta_2
+\frac{216}{5} \HA_{\{6,0\},1,-1}(1) \zeta_2
-\frac{432}{5} \HA_{\{6,1\},1,-1}(1) \zeta_2
-\frac{8071}{400} \zeta_2^2
\nonumber\\ &
-\frac{372161 \zeta_3}{2160}
-\frac{2615}{16} \zeta_2 \zeta_3
-\frac{169091}{320} \zeta_5
+\frac{40967}{108} \zeta_2 \ln (2)
-\frac{7399}{200} \zeta_2^2 \ln (2)
-\frac{1419}{5} \zeta_2 \ln^2 (2)
\nonumber\\ &
-86 \zeta_2 \ln^3 (2)
+\frac{329 c_1}{120}
+\frac{48}{5} \ln (2) c_1
+\frac{313 c_2}{100}
+\Bigl(
        \frac{22091 \zeta_2}{150}
        -228 \ln (2) \zeta_2
\Bigr) \ln (\beta )
\nonumber\\ &
-\frac{138}{5} \zeta_2 \ln ^2(\beta )
+ i \pi \bigg(
 \frac{1}{\beta^3} \Bigl(
        -\frac{1}{2} \zeta_2
        +\frac{63}{16} \zeta_2^2
        +12 \zeta_2 \ln (2)
        -\frac{27}{4} \zeta_2 \ln^2 (2)
        +\Bigl(
                12 \zeta_2
\nonumber\\ &
                -\frac{27}{2} \ln (2) \zeta_2
        \Bigr) \ln (\beta )
        -\frac{27}{4} \zeta_2 \ln ^2(\beta )
\Bigr)
+ \frac{1}{\beta^2} \Bigl(
        \frac{907}{12} \zeta_2
        -\frac{163}{2} \zeta_2 \ln (2)
        -\frac{163}{2} \zeta_2 \ln (\beta )
\Bigr)
\nonumber\\ &
+ \frac{1}{\beta} \Bigl(
        \frac{3085859}{32400}
        +\frac{55349 \zeta_2}{9000}
        +\frac{21363}{400} \zeta_2^2
        +\frac{4487}{600} \zeta_3
        -\frac{55247 \ln (2)}{540}
        +\frac{11443}{300} \zeta_2 \ln (2)
\nonumber\\ &
        -\frac{147}{20} \zeta_3 \ln (2)
        +\frac{356 \ln^2 (2)}{9}
        -\frac{387}{20} \zeta_2 \ln^2 (2)
        -\frac{7 c_1}{20}
        +\Bigl(
                -\frac{55247}{540}
                +\frac{10321}{300} \zeta_2
                -\frac{147}{20} \zeta_3
\nonumber\\ &
                +\frac{712 \ln (2)}{9}
                -\frac{513}{10} \zeta_2 \ln (2)
        \Bigr) \ln (\beta )
        +\Bigl(
                \frac{356}{9}
                -\frac{471 \zeta_2}{20}
        \Bigr) \ln ^2(\beta )
\Bigr)
-\frac{22091}{300} \zeta_2
+114 \zeta_2 \ln (2)
\nonumber\\ &
+\frac{138}{5} \zeta_2 \ln (\beta )
\bigg)
\bigg\}
+ C_F^2 n_l T_F \bigg\{
 \frac{1}{\beta^2} \Bigl(
        -\frac{1802}{9} \zeta_2
        +22 \zeta_2^2
        +\frac{880}{3} \zeta_2 \ln (2)
        -88 \zeta_2 \ln^2 (2)
\nonumber\\ &
        -88 \zeta_2 \ln ^2(\beta )
        +\Bigl(
                \frac{880 \zeta_2}{3}
                -176 \ln (2) \zeta_2
        \Bigr) \ln (\beta )        
\Bigr)
+ \frac{1}{\beta} \Bigl(
        \frac{160}{3} \zeta_2
        -32 \zeta_2 \ln (2)
        -32 \zeta_2 \ln (\beta )
\Bigr)
\nonumber\\ &
-\frac{5462}{135}
+\frac{2822836 \zeta_2}{3375}
-\frac{79502}{225} \zeta_2^2
+\frac{11492}{45} \zeta_3
-\frac{175672}{225} \zeta_2 \ln (2)
-\frac{3584}{15} \zeta_2 \ln^2 (2)
+\frac{2596 c_1}{135}
\nonumber\\ &
+\Bigl(
        -\frac{968 \zeta_2}{25}
        +\frac{128}{5} \ln (2) \zeta_2
\Bigr) \ln (\beta )
+\frac{64}{5} \zeta_2 \ln ^2(\beta )
+ i \pi \bigg(
 \frac{1}{\beta^2} \Bigl(
        -\frac{440}{3} \zeta_2
        +88 \zeta_2 \ln (2)
\nonumber\\ &
        +88 \zeta_2 \ln (\beta )
\Bigr)
+ \frac{1}{\beta} \Bigl(
        -\frac{389}{9}
        -\frac{34}{3} \zeta_2
        +16 \zeta_3
        +\frac{160 \ln (2)}{9}
        -\frac{16}{3} \ln^2 (2)
        +\Bigl(
                \frac{160}{9}
\nonumber\\ &
-\frac{32 \ln (2)}{3}\Bigr) \ln (\beta )
        -\frac{16}{3} \ln ^2(\beta )
\Bigr)
+\frac{484}{25} \zeta_2
-\frac{64}{5} \zeta_2 \ln (2)
-\frac{64}{5} \zeta_2 \ln (\beta )
\bigg)
\bigg\}
\nonumber\\ &
+ C_A C_F n_l T_F \bigg\{
 \frac{1}{\beta} \Bigl(
        \frac{4544}{9} \zeta_2
        -\frac{848}{3} \zeta_2 \ln (2)
        -\frac{848}{3} \zeta_2 \ln (\beta )
\Bigr)
+\frac{81784}{405}
-\frac{137236}{375} \zeta_2
+\frac{62884}{225} \zeta_2^2
\nonumber\\ &
-\frac{1348}{45} \zeta_3
+\frac{133808}{225} \zeta_2 \ln (2)
+\frac{1792}{15} \zeta_2 \ln^2 (2)
-\frac{256}{5} \zeta_2 \ln ^2(\beta )
-\frac{1592 c_1}{135}
+\Bigl(
        \frac{11936 \zeta_2}{75}
\nonumber\\ &
        -\frac{512}{5} \ln (2) \zeta_2
\Bigr) \ln (\beta )
+ i \pi \bigg(
\frac{1}{\beta} \Bigl(
        -\frac{12376}{81}
        +\frac{8}{3} \zeta_2
        -\frac{56}{3} \zeta_3
        +\frac{4544 \ln (2)}{27}
        -\frac{424}{9} \ln^2 (2)
        +\Bigl(
                \frac{4544}{27}
\nonumber\\ &
-\frac{848 \ln (2)}{9}\Bigr) \ln (\beta )
        -\frac{424}{9} \ln ^2(\beta )
\Bigr)
-\frac{5968}{75} \zeta_2
+\frac{256}{5} \zeta_2 \ln (2)
+\frac{256}{5} \zeta_2 \ln (\beta )
\bigg)
\bigg\}
\nonumber\\ &
+ C_F n_l^2 T_F^2 \bigg\{
 \frac{1}{\beta} \Bigl(
        -\frac{800}{9} \zeta_2
        +\frac{128}{3} \zeta_2 \ln (2)
        +\frac{128}{3} \zeta_2 \ln (\beta )
\Bigr)
-\frac{2576}{81}
-\frac{128}{9} \zeta_2
\nonumber\\ &
+ i \pi \bigg(
\frac{1}{\beta} \Bigl(
        \frac{2536}{81}
        -\frac{800 \ln (2)}{27}
        +\frac{64 \ln^2 (2)}{9}
        +\Bigl(
                -\frac{800}{27}+\frac{128 \ln (2)}{9}\Bigr) \ln (\beta )
        +\frac{64 \ln ^2(\beta )}{9}
\Bigr)
\bigg)
\bigg\}
\nonumber\\ &
+ C_F n_l n_h T_F^2 \bigg\{
-\frac{2848}{81}
+\frac{896}{45} \zeta_2
%
%
\bigg\}
\bigg].
\end{align}

\noindent
The constant $c_2$ is given by
\begin{eqnarray}
c_2 &=& 26 \zeta_2^2 \ln(2) - 20 \zeta_2 \ln^3(2) - \ln^5(2) + 120 \Li_5\left(\frac{1}{2}\right).
\end{eqnarray}
Also the following cyclotomic constants, i.e. the cyclotomic harmonic polylogarithms at $x=1$, 
\begin{eqnarray}
&& \{\HA_{0,\{6,0\}}(1), \HA_{\{6,0\},-1}(1), \HA_{\{6,1\},-1}(1), \HA_{\{6,0\},1,-1}(1), \HA_{\{6,1\},1,-1}(1),
\nonumber\\ &&
\HA_{0,0,\{3,0\},1}(1), \HA_{0,0,\{3,1\},1}(1), \HA_{0,0,\{6,0\},1,-1}(1), \HA_{0,0,\{6,1\},1,-1}(1)\}
\end{eqnarray}
contribute. Here the letters of cyclotomy 3 and 6 are
\begin{eqnarray}
\label{eq:CYCL}
\Bigl\{
f_{\{3,0\}}(x) &=&
\frac{1}{1+x+x^2},~~
f_{\{3,1\}}(x) =
\frac{x}{1+x+x^2},
\nonumber\\
f_{\{6,0\}}(x) 
&=& \frac{1}{1-x+x^2},~~
f_{\{6,1\}}(x) = \frac{x}{1-x+x^2}\Bigr\}.
\end{eqnarray}
The simpler cyclotomic constants  have been mapped to 
\begin{eqnarray}
\HA_{0,\{6,0\}}(1) &=&  \frac{2}{\sqrt{3}} {\rm Cl}_2\left(\frac{\pi}{3}\right)
\\
\HA_{\{6,1\},-1}(1) &=& \frac{1}{4} \Li_2\left(\frac{1}{4}\right) + \frac{\pi^2}{72} + 
\frac{1}{2} \ln^2(2) - \frac{1}{2} \ln(2) \ln(3) + \frac{1}{2} \HA_{\{6,0\},-1}(1)
\\
\HA_{\{6,1\},1,-1}(1) &=& - \frac{5 \pi}{18} {\rm Cl}_2\left(\frac{\pi}{3}\right) + 2 {\sf 
Re}\left(\Li_3\left(\frac{1+i \sqrt{3}}{4}\right)\right) + \frac{1}{4} \Li_2\left(\frac{1}{4}\right) \ln(2)
+\frac{17}{72} \zeta_3 +\frac{1}{6} \ln^3(2) 
\nonumber\\ &&
-\frac{\pi^2}{72} \ln(2)
+ \frac{1}{2} \HA_{\{6,0\},1,-1}(1),
\end{eqnarray}
\cite{Broadhurst:1998rz}, see also 
\cite{Henn:2015sem},\footnote{We gave $\pi$ here the preference instead of $\zeta_2 = \pi^2/6$ as it appears
individually.}
where ${\rm Cl}_2\left(\frac{\pi}{3}\right), \Li_2\left(\frac{1}{4}\right), \ln(3)$ and ${\sf
Re}\left(\Li_3\left(\frac{1+i\sqrt{3}}{4}\right)\right)$ seem to be all new independent constants 
\cite{Ablinger:2014bra,Ablinger:2014yaa} beyond the MZVs \cite{Blumlein:2009cf}, respectively, 
referring to functional representations w.r.t. the polylogarithms \cite{DUDE,DILOG4}. The Clausen functions 
\cite{CLAUSEN} are defined by
\begin{eqnarray}
{\rm Cl}_k(z) = {\sf Im} \Li_k\left(e^{iz}\right). 
\end{eqnarray}

\subsection{Checks}

\vspace*{1mm}
\noindent
By maintaining the gauge parameter $\xi$ to first order, a partial check on gauge invariance has been obtained.
After appropriately considering $\alpha_s$-decoupling, the UV renormalized results satisfy the universal IR 
structure, confirming again the correctness of all pole terms, see \cite{Blumlein:2018tmz}. Finally, we have 
compared our results with those of Ref.~\cite{Henn:2016tyf, Lee:2018nxa}, in the region $x \in [0,1]$ which 
have 
been obtained using a different method, and agree by adjusting the respective conventions. We also agree now 
with the results in \cite{Henn:2016tyf, Lee:2018nxa} for the expansions given in \cite{KIT}.

\subsection{Numerical Results}

\begin{figure}[H]
\centerline{%
\includegraphics[width=0.49\textwidth]{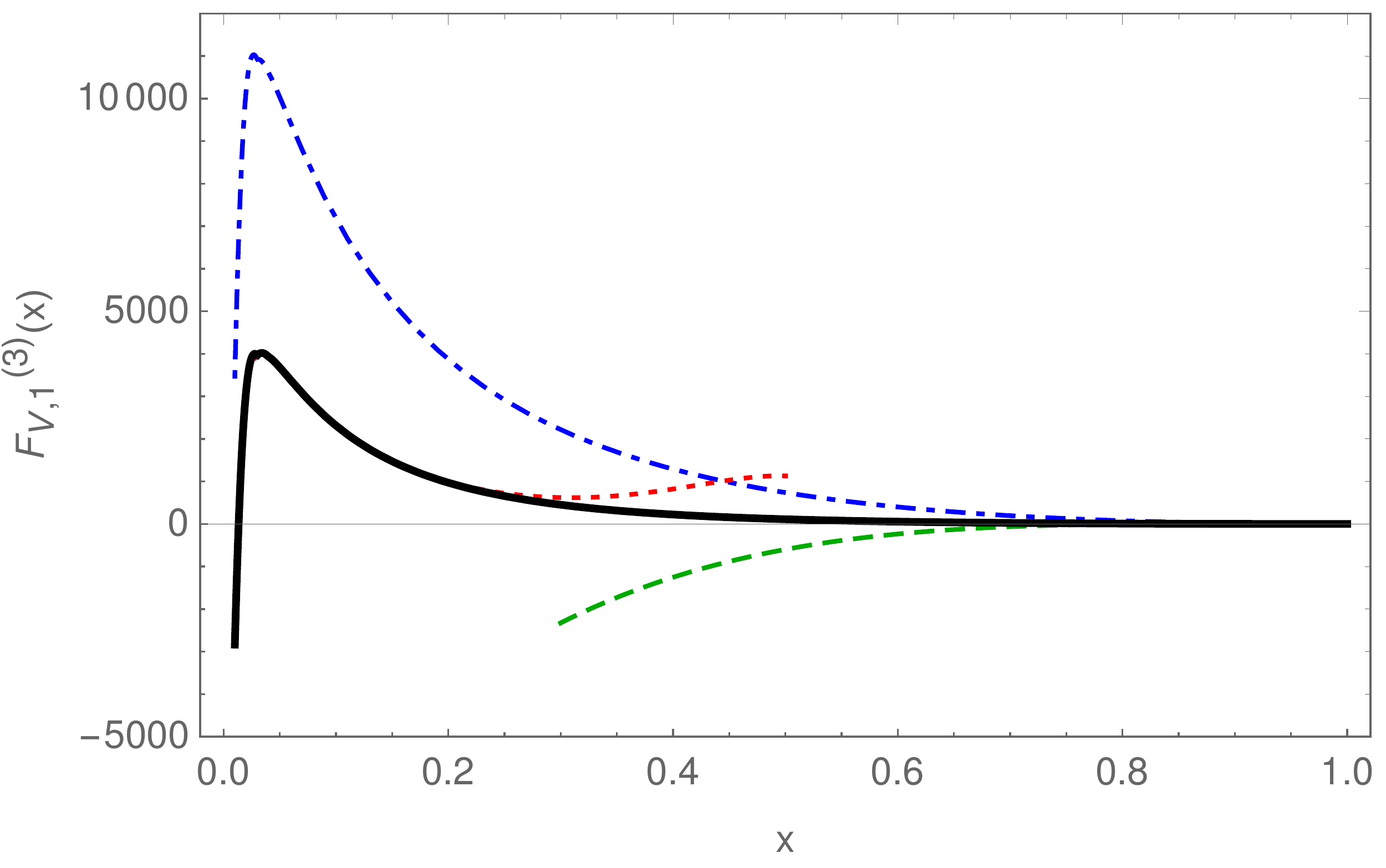}
\includegraphics[width=0.49\textwidth]{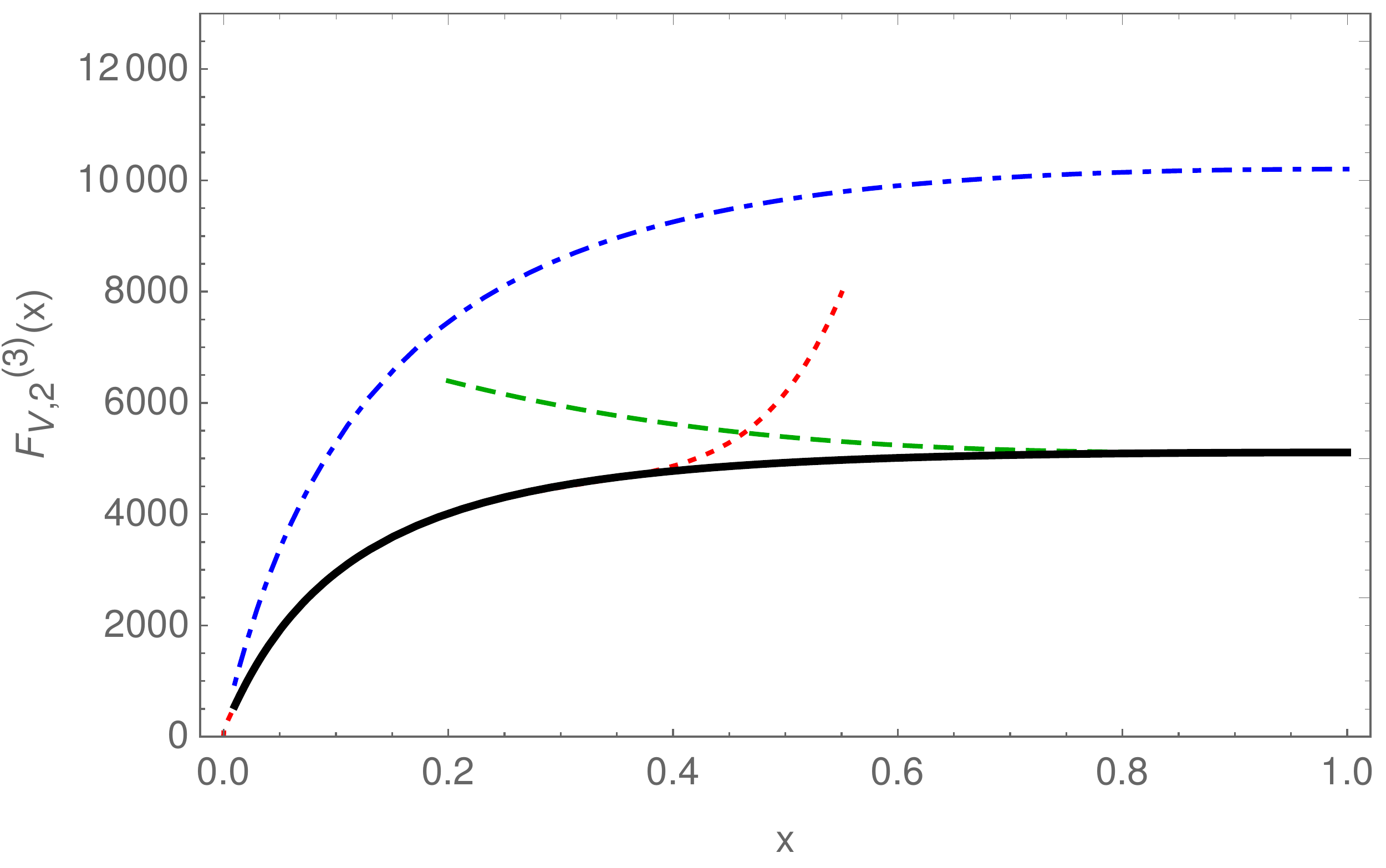}}
\caption{\sf The $O(\varepsilon^0)$ contribution to the vector three-loop form factors 
$F_{V,1}^{(3)}$ (left) and $F_{V,2}^{(3)}$ (right) as a function of $x \in [0,1]$. 
Dash-dotted line: leading color 
contribution of the non--singlet form factor; Full line: sum of the complete non--singlet $n_l$-contributions 
for 
$n_l =5$ and the color--planar non--singlet form factor; Dashed line: large $x$ expansion; Dotted line: small 
$x$
expansion.}
\label{fig:VF12ep0}
\end{figure}

\vspace*{1mm}
\noindent
The color planar parts to the three-loop vector form factors $F_{V,1}^{(3)}$ and $F_{V,2}^{(3)}$ are 
illustrated in Figure~\ref{fig:VF12ep0} in the range $x \in [0,1]$, showing the complete $n_l$ 
contributions as well. In this region the form factors are real. We also indicate  a series of 
expansion terms around $x=0$ and $x=1$, which are working in a wider kinematic region.

\begin{figure}[H]
\centerline{%
\includegraphics[width=0.49\textwidth]{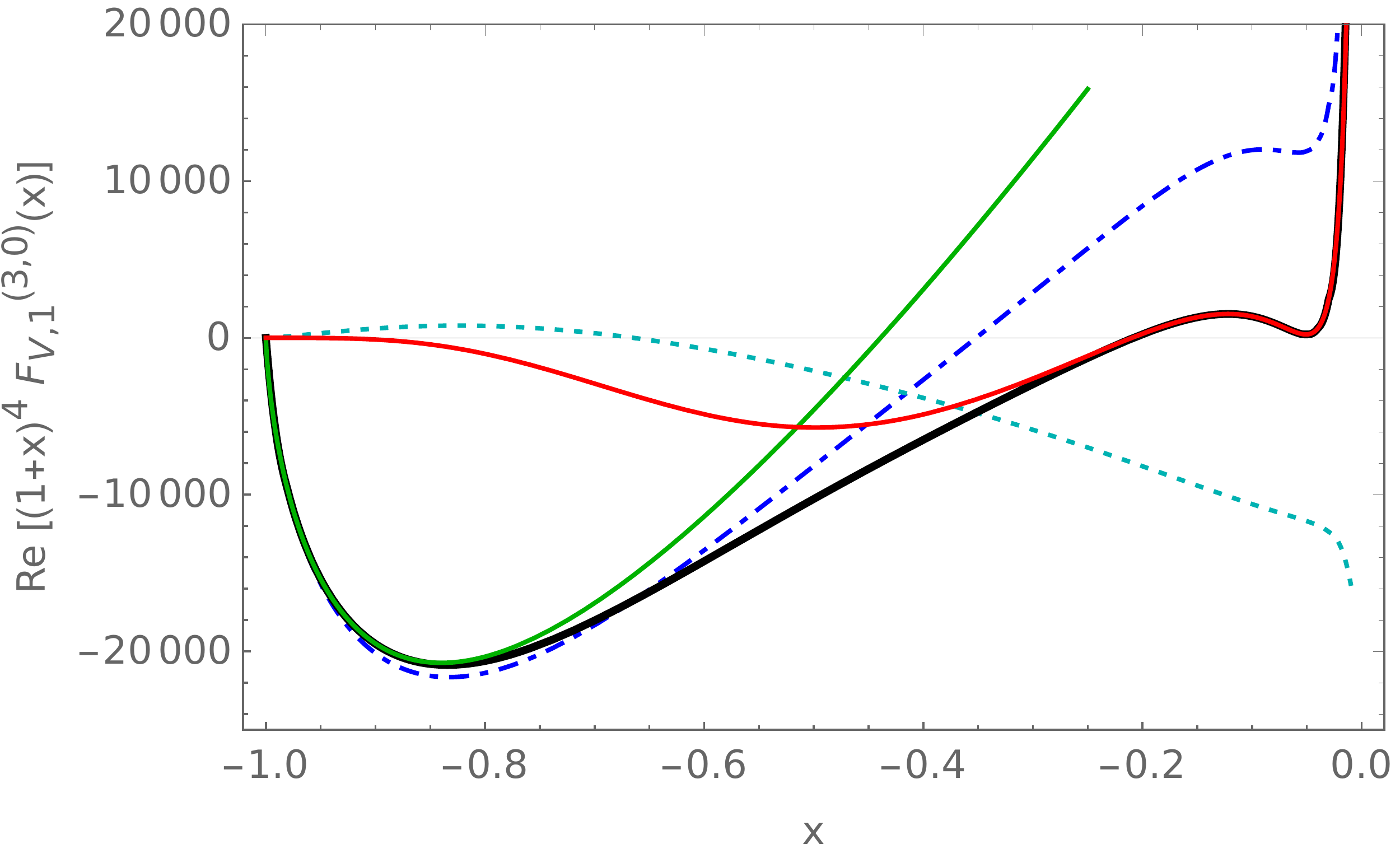}
\includegraphics[width=0.49\textwidth]{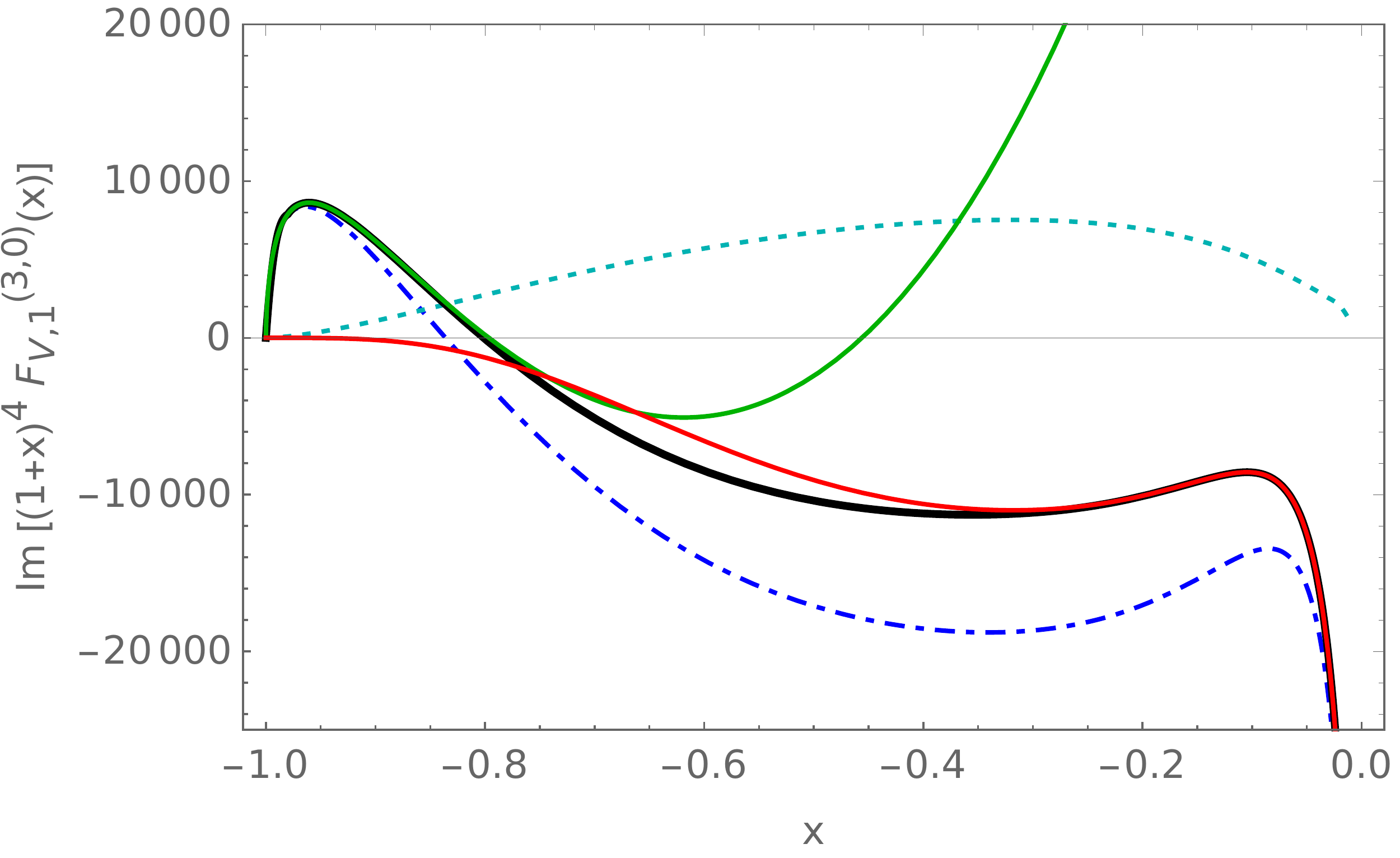}}
\caption{\sf The $O(\varepsilon^0)$ contribution to the vector three-loop form factors 
${\sf Re}[F]_{V,1}^{(3)}$ (left) and ${\sf Im}[F]_{V,1}^{(3)}$ (right) as a function of $x \in [-1,0]$.
Full red line: expansion around $x=0$; Full green line: expansion around $x = -1$; Dotted line: $n_l$-contributions.
Dash-dotted line: leading color contribution of the non--singlet form factor; Full black line: sum of the complete 
non--singlet $n_l$-contributions for $n_l =5$ and the color--planar non--singlet form factor.}
\label{fig:VF34ep0}
\end{figure}

\begin{figure}[H]
\centerline{%
\includegraphics[width=0.49\textwidth]{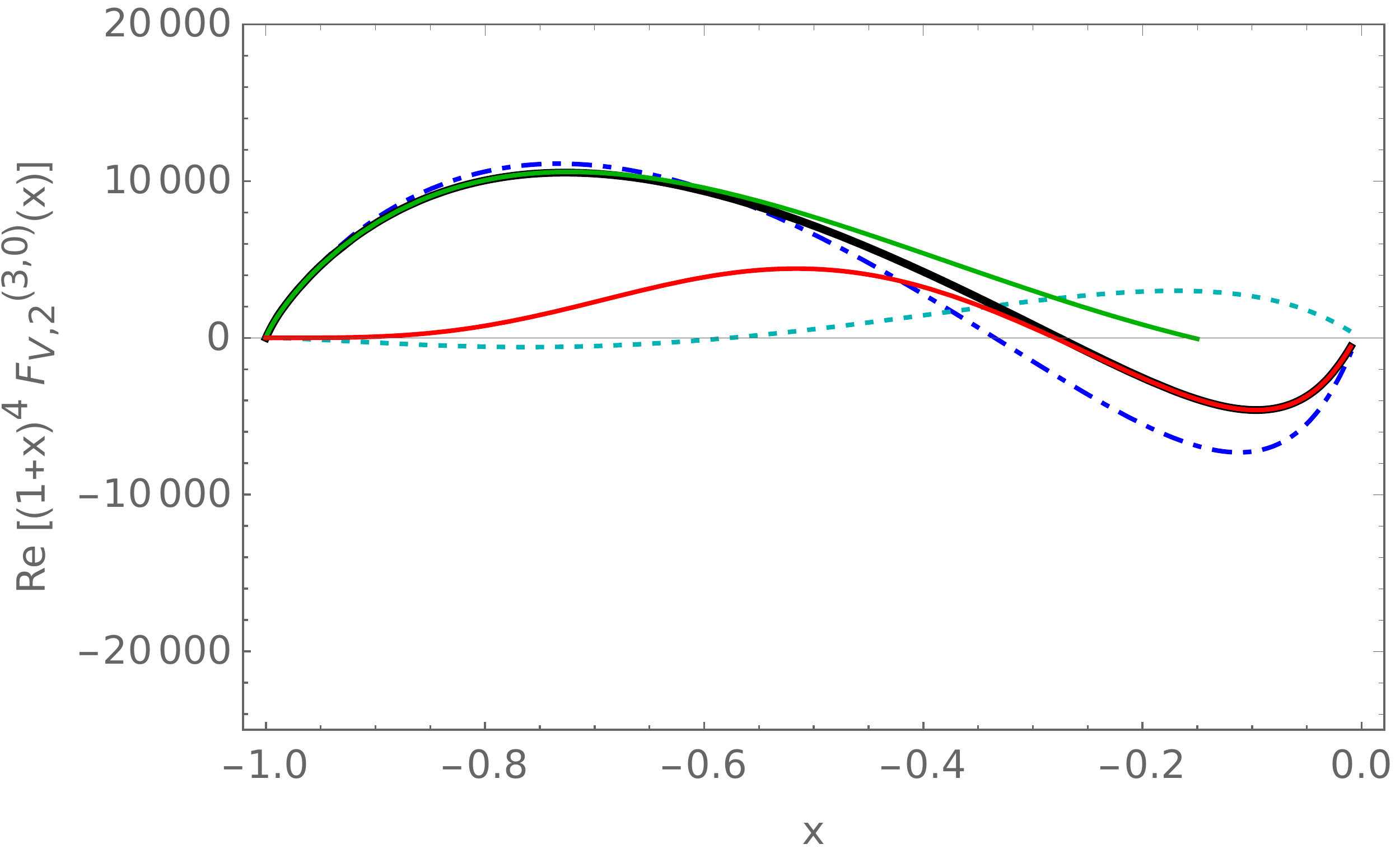}
\includegraphics[width=0.49\textwidth]{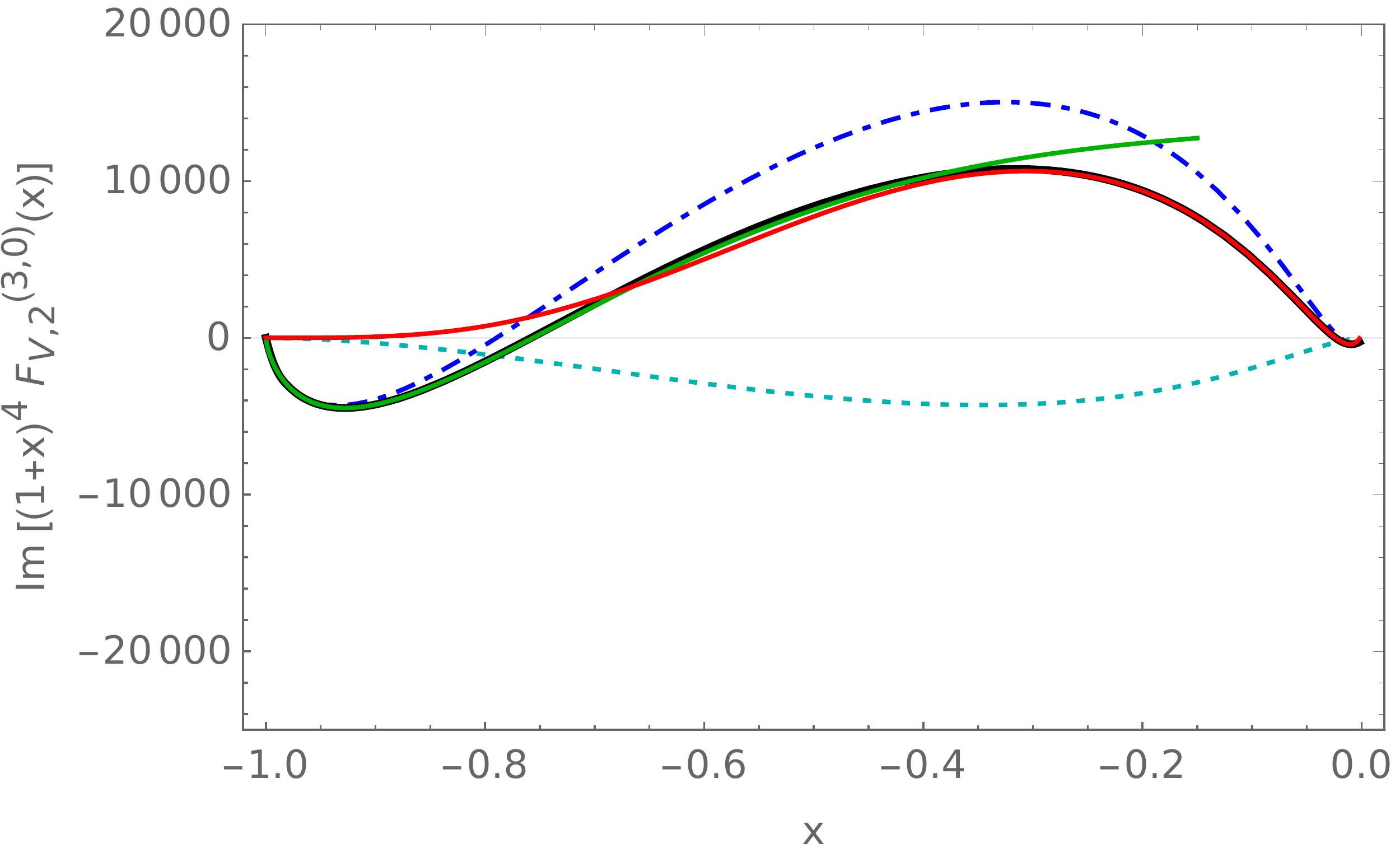}}
\caption{\sf The $O(\varepsilon^0)$ contribution to the vector three-loop form factors 
${\sf Re}[F]_{V,2}^{(3)}$ (left) and ${\sf Im}[F]_{V,2}^{(3)}$ (right) as a function of $x \in [-1,0]$.
Full red line: expansion around $x=0$; Full green line: expansion around $x = -1$; Dotted line: $n_l$-contributions.
Dash-dotted line: leading color contribution of the non--singlet form factor; Full black line: sum of the complete 
non--singlet $n_l$-contributions for $n_l =5$ and the color--planar non--singlet form factor.}
\label{fig:VF56ep0}
\end{figure}

The behaviour of the vector form factors in the region $x \in [-1,0]$ is illustrated in 
Figures~\ref{fig:VF34ep0} and \ref{fig:VF56ep0}. Here the two form factors have a real and imaginary part. The 
threshold expansions around $x=-1$ and the expansion around $x=0$ are also shown. They work in the regions
$x \in [-1,-0.7]$ and $x \in [-0.3,-0]$ respectively. In all cases the $n_l$-contributions are non-negligible
in a wide kinematical range. For the form factor $F_{V,2}^{(3)}$ the threshold expansion works well even
in the region $x \in [-1,-0.5]$ and the expansion around $x=0$ in $[-0.35,0]$.

\vspace*{1cm}

\vspace*{2cm}
%
\section{Numerical implementation for Harmonic and Cylotomic Harmonic Polylogarithms}
\label{sec:5}

\vspace*{1mm}
\noindent
The color--planar part of the three-loop massive form factors depends on 206 cyclotomic harmonic polylogarithms 
(HPLs)
\cite{Ablinger:2011te} up to weight {\sf w=6} and correspondingly the harmonic polylogarithms also up to 
weight {\sf w = 6}. In intermediary 
results for both cases HPLs of {\sf w = 8} appear. A {\tt FORTRAN}-implementation of the harmonic polylogarithms 
to 
{\sf w = 8} has been given in \cite{Ablinger:2018sat}.\footnote{A corresponding 
{\tt Fortran}-program to {\sf w = 5} 
has been given in Ref.~\cite{Gehrmann:2001pz}.} The space of the cyclotomic HPLs already up to  {\sf w = 6}
is very large and therefore we will rather represent the contributing individual functions numerically, 
and do not refer to an associated basis representation. 

The  main argument of the cyclotomic HPLs, $x$, is located in the interval $[-1,1]$ in the present physical 
application. This is going to be the range we are considering in 
the following. In Ref.~\cite{Ablinger:2011te} the range $x \in [0,1]$ was considered. Here the cyclotomic 
HPLs are real-valued. In the extension to $x \in [-1,0[$ some of the cyclotomic HPLs will become complex, as we will 
show below. 
The cyclotomic HPLs are given as iterated integrals over the  letters  (\ref{eq:CYCL}) and those present in the 
usual HPLs, cf.~(\ref{eq:HPL1}).
In the following $\{6,0\}$ and $\{6,1\}$ (resp.\ $\{3,0\}$ and $\{3,1\}$) encode the corresponding cyclotomic 
letters. One obtains e.g.
\begin{eqnarray}
\HA[0, 1, \{6, 1\}, x] = \int_0^x \frac{dx_1}{x_1}  \int_0^{x_1} \frac{dx_2}{1-x_2} \int_0^{x_2} dx_3 
\frac{x_3}{1-x_3+x_3^2}.
\end{eqnarray}

\begin{figure}[H]\centering
\includegraphics[width=0.47\textwidth]{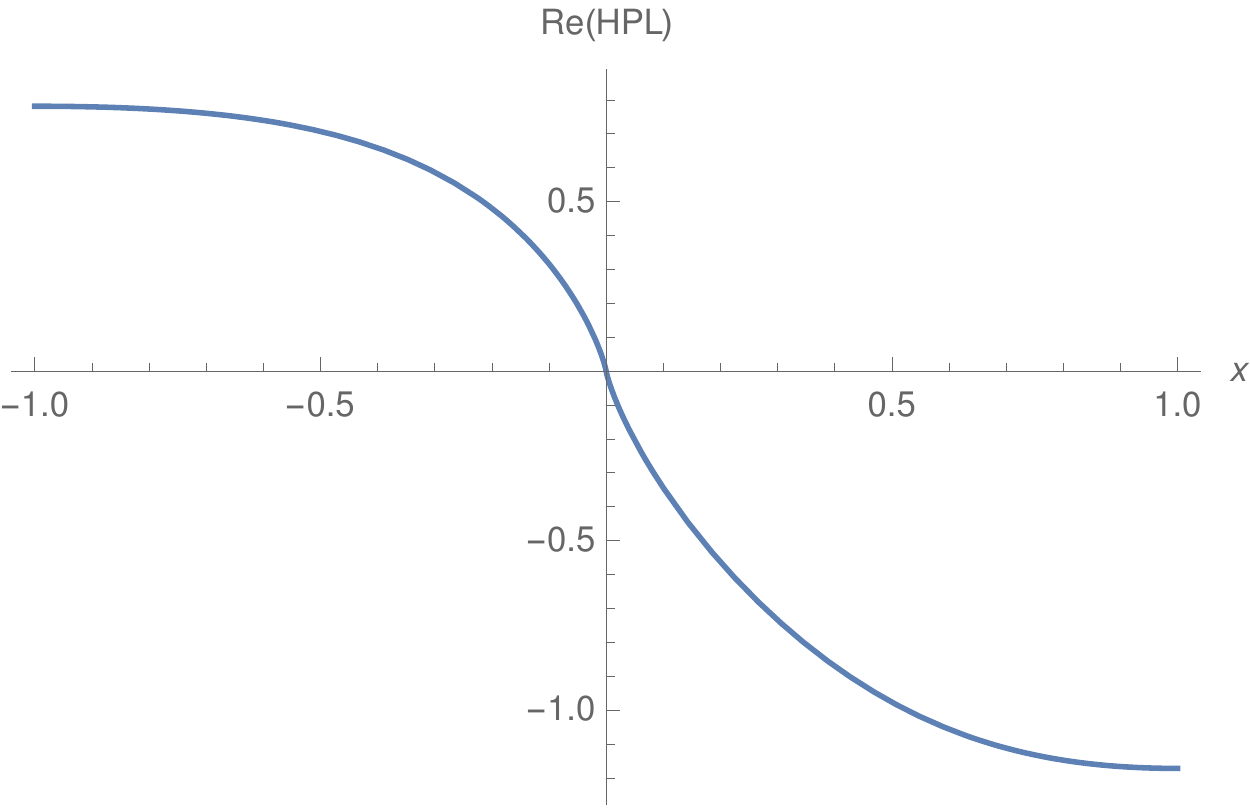}
\includegraphics[width=0.47\textwidth]{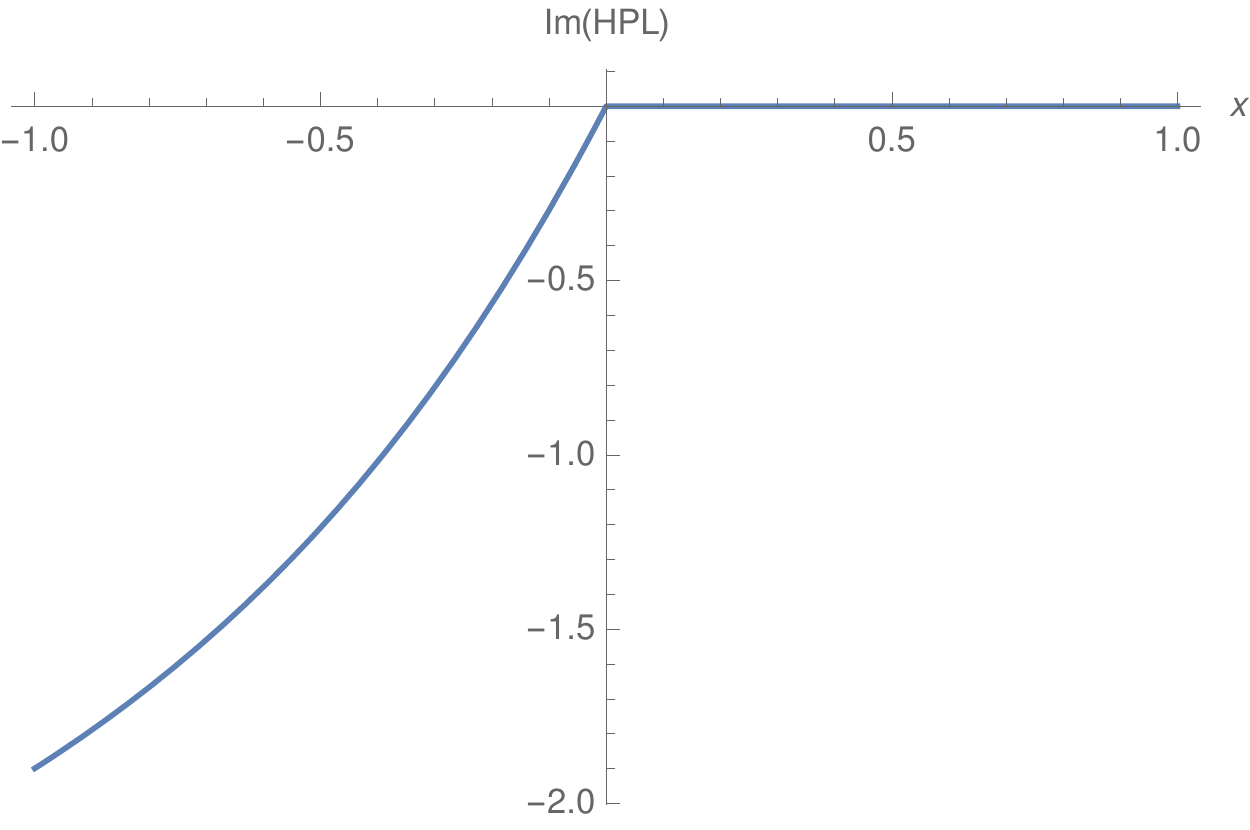}
\caption[]{\sf \small The real (left panel) and imaginary part (right panel) of the function 
{\tt \HA[\{6,0\},0,x]} in the region $x \in [-1,1]$.} \label{Fig_CYCL1}
\end{figure}

\noindent
The following cyclotomic HPLs contribute:
\begin{eqnarray}
&&{\sf w = 1:} \nonumber\\
&&\HA[\{6, 0\}, x], \HA[\{6, 1\}, x] 
\\
&&{\sf w = 2:} \nonumber\\
&&\HA[0, \{6, 0\}, x], \HA[0, \{6, 1\}, x], \HA[\{6, 0\}, 0, x], \HA[\{6, 0\}, 1, x], \HA[\{6, 1\}, 0, x], \HA[\{6, 1\}, 1, x]
\\
&&{\sf w = 3:} \nonumber\\
&&\HA[0, 0, \{6, 0\}, x], \HA[0, 0, \{6, 1\}, x], \HA[0, 1, \{6, 0\}, x],
 \HA[0, 1, \{6, 1\}, x], \HA[0, \{6, 0\}, 1, x], 
\nonumber\\
&&\HA[0, \{6, 1\}, 1, x], \HA[\{6, 0\}, 0, -1, x], \HA[\{6, 0\}, 0, 0, x], \HA[\{6, 0\}, 0, 1, x],
 \HA[\{6, 0\}, 0, \{6, 0\}, x], 
\nonumber\\
&&\HA[\{6, 0\}, 0, \{6, 1\}, x], \HA[\{6, 0\}, 1, 0, x],
 \HA[\{6, 1\}, 0, -1, x], \HA[\{6, 1\}, 0, 0, x], \HA[\{6, 1\}, 0, 1, x],
\nonumber\\ &&
 \HA[\{6, 1\}, 0, \{6, 0\}, x], \HA[\{6, 1\}, 0, \{6, 1\}, x], \HA[\{6, 1\}, 1, 0, x]
\\
&&{\sf w = 4:} \nonumber\\
&& \HA[0, \{6, 0\}, 0, 0, x], \HA[0, \{6, 1\}, 0, 0, x], \HA[\{6, 0\}, 0, -1, 0, x],
 \HA[\{6, 0\}, 0, 0, -1, x], 
\nonumber\\
&& \HA[\{6, 0\}, 0, 0, 0, x], 
\HA[\{6, 0\}, 0, 0, 1, x],
 \HA[\{6, 0\}, 0, 1, 0, x], \HA[\{6, 0\}, 1, 0, 0, x], 
\nonumber\\
&& \HA[\{6, 0\}, 1, 1, 0, x],
 \HA[\{6, 1\}, 0, -1, 0, x], 
\HA[\{6, 1\}, 0, 0, -1, x], \HA[\{6, 1\}, 0, 0, 0, x],
\nonumber\\ 
&&  \HA[\{6, 1\}, 0, 0, 1, x], \HA[\{6, 1\}, 0, 1, 0, x], \HA[\{6, 1\}, 1, 0, 0, x],
\HA[\{6, 1\}, 1, 1, 0, x], 
\\
&&{\sf w = 5:} \nonumber\\
&& \HA[0, \{6, 0\}, 0, -1, 0, x], \HA[0, \{6, 0\}, 0, 0, 0, x]
 \HA[0, \{6, 0\}, 0, 1, 0, x], \HA[0, \{6, 0\}, 1, 0, 0, x],
\nonumber\\ &&
 \HA[0, \{6, 1\}, 0, -1, 0, x], \HA[0, \{6, 1\}, 0, 0, 0, x],
 \HA[0, \{6, 1\}, 0, 1, 0, x], \HA[0, \{6, 1\}, 1, 0, 0, x],
\nonumber\\ &&
 \HA[\{6, 0\}, 0, -1, 0, 0, x], \HA[\{6, 0\}, 0, -1, 0, 1, x],
 \HA[\{6, 0\}, 0, -1, 1, 0, x], \HA[\{6, 0\}, 0, 0, -1, 0, x],
\nonumber\\ &&
 \HA[\{6, 0\}, 0, 0, 0, 0, x], \HA[\{6, 0\}, 0, 0, 0, 1, x],
 \HA[\{6, 0\}, 0, 0, 1, 0, x], \HA[\{6, 0\}, 0, 1, -1, 0, x],
\nonumber\\ &&
 \HA[\{6, 0\}, 0, 1, 0, 0, x], \HA[\{6, 0\}, 0, 1, 0, 1, x],
 \HA[\{6, 0\}, 0, 1, 1, 0, x], \HA[\{6, 0\}, 1, 0, -1, 0, x],
\nonumber\\ &&
 \HA[\{6, 0\}, 1, 0, 0, 0, x], \HA[\{6, 0\}, 1, 0, 0, 1, x],
 \HA[\{6, 0\}, 1, 0, 1, 0, x], \HA[\{6, 0\}, 1, 1, 0, 0, x],
\nonumber\\ &&
 \HA[\{6, 1\}, 0, -1, 0, 0, x], \HA[\{6, 1\}, 0, -1, 0, 1, x],
 \HA[\{6, 1\}, 0, -1, 1, 0, x], \HA[\{6, 1\}, 0, 0, -1, 0, x],
\nonumber\\ &&
 \HA[\{6, 1\}, 0, 0, 0, 0, x], \HA[\{6, 1\}, 0, 0, 0, 1, x],
 \HA[\{6, 1\}, 0, 0, 1, 0, x], \HA[\{6, 1\}, 0, 1, -1, 0, x],
\nonumber\\ &&
 \HA[\{6, 1\}, 0, 1, 0, 0, x], \HA[\{6, 1\}, 0, 1, 0, 1, x],
 \HA[\{6, 1\}, 0, 1, 1, 0, x], \HA[\{6, 1\}, 1, 0, -1, 0, x],
\nonumber\\ &&
 \HA[\{6, 1\}, 1, 0, 0, 0, x], \HA[\{6, 1\}, 1, 0, 0, 1, x],
 \HA[\{6, 1\}, 1, 0, 1, 0, x], \HA[\{6, 1\}, 1, 1, 0, 0, x]
\\
&&{\sf w = 6:} \nonumber\\
&& \HA[0, 0, \{6, 0\}, 0, -1, 0, x], \HA[0, 0, \{6, 0\}, 0, 0, 0, x], 
 \HA[0, 0, \{6, 0\}, 0, 1, 0, x], 
\nonumber\\ &&
\HA[0, 0, \{6, 0\}, 1, 0, 0, x],
 \HA[0, 0, \{6, 1\}, 0, -1, 0, x], \HA[0, 0, \{6, 1\}, 0, 0, 0, x], 
\nonumber\\ &&
 \HA[0, 0, \{6, 1\}, 0, 1, 0, x], \HA[0, 0, \{6, 1\}, 1, 0, 0, x], 
 \HA[0, 1, \{6, 0\}, 0, -1, 0, x], 
\nonumber\\ &&
\HA[0, 1, \{6, 0\}, 0, 0, 0, x], \HA[0, 1, \{6, 0\}, 0, 1, 0, x], \HA[0, 1, \{6, 0\}, 1, 0, 0, x], 
\nonumber\\ &&
 \HA[0, 1, \{6, 1\}, 0, -1, 0, x], \HA[0, 1, \{6, 1\}, 0, 0, 0, x], 
 \HA[0, 1, \{6, 1\}, 0, 1, 0, x], 
\nonumber\\ &&
\HA[0, 1, \{6, 1\}, 1, 0, 0, x], 
 \HA[0, \{6, 0\}, 0, -1, 0, 0, x], \HA[0, \{6, 0\}, 0, -1, 0, 1, x], 
\nonumber\\ &&
 \HA[0, \{6, 0\}, 0, -1, 1, 0, x], \HA[0, \{6, 0\}, 0, 0, -1, 0, x], 
 \HA[0, \{6, 0\}, 0, 0, 0, 0, x], 
\nonumber\\ &&
\HA[0, \{6, 0\}, 0, 0, 0, 1, x], 
 \HA[0, \{6, 0\}, 0, 0, 1, 0, x], \HA[0, \{6, 0\}, 0, 1, -1, 0, x], 
\nonumber\\ &&
 \HA[0, \{6, 0\}, 0, 1, 0, 0, x], \HA[0, \{6, 0\}, 0, 1, 0, 1, x], 
 \HA[0, \{6, 0\}, 0, 1, 1, 0, x], 
\nonumber\\ &&
\HA[0, \{6, 0\}, 1, 0, -1, 0, x], 
 \HA[0, \{6, 0\}, 1, 0, 0, 0, x], \HA[0, \{6, 0\}, 1, 0, 0, 1, x], 
\nonumber\\ &&
 \HA[0, \{6, 0\}, 1, 0, 1, 0, x], \HA[0, \{6, 0\}, 1, 1, 0, 0, x], 
 \HA[0, \{6, 1\}, 0, -1, 0, 0, x], 
\nonumber\\ &&
\HA[0, \{6, 1\}, 0, -1, 0, 1, x], 
 \HA[0, \{6, 1\}, 0, -1, 1, 0, x], \HA[0, \{6, 1\}, 0, 0, -1, 0, x], 
\nonumber\\ &&
 \HA[0, \{6, 1\}, 0, 0, 0, 0, x], \HA[0, \{6, 1\}, 0, 0, 0, 1, x], 
 \HA[0, \{6, 1\}, 0, 0, 1, 0, x], 
\nonumber\\ &&
\HA[0, \{6, 1\}, 0, 1, -1, 0, x], 
 \HA[0, \{6, 1\}, 0, 1, 0, 0, x], \HA[0, \{6, 1\}, 0, 1, 0, 1, x], 
\nonumber\\ &&
 \HA[0, \{6, 1\}, 0, 1, 1, 0, x], \HA[0, \{6, 1\}, 1, 0, -1, 0, x], 
 \HA[0, \{6, 1\}, 1, 0, 0, 0, x], 
\nonumber\\ &&
\HA[0, \{6, 1\}, 1, 0, 0, 1, x], 
 \HA[0, \{6, 1\}, 1, 0, 1, 0, x], \HA[0, \{6, 1\}, 1, 1, 0, 0, x], 
\nonumber\\ &&
 \HA[\{6, 0\}, 0, -1, -1, 0, 0, x], \HA[\{6, 0\}, 0, -1, 0, -1, 0, x], 
 \HA[\{6, 0\}, 0, -1, 0, 0, 0, x], 
\nonumber\\ &&
\HA[\{6, 0\}, 0, -1, 0, 0, 1, x], 
 \HA[\{6, 0\}, 0, -1, 0, 1, 0, x], \HA[\{6, 0\}, 0, -1, 1, 0, 0, x], 
\nonumber\\ &&
 \HA[\{6, 0\}, 0, 0, -1, -1, 0, x], \HA[\{6, 0\}, 0, 0, -1, 0, 0, x], 
 \HA[\{6, 0\}, 0, 0, -1, 0, 1, x], 
\nonumber\\ &&
\HA[\{6, 0\}, 0, 0, -1, 1, 0, x], 
 \HA[\{6, 0\}, 0, 0, 0, -1, 0, x], \HA[\{6, 0\}, 0, 0, 0, 0, 0, x], 
\nonumber\\ &&
 \HA[\{6, 0\}, 0, 0, 0, 0, 1, x], \HA[\{6, 0\}, 0, 0, 0, 1, 0, x], 
 \HA[\{6, 0\}, 0, 0, 1, -1, 0, x], 
\nonumber\\ &&
\HA[\{6, 0\}, 0, 0, 1, 0, 0, x], 
 \HA[\{6, 0\}, 0, 0, 1, 0, 1, x], \HA[\{6, 0\}, 0, 0, 1, 1, 0, x], 
\nonumber\\ &&
 \HA[\{6, 0\}, 0, 1, -1, 0, 0, x], \HA[\{6, 0\}, 0, 1, 0, -1, 0, x], 
 \HA[\{6, 0\}, 0, 1, 0, 0, 0, x], 
\nonumber\\ &&
\HA[\{6, 0\}, 0, 1, 0, 0, 1, x], 
 \HA[\{6, 0\}, 0, 1, 0, 1, 0, x], \HA[\{6, 0\}, 0, 1, 1, 0, 0, x], 
\nonumber\\ &&
 \HA[\{6, 0\}, 0, \{6, 0\}, 0, -1, 0, x], \HA[\{6, 0\}, 0, \{6, 0\}, 0, 0, 0, x], 
 \HA[\{6, 0\}, 0, \{6, 0\}, 0, 1, 0, x], 
\nonumber\\ &&
\HA[\{6, 0\}, 0, \{6, 0\}, 1, 0, 0, x], 
 \HA[\{6, 0\}, 0, \{6, 1\}, 0, -1, 0, x], \HA[\{6, 0\}, 0, \{6, 1\}, 0, 0, 0, x], 
\nonumber\\ &&
 \HA[\{6, 0\}, 0, \{6, 1\}, 0, 1, 0, x], \HA[\{6, 0\}, 0, \{6, 1\}, 1, 0, 0, x], 
 \HA[\{6, 0\}, 1, 0, -1, 0, 0, x], 
\nonumber\\ &&
\HA[\{6, 0\}, 1, 0, 0, -1, 0, x], 
 \HA[\{6, 0\}, 1, 0, 0, 0, 0, x], \HA[\{6, 0\}, 1, 0, 0, 1, 0, x], 
\nonumber\\ &&
 \HA[\{6, 0\}, 1, 0, 1, 0, 0, x], \HA[\{6, 0\}, 1, 1, 0, 0, 0, x], 
 \HA[\{6, 1\}, 0, -1, -1, 0, 0, x], 
\nonumber\\ &&
\HA[\{6, 1\}, 0, -1, 0, -1, 0, x], 
 \HA[\{6, 1\}, 0, -1, 0, 0, 0, x], \HA[\{6, 1\}, 0, -1, 0, 0, 1, x], 
\nonumber\\ &&
 \HA[\{6, 1\}, 0, -1, 0, 1, 0, x], \HA[\{6, 1\}, 0, -1, 1, 0, 0, x], 
 \HA[\{6, 1\}, 0, 0, -1, -1, 0, x], 
\nonumber\\ &&
\HA[\{6, 1\}, 0, 0, -1, 0, 0, x], 
 \HA[\{6, 1\}, 0, 0, -1, 0, 1, x], \HA[\{6, 1\}, 0, 0, -1, 1, 0, x], 
\nonumber\\ &&
 \HA[\{6, 1\}, 0, 0, 0, -1, 0, x], \HA[\{6, 1\}, 0, 0, 0, 0, 0, x], 
 \HA[\{6, 1\}, 0, 0, 0, 0, 1, x], 
\nonumber\\ &&
\HA[\{6, 1\}, 0, 0, 0, 1, 0, x], 
 \HA[\{6, 1\}, 0, 0, 1, -1, 0, x], \HA[\{6, 1\}, 0, 0, 1, 0, 0, x], 
\nonumber\\ &&
 \HA[\{6, 1\}, 0, 0, 1, 0, 1, x], \HA[\{6, 1\}, 0, 0, 1, 1, 0, x], 
 \HA[\{6, 1\}, 0, 1, -1, 0, 0, x], 
\nonumber\\ &&
\HA[\{6, 1\}, 0, 1, 0, -1, 0, x], 
 \HA[\{6, 1\}, 0, 1, 0, 0, 0, x], \HA[\{6, 1\}, 0, 1, 0, 0, 1, x], 
\nonumber\\ &&
 \HA[\{6, 1\}, 0, 1, 0, 1, 0, x], \HA[\{6, 1\}, 0, 1, 1, 0, 0, x], 
 \HA[\{6, 1\}, 0, \{6, 0\}, 0, -1, 0, x], 
\nonumber\\ &&
\HA[\{6, 1\}, 0, \{6, 0\}, 0, 0, 0, x], 
 \HA[\{6, 1\}, 0, \{6, 0\}, 0, 1, 0, x], \HA[\{6, 1\}, 0, \{6, 0\}, 1, 0, 0, x], 
\nonumber\\ &&
 \HA[\{6, 1\}, 0, \{6, 1\}, 0, -1, 0, x], \HA[\{6, 1\}, 0, \{6, 1\}, 0, 0, 0, x], 
 \HA[\{6, 1\}, 0, \{6, 1\}, 0, 1, 0, x], 
\nonumber\\ &&
\HA[\{6, 1\}, 0, \{6, 1\}, 1, 0, 0, x], 
 \HA[\{6, 1\}, 1, 0, -1, 0, 0, x], \HA[\{6, 1\}, 1, 0, 0, -1, 0, x], 
\nonumber\\ &&
 \HA[\{6, 1\}, 1, 0, 0, 0, 0, x], \HA[\{6, 1\}, 1, 0, 0, 1, 0, x], 
 \HA[\{6, 1\}, 1, 0, 1, 0, 0, x], 
\nonumber\\ && \HA[\{6, 1\}, 1, 1, 0, 0, 0, x]. 
\end{eqnarray} 
As we mentioned already, some of the above cyclotomic HPLs become complex in the region $x \in [-1,0[$. An example 
is {\tt \HA[\{6, 0\}, 0, x]}, which we illustrate in Figure~\ref{Fig_CYCL1}. In the series expansion the imaginary part
results from the constant $\HA[0,-1] = i \pi$.

To prepare an arbitrary polynomial expression out of usual HPLs \cite{Remiddi:1999ew} and the above cyclotomic HPLs 
one runs the {\tt Mathematica} notebook {\tt CHPL\_prepare.nb}. It rewrites the usual HPLs and the cyclotomic 
HPLs from the form given by {\tt HarmonicSums} into the form the required for {\tt HPOLY.f} 
\cite{Ablinger:2018sat} up to {\sf w = 8} and the cyclotomic HPLs used in {\tt CPOLY.f}. The latter ones are 
given in the form
\begin{eqnarray}
{\tt CHPn(J,X),~~~n = 1...6},
\end{eqnarray}
with the argument ${\tt X} \in [-1,1]$. The index {\tt J} denotes the place of the respective cyclotomic HPL in the 
above lists, i.e. ${\tt \HA[0, \{6, 0\}, 1, x] = CHP3(5,X)}$. 
The numerical representation follows Ref.~\cite{Ablinger:2018sat}, Section~4. For {\tt n=1} there are no 
logarithmic 
contributions. In all other cases there are contributions of $\ln^{\tt n-1}(x)$ for the expansion around $x=0$, also 
implying imaginary parts for $x < 0$. 
For ${\tt 6 \geq n \geq 2}$ there are contributions of $\ln(1-x)$ for the expansion around $x=1$. Likewise, 
one has logarithmic contributions up to $O(\ln^2(1+x))$ expanding around $x = -1$.
The numerical representations have been derived by mutual 
use of
the package {\tt HarmonicSums}
\cite{Ablinger:2014rba, Ablinger:2010kw, Ablinger:2013hcp, Ablinger:2011te, Ablinger:2013cf, Ablinger:2014bra}, 
referring to the 
MZV data mine \cite{Blumlein:2009cf} and evaluating other special constants using {\tt Ginac} \cite{Vollinga:2004sn} 
numerically. The numerical performance of the cyclotomic HPLs in {\tt FORTRAN} turns out to be faster than a 
corresponding (complex-valued) representation in {\tt Ginac} \cite{Vollinga:2004sn,Bauer:2000cp}. 
\noindent
On the other 
hand, 
the former one is limited to double precision, while the latter one can be extended to arbitrary precision. 
We have tested the numerical implementation of the real-valued cyclotomic HPLs in {\tt CPOLY.f} comparing to the 
corresponding results obtained by corresponding numerical results provided by {\tt Ginac}. 
The code is compiled by {\tt gfortran CPOLY.f}.
The representation has an accuracy of
\begin{equation}
\sim 2 \cdot 10^{-15}
\end{equation}
and better. The reading of the data needed in {\tt CPOLY.f} requires $5.8 \cdot 10^{-2}$ sec. The calculation 
of all 206 cyclotomic HPLs at a given value of $x$ is performed in $2.3 \cdot 10^{-3}$ sec or faster.
In using the code 
{\tt CPOLY.f} only the subroutines {\tt UCPOLYIN} and {\tt UCPOLY} are user routines to provide further input 
and to perform the calculation, respectively. Any use of the code {\tt CPOLY.f} requires to quote the present 
paper.\footnote{After completion of this paper another numeric implementation in {\tt Mathematica} of cyclotomic 
harmonic polylogarithms appeared in \cite{KNIEHL1}.}
\section{Conclusion}
\label{sec:6}

\vspace*{1mm}
\noindent
We presented an algorithm to solve single-variate systems of differential equations, factorizing at first 
order and depending on the dimensional parameter $\ep$, analytically. Here no choice of a special basis 
representation is required. The Laurent expansion in the parameter $\ep$ leads to a one-variable problem.
We considered differential equations with rational coefficients in $x$ and $\ep$. The
algorithm solves these systems in terms of iterative integrals over finite alphabets and 
rational terms to any order in the dimensional parameter $\ep$. This method can be applied to a wide range of 
problems in Quantum Field Theory, after one knows whether the corresponding systems factorize to first order, which is 
checked by the present algorithm. 

In the example of the massive three--loop form factors the emerging letters are those forming the HPLs and the 
cyclotomic
HPLs at cyclotomy {\sf c = 3,4} and {\sf 6}. The homogeneous solutions are the same for any order in $\ep$. The 
corresponding inhomogeneities then determine the respective inhomogeneous solutions using the variation of constants.
The iterative-integral structure is preserved by the latter operation, as can be shown by integration-by-parts.
Besides the harmonic and cyclotomic harmonic polylogarithms up to weight {\sf w = 6} also associated special 
constants appear. In the cyclotomic case not all their relations have been proven yet by analytic methods. However,
a series of relations has been conjectured by using {\tt PSLQ} \cite{Henn:2015sem}. Assuming that these 
relations would hold, the results at three--loop order presented in this paper can finally be expressed by 
very few multiple zeta values only
\begin{equation}
\left\{ \ln(2), \zeta_2, \zeta_3, \Li_4\left(\frac{1}{2}\right), \zeta_5 \right\}
\end{equation}
and no special cyclotomic constants contribute. However, cyclotomic constants remain in the expansion around
$x = -1$.

Our result for the vector form factors agree with those given in Ref.~\cite{Henn:2016kjz,Henn:2016tyf,Lee:2018nxa}.
We provide the {\tt FORTRAN}-code {\tt CPOLY.f} which allows to calculate the cyclotomic harmonic polylogarithms 
contributing to all massive three-loop form factors in the color--planar limit.

\vspace{2ex}
\noindent
{\bf Acknowledgment.}~We would like to thank M.~Round and K.~Sch\"onwald for discussions. This work was 
supported in part by the Austrian Science Fund (FWF) grant SFB F50 (F5009-N15), by the bilateral project 
DNTS-Austria 01/3/2017 (WTZ BG03/2017), funded by the Bulgarian National Science Fund and OeAD (Austria), 
by the EU TMR network SAGEX Marie Sk\l{}odowska-Curie grant agreement No. 764850 and COST action CA16201: 
Unraveling new physics at the LHC through the precision frontier. The Feynman diagrams have been drawn 
using {\tt Axodraw} \cite{Vermaseren:1994je}.

\newpage

\end{document}